\documentclass[lettersize,journal]{IEEEtran}

\usepackage{amsmath,amsfonts}
\usepackage{amssymb}
\usepackage{amsbsy}
\usepackage{amsthm}
\usepackage{array}

\usepackage{graphicx}
\usepackage{graphbox}
\usepackage{multirow}
\usepackage{booktabs}
\usepackage{makecell}
\usepackage{float}
\usepackage[caption=false,font=normalsize,labelfont=sf,textfont=sf]{subfig}

\usepackage{algorithm}
\usepackage{algpseudocode}

\usepackage{textcomp}
\usepackage{stfloats}
\usepackage{url}
\usepackage{verbatim}
\usepackage{lipsum}
\usepackage{tikz}
\usepackage{xcolor}
\usepackage{balance}
\usepackage{cite}
\usepackage{hyperref}
\usepackage{cleveref}
\usepackage{caption}
\hypersetup{
	colorlinks=false,
	pdfborder={0 0 1},
	linkbordercolor={1 0 0},
	citebordercolor={1 0 0},
	urlbordercolor={1 0 0}
}

\def\BibTeX{{\rm B\kern-.05em{\sc i\kern-.025em b}\kern-.08em
		T\kern-.1667em\lower.7ex\hbox{E}\kern-.125emX}}

\newcommand{\thickrule}{\specialrule{\heavyrulewidth}{0pt}{0pt}}
\newcommand{\doublerule}{%
	\specialrule{\heavyrulewidth}{0pt}{0pt}%
	\specialrule{\heavyrulewidth}{3pt}{0pt}%
}

\makeatletter
\renewenvironment{proof}[1][\proofname]{\par
	\normalfont
	\topsep6\p@\@plus6\p@
	\trivlist
	\item[\hskip\labelsep\hskip\parindent\itshape #1:] \ignorespaces
}{%
	\qed\endtrivlist
}
\makeatother

\newcounter{theorem}
\renewcommand{\thetheorem}{\arabic{theorem}}

\newenvironment{theorem}[1][]{%
	\refstepcounter{theorem}%
	\par\vspace{3pt}%
	\noindent\hspace*{\parindent}%
	{\textit{Theorem~\thetheorem:}}\hspace{0.5em}%
	\normalfont
}{%
	\par\vspace{3pt}
}

\begin{document}
	\title{SVD-Based UGRM-GFT on Directed Product Graphs}
	\author{Guoyun~Xie and Zhichao~Zhang,~\IEEEmembership{Member,~IEEE}
		\thanks{This work was supported in part by the Open Foundation of Hubei Key Laboratory of Applied Mathematics (Hubei University) under Grant HBAM202404; and in part by the Foundation of Key Laboratory of System Control and Information Processing, Ministry of Education under Grant Scip20240121. \emph{(Corresponding author: Zhichao~Zhang.)}}
		\thanks{Guoyun~Xie is with the School of Mathematics and Statistics, Nanjing University of Information Science and Technology, Nanjing 210044, China (e-mail: xgy10260@163.com).}
		\thanks{Zhichao~Zhang is with the School of Mathematics and Statistics, Nanjing University of Information Science and Technology, Nanjing 210044, China, with the Hubei Key Laboratory of Applied Mathematics, Hubei University, Wuhan 430062, China, and also with the Key Laboratory of System Control and Information Processing, Ministry of Education, Shanghai Jiao Tong University, Shanghai 200240, China (e-mail: zzc910731@163.com).}}

	\maketitle
	\begin{abstract}
		Traditional directed graph signal processing generally depends on fixed representation matrices, whose rigid structures limit the model's ability to adapt to complex graph topologies. To address this issue, this study employs the unified graph representation matrix (UGRM) to propose a generalized graph Fourier transform (UGRM-GFT) based on singular value decomposition (SVD) for signal analysis on directed graphs and Cartesian product graphs. We define UGRM-GFT for general directed graphs by introducing a parameterized UGRM that incorporates traditional representations such as the Laplacian and adjacency matrices. The SVD is used to construct spectral transform pairs with both left and right singular vectors, ensuring numerical stability. We extend this approach to two types of UGRM-GFTs applied to directed Cartesian product graphs: UGRM-GFT-I performs SVD directly on the composite UGRM matrix, suitable for globally coupled signals, while UGRM-GFT-II separately applies SVD to the factor graph UGRMs, significantly reducing computational complexity while preserving spectral expressiveness. Theoretical analysis establishes approximation error bounds and characterizes the spectral behavior of the proposed method with respect to the parameters $\alpha$ and $k$ embedded in the UGRM. Experimental results on real-world datasets demonstrate that the proposed method achieves superior energy compaction and significantly outperforms traditional fixed-matrix approaches in denoising tasks. Specifically, it attains a higher signal-to-noise ratio, and a lower bandlimiting approximation error.
	\end{abstract}
	
	
	\section{Introduction}
	\IEEEPARstart 
    {W}{ith} {the continuous advancement of big data and complex system research, graph-structured data have been widely applied in numerous fields such as sensor networks\cite{ref1,ref2}, social networks\cite{ref3}, genetics \cite{ref4}, neuroscience\cite{ref5}, and image processing\cite{ref6,ref7}. Such data typically exist in non-Euclidean spaces, where internal relationships are depicted by nodes and edges, complicating the direct application of traditional signal processing methods. Graph signal processing (GSP) has emerged as an extension of classical tools such as Fourier transforms, filtering, and wavelet analysis to the graph domain, offering a unified framework for the representation and analysis of irregularly structured data\cite{ref8,ref9,ref10,ref11,ref12,ref13,ref14,ref15}.

    In directed graphs, the directionality of edges signifies asymmetric relationships, including information flow, social influence, or causal dependencies. Directed graphs are extensively employed to represent network interaction frameworks characterized by asymmetrical relationships among agents, such as individuals and organizations in social networks, as well as leaders and followers in citation networks\cite{ref16,ref17,ref18}. Therefore, signal processing on directed graphs possesses considerable theoretical and practical significance.

    The graph Fourier transform (GFT) is a core tool in graph signal processing, enabling the decomposition of graph signals into frequency components that represent different variation patterns of the graph structure\cite{ref19,ref20,ref21,ref22}. Existing GFT methods primarily categorize into two types: (i) adjacency matrix-based eigen decomposition, wherein eigenvectors function as the Fourier basis and eigenvalues denote frequencies\cite{ref20}. (ii) Laplacian matrix-based eigen decomposition, employing its eigenvectors to establish an orthogonal frequency domain basis. Furthermore, in-degree matrices, signless Laplacian matrices, and normalized Laplacian matrices can also be utilized to formulate GFTs\cite{ref9}. Despite subsequent research suggesting parameterized graph representation matrices to improve flexibility, these approaches typically do not effectively apply to directed graph contexts\cite{ref23,ref24,ref25}.

    The GFT on directed graphs is crucial for identifying patterns in social networks, measuring the influence of individuals and communities, and understanding network dynamics\cite{ ref26,ref27,ref28,ref29,ref30,ref31,ref32,ref33}. In recent years, researchers have proposed multiple definitions of GFT, predominantly derived from the graph's Laplacian or adjacency matrices. Among these, the Jordan decomposition of the Laplacian matrix is a prevalent method for defining GFT; however, it frequently experiences issues with numerical stability and elevated computational complexity\cite{ref19,ref20,ref21,ref27}. To address these issues, subsequent research presented the magnetic Laplacian matrix $\mathbf{L}_q$ ($q \geq 0$), a positive semi-definite Hermitian matrix that effectively encapsulates edge directional relationships in directed graphs via phases in the complex plane\cite{ref34,ref35}. Furthermore, techniques utilizing the singular value decomposition (SVD) of the Laplacian matrix have garnered interest owing to their superior numerical stability and reduced computational complexity\cite{ref26}.

    Spatiotemporal signals, which inherently exist on product graphs, frequently do not adequately represent the correlation attributes across various dimensions when modeled on a singular directed graph\cite{ ref36, ref37, ref38, ref39,ref40,ref41,ref42,ref43}. To address this issue, recent research has commenced the extension of the GFT framework to directed product graphs. Cheng et al. proposed two novel SVD-based GFTs for directed Cartesian product graphs\cite{ref44}. These methods employ the singular value decomposition of the Laplacian matrices of the constituent graphs, thereby maintaining Parseval's identity and numerical stability while also reverting to the classical joint GFT for undirected graphs. Moreover, they exhibit effective energy compaction when handling signals characterized by robust spatiotemporal correlations and surpass GFT methods reliant on the magnetic Laplacian in denoising applications. Nevertheless, the majority of current GFT methods rely on static graph shift operators, such as Laplacian or adjacency matrices, which lack the adaptability to accommodate alterations in graph structure, thereby constraining the model's expressive capacity and applicability in practical graph data contexts.

    To address these limitations, the unified graph representation matrix (UGRM) has been proposed as a parameterized generalized graph representation framework\cite{ref25}. Through the introduction of tunable parameters $\alpha$ and $k$, UGRM integrates traditional representation matrices, including the Laplacian matrix, adjacency matrix, and degree matrix, into a unified expression, thereby enhancing its adaptability to complex graph structures while maintaining mathematical precision. Inspired by this, this study proposes two GFTs on directed Cartesian product graphs. The method utilizes the SVD of UGRM, combining left and right singular vectors to create spectral transform pairs, thereby ensuring numerical stability and improving adaptive representation capability for graph topology.

The contributions of this paper are as follows:

\begin{itemize}
	\item We utilize the UGRM to construct the SVD-based generalized GFT (UGRM-GFT), 
	which is compatible with traditional graph matrices and accommodates complex directed 
	graph structures. We further proposed two implementation schemes for directed Cartesian 
	product graphs: UGRM-GFT-I, a global decomposition based on the holistic UGRM, and 
	UGRM-GFT-II, a separable decomposition based on the SVDs of the factor graph UGRMs. 
	In particular, UGRM-GFT-II significantly reduces computational complexity from 
	$\mathcal{O}(N_1^3 N_2^3)$ to $\mathcal{O}(N_1^3 + N_2^3)$ while maintaining 
	spectral expressive capability.
	
	\item Established approximation error bounds for UGRM-GFT and characterized the 
	spectral behavior with respect to parameters $\alpha$ and $k$, providing a theoretical 
	foundation for spectral ordering and parameter selection; derived explicit closed-form 
	expressions for both forward and inverse transformations, facilitating practical 
	applications such as graph signal denoising.
	
	\item Experiments conducted on multiple real-world datasets (sea surface temperature~(SST), PM-25, and 
	COVID) revealed that UGRM-GFT consistently outperforms traditional fixed-matrix 
	methods in denoising tasks, attaining a higher signal-to-noise ratio (SNR) and a lower 
	bandlimiting approximation error (BAE) across varying noise levels.
\end{itemize}
	This study is organized as follows: Section \ref{sec:chapter2} presents fundamental background information, encompassing UGRM, directed Cartesian product graphs, and current SVD-based GFT methodologies. Section \ref{sec:chapter3} presents the principal contribution: an innovative SVD-based UGRM-GFT framework applicable to general directed graphs, providing comprehensive definitions and theoretical properties of its forward and inverse transformations. Section \ref{sec:chapter4} explicitly applies this framework to directed Cartesian product graphs, executing SVD decomposition directly on the UGRM of the composite graph structure. This method is appropriate for globally coupled signals, and its computational complexity has been examined. Section \ref{sec:chapter5} proposes decomposing the UGRM of factor graphs into distinct SVD components before recombination, thereby markedly reducing computational complexity. Section \ref{sec:chapter6} substantiates the proposed methodology through comprehensive experiments on real-world datasets (SST, PM-25, and COVID), illustrating its considerable superiority over conventional fixed-matrix techniques in denoising tasks. Section \ref{sec:chapter7}  summarizes the present study and delineates prospective avenues for future research. Supporting evidence is provided in the appendix.
	
	Fig. \ref{fig:fullwidthpdf} systematically elaborates on the developmental trajectory, theoretical framework, and practical applications of UGRM-GFT. 
	
	\begin{figure*}[htbp]
		\centering 
		
		\includegraphics[width=1.0\textwidth]{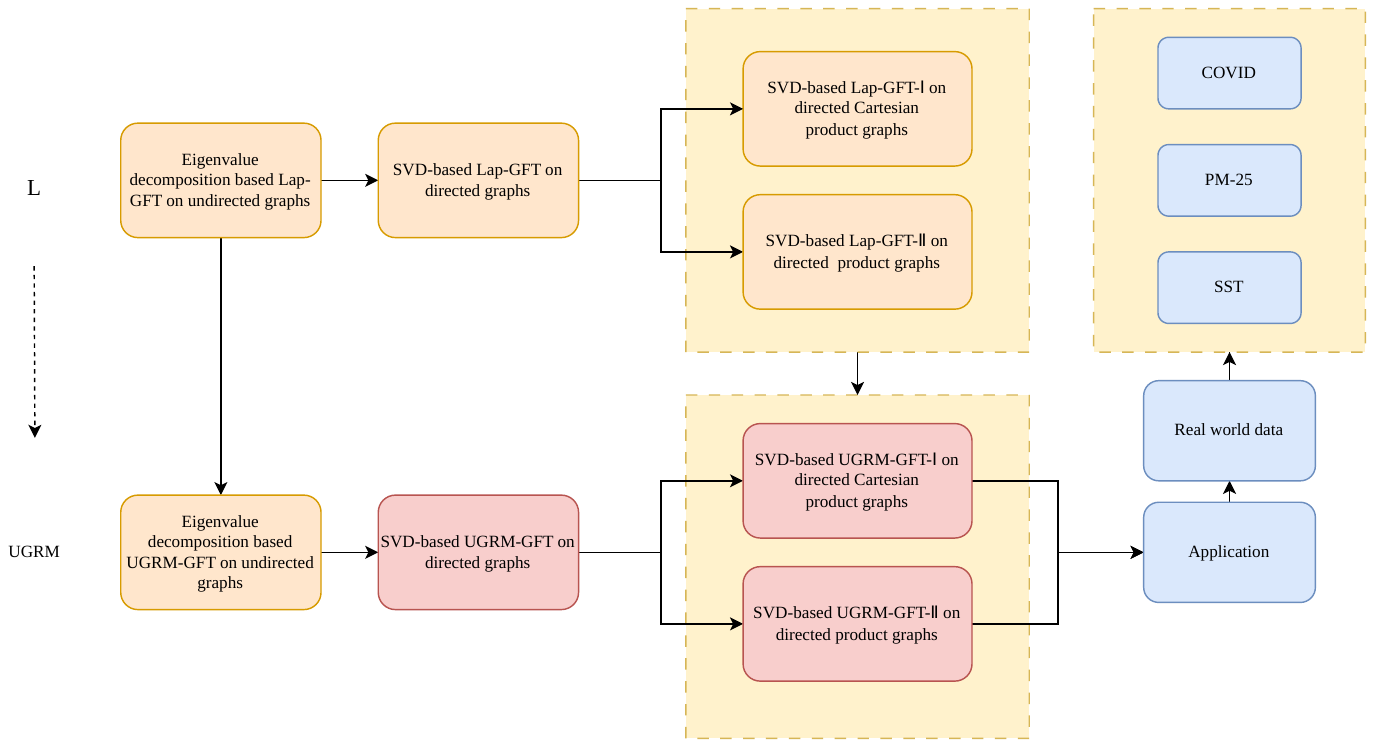}
		
		\vspace{-110pt}   
		
		\captionsetup{justification=centering} 
		
		\caption{Overview of UGRM-GFT method development and application scenarios.}
		\label{fig:fullwidthpdf}
	\end{figure*}
	
	\section{Preliminaries}
	\label{sec:chapter2}
	In this section, we provide preliminary knowledge regarding the UGRM, the directed Cartesian product graph, and two GFTs $\mathcal{F}_\boxtimes$ and $\mathcal{F}_\otimes$ based on the Laplacian matrix on the Cartesian product graph $\mathcal{G}_1 \boxtimes \mathcal{G}_2$. 
	\subsection{Unified Graph Representation Matrix}
	Traditional GSP methods usually rely on fixed-structure matrix representations, including adjacency matrix $\mathbf{A}$, in-degree matrix $\mathbf{D}$, Laplacian matrix $\mathbf{L}$, and signless Laplacian matrix $\mathbf{Q}$. However, such rigid representations are difficult to adapt to complex and changing graph topologies. For this reason, the proposal of counting graph matrices provides new research ideas for graph representation learning. Nikiforov et al. proposed the \(\alpha\)-adjacency matrix \(\mathbf{A}\), which, for particular values of \(\alpha\), recovers the classical representation matrices \(\mathbf{A}\), \(\mathbf{Q}\), and \(\mathbf{D}\)~\cite{ref23}. However, matrix $\mathbf{L}$ cannot be obtained from the matrix $\mathbf{A}_\alpha$. Wang et al. proposed a class of graph representations $\mathbf{L}_\alpha$ called $\alpha$-Laplacian, which is capable of recovering the matrix $\mathbf{L}$, but is unable to obtain the matrix $\mathbf{A}$ exactly\cite{ref45}. Therefore, Averty et al. proposed the UGRM, defined as follows\cite{ref25}
	\begin{equation} 
		\mathbf{P}^{\alpha,k} := \alpha \mathbf{D} + (2k-1)(\alpha-1)\mathbf{A}. 
		\label{1}
	\end{equation} 
	By adjusting the values of the parameters $\alpha$ and $k$, 
	the UGRM is able to adaptively degenerate into the classical 
	representations of the traditional adjacency matrix $\mathbf{A}$, 
	Laplacian matrix $\mathbf{L}$, in-degree matrix $\mathbf{D}$, 
	and signless Laplacian matrix $\mathbf{Q}$. 
	Specifically, the parameters $\alpha$ and $k$ 
		are restricted to $[0,1]$, as this is the domain over which 
		$\mathbf{P}^{\alpha,k}$ recovers all classical graph 
		representations; values outside this range lack standard 
		graph-theoretic interpretation.
	The UGRM can function both as a traditional matrix model and 
	as a flexible parameterized framework. This dual capability 
	maintains the mathematical rigor of traditional matrices while 
	substantially expanding the expressive capacity of the model.
	\subsection{Directed Cartesian Product Graph}
	Cartesian product graph is a type of graph multiplication\cite{ref46}. Consider two directed graphs $\mathcal{G}_1 = (\mathcal{V}_1, \mathcal{E}_1, \mathbf{A}_1)$ and $\mathcal{G}_2 = (\mathcal{V}_2, \mathcal{E}_2, \mathbf{A}_2)$. $\mathbf{L}_l$  are the Laplacian matrix of $\mathcal{G}_l, l=1,2$. First, $\boxtimes$ represents the Cartesian product between graphs. $ \mathcal{G}_1 \boxtimes \mathcal{G}_2$ is the Cartesian product graph with vertex set $ \mathcal{V}_1 \times \mathcal{V}_2$ where $\mathcal{V}_1 = \{0, \cdots, N_1 - 1\},\mathcal{V}_2 = \{0, \cdots, N_2 - 1\}$. Then, the set of edges of $\mathcal{E}$ satisfies: $ [[i_1, j_1] \in \mathcal{E}_1, i_2 = j_2] \text{ or } [i_1 = j_1, [i_2, j_2] \in \mathcal{E}_2] $, and the weighting function $\mathbf{A}$ is given by 
	\begin{equation}
		\mathbf{A}((i_1, i_2), (j_1, j_2)) = \mathbf{A}_1(i_1, j_1) \delta(i_2, j_2) + \delta(i_1, j_1) \mathbf{A}_2(i_2, j_2),
		\label{2}
	\end{equation}
	where $\delta$ is the delta function, i.e., $\delta_{ij} = 1$ when $i = j$, otherwise $\delta_{ij} = 0$. Fig. \ref{fig:Directed Cartesian product graph} shows one example of a directed Cartesian product operation.
	
    The graphs $\mathcal{G}_1$ and $\mathcal{G}_2$ are factor graphs of $\mathcal{G}_1 \boxtimes \mathcal{G}_2$. The Laplacian matrix $\mathbf{L}_\boxtimes$, adjacency matrix $\mathbf{A}_\boxtimes$, in-degree matrix $\mathbf{D}_\boxtimes$, signless Laplacian matrix $\mathbf{Q}_\boxtimes$, and UGRM $\mathbf{P}_\boxtimes^{\alpha,k}$ of the graph $\mathcal{G}_1 \otimes \mathcal{G}_2$ can be determined from  adjacency matrices $\mathbf{A}_1, \mathbf{A}_2$, in-degree matrices $\mathbf{D}_1, \mathbf{D}_2$ , signless Laplacian matrices $\mathbf{Q}_1, \mathbf{Q}_2$, Laplacian matrices $\mathbf{L}_1, \mathbf{L}_2$, and UGRMs $\mathbf{P}_1^{\alpha,k}, \mathbf{P}_2^{\alpha,k}$ of the factor graphs, those are,
	\begin{equation}
		\mathbf{L}_\boxtimes = \mathbf{L}_1 \otimes \mathbf{I}_{N_2} + \mathbf{I}_{N_1} \otimes \mathbf{L}_2,
		\label{3}
	\end{equation}
	\begin{equation}
		\mathbf{A}_\boxtimes = \mathbf{A}_1 \otimes \mathbf{I}_{N_2} + \mathbf{I}_{N_1} \otimes \mathbf{A}_2, 
		\label{4} 
	\end{equation}
		\begin{equation}
		\mathbf{D}_\boxtimes = \mathbf{D}_1 \otimes \mathbf{I}_{N_2} + \mathbf{I}_{N_1} \otimes \mathbf{D}_2,
		\label{5}
	\end{equation}
		\begin{equation}
		\mathbf{Q}_\boxtimes = \mathbf{Q}_1 \otimes \mathbf{I}_{N_2} + \mathbf{I}_{N_1} \otimes \mathbf{Q}_2,
		\label{6}
	\end{equation}
	\begin{equation} 
		\mathbf{P}_\boxtimes^{\alpha,k} = \mathbf{P}_1^{\alpha,k} \otimes \mathbf{I}_{N_2} + \mathbf{I}_{N_1} \otimes \mathbf{P}_2^{\alpha,k}. 
		\label{7} 
	\end{equation}
	where \( \otimes \) denotes the Kronecker product, and \( \mathbf{I}_n \) is an \(n \times n \) dimensional unit matrix. Note that if \(\mathbf{A} \) is an \(m \times n \) matrix and \(\mathbf{B} \) is a \(p \times q \) matrix, then the Kronecker product \(\mathbf{A} \otimes \mathbf{B} \) is a \(pm \times qn \) block matrix\cite{ref46}:
	\begin{equation} 
		\mathbf{A} \otimes \mathbf{B} = \begin{pmatrix} 
			a_{11}\mathbf{B} & \cdots & a_{1n}\mathbf{B} \\\ 
			\vdots & \ddots & \vdots \\\ 
			a_{m1}\mathbf{B} & \cdots & a_{mn}\mathbf{B} 
		\end{pmatrix}.
		\label{8}  
	\end{equation}
	\begin{figure*}[!t] 
		\centering
		\begin{tikzpicture}
			\node[anchor=center] (ring) at (0,0) {\includegraphics[width=0.3\textwidth]{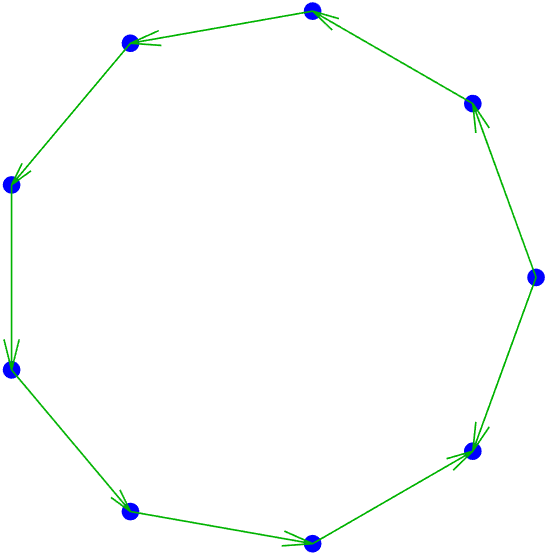}};
			\node[anchor=north] at (ring.south) {(a)}; 
			
			\node[anchor=center] at (4,0) {$\times$};
			
			\node[anchor=center] (line1) at (6,0) {\includegraphics[width=0.2\textwidth, angle=90]{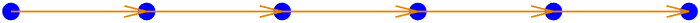}};
			\node[anchor=north] at (line1.south) {(b)}; 
			
			\node[anchor=center] at (8,0) {$=$};
			
			\node[anchor=center] (line2) at (12,0) {\includegraphics[width=0.5\textwidth, angle=90]{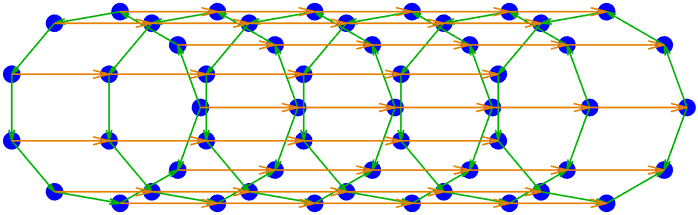}};
			\node[anchor=north] at (line2.south) {(c)}; 
		\end{tikzpicture}
		\caption{(a)(b) Graphs and (c) their directed Cartesian product graphs. The green solid edges in figure (c) are from (a) and the orange solid edges are from (b).}
		\label{fig:Directed Cartesian product graph}
	\end{figure*}
	\vspace{-30pt}
	\subsection{SVD-BASED Lap-GFT-I ON DIRECTED CARTESIAN PRODUCT GRAPHS}
	 Let $\mathcal{G}_1 = (\mathcal{V}_1, \mathcal{E}_1)$ and $\mathcal{G}_2 = (\mathcal{V}_2, \mathcal{E}_2)$ be two directed graphs of orders $N_1$ and $N_2$. For the Laplacian  $\mathbf{L}_\boxtimes$ on a directed Cartesian product graph $\mathcal{G}$, given by (\ref{3}), one can take its SVD as follows
    \begin{equation}
	\mathbf{L}_{\boxtimes} = \mathbf{U}_{\boxtimes} \boldsymbol{\Sigma} \mathbf{V}_{\boxtimes}^T = \sum_{t=0}^{N - 1} \sigma_t\mathbf{u}_t \mathbf{v}_t^T. 
	\label{9}
    \end{equation}
    
    Based on the SVD  of the Laplacian matrix $\mathbf{L}_\boxtimes$,  the GFT $\mathcal{F}_{\boxtimes}: \mathbb{R}^N \to \mathbb{R}^{2N}$  and inverse GFT $\mathcal{F}_{\boxtimes}^{-1}: \mathbb{R}^{2N} \to \mathbb{R}^N$ on the Cartesian product graph $\mathcal{G}$ are defined as
    \begin{equation}
    \begin{split}
    	\mathcal{F}_{\boxtimes} \mathbf{x} :&= \frac{1}{2} \begin{pmatrix}
    		(\mathbf{U}_{\boxtimes} + \mathbf{V}_{\boxtimes})^T \mathbf{x} \\
    		(\mathbf{U}_{\boxtimes} - \mathbf{V}_{\boxtimes})^T \mathbf{x}
    	\end{pmatrix} \\
    	&= \frac{1}{2} \begin{pmatrix}
    		(\mathbf{u}_0 + \mathbf{v}_0)^T \mathbf{x} \\
    		\vdots \\
    		(\mathbf{u}_{N-1} + \mathbf{v}_{N-1})^T \mathbf{x} \\
    		(\mathbf{u}_0 - \mathbf{v}_0)^T \mathbf{x} \\
    		\vdots \\
    		(\mathbf{u}_{N-1} - \mathbf{v}_{N-1})^T \mathbf{x}
    	\end{pmatrix}
    \end{split}
    \label{10}
    \end{equation}
    and
    \begin{equation}
    	\mathcal{F}_{\boxtimes}^{-1}\begin{pmatrix} \mathbf{z}_1 \\ \mathbf{z}_2 \end{pmatrix} := \frac{1}{2}\left( \mathbf{U}_{\boxtimes}(\mathbf{z}_1 + \mathbf{z}_2) + \mathbf{V}_{\boxtimes}(\mathbf{z}_1 - \mathbf{z}_2) \right),
    	\label{11}
    \end{equation}
    respectively, where $\mathbf{x}$ is a graph signal on the Cartesian product graph $\mathcal{G}$, and $\mathbf{z}_1, \mathbf{z}_2$ are vectors in $\mathbb{R}^{N}$.

    \subsection{SVD-BASED Lap-GFT-II ON DIRECTED PRODUCT GRAPHS}
    Let \( \mathcal{G}_1 = (\mathcal{V}_1, \mathcal{E}_1) \) and \( \mathcal{G}_2 = (\mathcal{V}_2, \mathcal{E}_2) \) be two directed graphs, and denote their Laplacians and orders by \( \mathbf{L}_l \) and \( N_l \), \( l = 1, 2 \) respectively. For the Laplacian matrices \( \mathbf{L}_l \), \( l = 1, 2 \), we take their SVDs.
    
    \begin{equation}
    \mathbf{L}_l = \mathbf{U}_l \boldsymbol{\Sigma}_l \mathbf{V}_l^T = \sum_{i=0}^{N_l - 1} \sigma_{l,i} \mathbf{u}_{l,i} \mathbf{v}_{l,i}^T ,
    \label{12}
    \end{equation}
    where \( \sigma_{l,i}, 0 \leq i \leq N_l - 1 \), are singular values of the Laplacian matrix \( \mathbf{L}_l \) with a nondecreasing order, \( \mathbf{U}_l = [\mathbf{u}_{l,0}, \ldots, \mathbf{u}_{l,N_l - 1}] \) and \( \mathbf{V}_l = [\mathbf{v}_{l,0}, \ldots, \mathbf{v}_{l,N_l - 1}] \) are orthonormal matrices. Set
    
    \begin{equation}
    \mathbf{U}_{\otimes} = \mathbf{U}_1 \otimes \mathbf{U}_2 \quad \text{and} \quad \mathbf{V}_{\otimes} = \mathbf{V}_1 \otimes \mathbf{V}_2 ,
    \label{13}
    \end{equation}
    
    Based on the SVD of the Laplacian matrices $\mathbf{L}_l$ ($l=1,2$), 
    the second GFT $\mathcal{F}_{\otimes}: \mathbb{R}^N \to \mathbb{R}^{2N}$  and inverse GFT $\mathcal{F}_{\boxtimes}^{-1}: \mathbb{R}^{2N} \to \mathbb{R}^N$ on the product graph $\mathcal{G}$ are defined as
    \begin{equation}
    	\mathcal{F}_{\otimes} \mathbf{x} := \frac{1}{2} \begin{pmatrix}
    		(\mathbf{U}_{\otimes} + \mathbf{V}_{\otimes})^T \mathbf{x} \\
    		(\mathbf{U}_{\otimes} - \mathbf{V}_{\otimes})^T \mathbf{x}
    	\end{pmatrix}
    \end{equation}
    and
     \begin{equation}
    	\mathcal{F}_{\otimes}^{-1}\begin{pmatrix} \mathbf{z}_1 \\ \mathbf{z}_2 \end{pmatrix} := \frac{1}{2}\left( \mathbf{U}_{\otimes}(\mathbf{z}_1 + \mathbf{z}_2) + \mathbf{V}_{\otimes}(\mathbf{z}_1 - \mathbf{z}_2) \right),
    \end{equation} respectively. 

     \section{SVD-BASED UGRM-GFT ON DIRECTED GRAPHS}
      \label{sec:chapter3}
    Existing GFT methods on directed graphs predominantly rely on fixed graph shift operators—most notably the Laplacian or adjacency matrix whose static structure limits their capacity to adapt to diverse and complex graph topologies. To overcome this fundamental limitation, we introduce the unified graph representation matrix (UGRM) $\mathbf{P}^{\alpha,k} = \alpha\mathbf{D} + (2k-1)(\alpha-1)\mathbf{A}$ as the spectral operator, which subsumes classical representations as special cases through the tunable parameters $\alpha$ and $k$. For a directed graph $\mathcal{G}$ with $N$ nodes, the SVD of $\mathbf{P}^{\alpha,k}$ is given by
    \begin{equation}
    	\mathbf{P}^{\alpha,k} = \mathbf{U}^{\alpha,k}
    	\boldsymbol{\Sigma}^{\alpha,k} (\mathbf{V}^{\alpha,k})^T
    	= \sum_{t=0}^{N-1} \sigma_t^{\alpha,k} \mathbf{u}_t^{\alpha,k}
    	(\mathbf{v}_t^{\alpha,k})^T,
    	\label{16}
    \end{equation}
    where $\mathbf{U}^{\alpha,k} = [\mathbf{u}_0^{\alpha,k}, \ldots, \mathbf{u}_{N-1}^{\alpha,k}]$ and $\mathbf{V}^{\alpha,k} = [\mathbf{v}_0^{\alpha,k}, \ldots, \mathbf{v}_{N-1}^{\alpha,k}]$ are orthogonal matrices whose columns constitute the left and right singular vectors, respectively, and $\boldsymbol{\Sigma}^{\alpha,k} = \operatorname{diag}(\sigma_0^{\alpha,k}, \ldots, \sigma_{N-1}^{\alpha,k})$ contains the corresponding non-negative singular values. The computational complexity of performing this SVD is $\mathcal{O}(N^3)$ \cite{ref47}.
    
    Based on the SVD \eqref{16} of the UGRM $P_{\boxtimes}^{\alpha,k}$ and following the approach of \cite{ref26}, we define the UGRM-GFT and its inverse as follows.
    
    \textit{Definition 1}: Let $\mathcal{G}$ be a directed graph with UGRM $\mathbf{P}^{\alpha,k}$ given by \eqref{1}. The UGRM-GFT $\mathcal{F}^{\alpha,k}: \mathbb{R}^N \mapsto \mathbb{R}^{2N}$ on $\mathcal{G}$ is defined as
    \begin{equation}
    	\begin{split}
    		\mathcal{F}^{\alpha,k} \mathbf{x} :&= \frac{1}{2}
    		\begin{pmatrix}
    			(\mathbf{U}^{\alpha,k} + \mathbf{V}^{\alpha,k})^T \mathbf{x} \\
    			(\mathbf{U}^{\alpha,k} - \mathbf{V}^{\alpha,k})^T \mathbf{x}
    		\end{pmatrix} \\
    		&= \frac{1}{2}
    		\begin{pmatrix}
    			(\mathbf{u}_0^{\alpha,k} + \mathbf{v}_0^{\alpha,k})^T \mathbf{x} \\
    			\vdots \\
    			(\mathbf{u}_{N-1}^{\alpha,k} + \mathbf{v}_{N-1}^{\alpha,k})^T \mathbf{x} \\
    			(\mathbf{u}_0^{\alpha,k} - \mathbf{v}_0^{\alpha,k})^T \mathbf{x} \\
    			\vdots \\
    			(\mathbf{u}_{N-1}^{\alpha,k} - \mathbf{v}_{N-1}^{\alpha,k})^T \mathbf{x}
    		\end{pmatrix},
    	\end{split}
    	\label{17}
    \end{equation}
    and its inverse $(\mathcal{F}^{\alpha,k})^{-1}: \mathbb{R}^{2N} \mapsto \mathbb{R}^N$ is given by
    \begin{equation}
    	(\mathcal{F}^{\alpha,k})^{-1}
    	\begin{pmatrix} \mathbf{z}_1^{\alpha,k} \\ \mathbf{z}_2^{\alpha,k} \end{pmatrix}
    	:= \frac{1}{2} \Bigl(
    	\mathbf{U}^{\alpha,k}(\mathbf{z}_1^{\alpha,k} + \mathbf{z}_2^{\alpha,k})
    	+ \mathbf{V}^{\alpha,k}(\mathbf{z}_1^{\alpha,k} - \mathbf{z}_2^{\alpha,k})
    	\Bigr),
    	\label{18}
    \end{equation}
    where $\mathbf{z}_1^{\alpha,k}, \mathbf{z}_2^{\alpha,k} \in \mathbb{R}^N$. By the orthogonality of $\mathbf{U}^{\alpha,k}$ and $\mathbf{V}^{\alpha,k}$, one immediately verifies the perfect reconstruction identity
    \begin{equation}
    	(\mathcal{F}^{\alpha,k})^{-1} \mathcal{F}^{\alpha,k} \mathbf{x} = \mathbf{x},
    	\quad \mathbf{x} \in \mathbb{R}^N,
    	\label{19}
    \end{equation}
    and Parseval's identity
    \begin{equation}
    	\|\mathcal{F}^{\alpha,k} \mathbf{x}\|_2 = \|\mathbf{x}\|_2,
    	\quad \mathbf{x} \in \mathbb{R}^N,
    \end{equation}
    confirming that UGRM-GFT constitutes an energy-preserving isometric transformation on the signal space $\mathbb{R}^N$.
    
    A central theoretical question in the proposed framework concerns how 
    the spectral components induced by $\mathbf{P}^{\alpha,k}$ should be 
    interpreted and ordered, so that the leading retained components capture 
    the dominant structure of graph signals. Unlike the classical undirected 
    Laplacian, whose eigenvalues admit a universal smoothness interpretation, 
    the operator $\mathbf{P}^{\alpha,k}$ may interpolate between qualitatively 
    distinct spectral behaviors depending on the parameter regime $(\alpha, k)$. 
    To make this precise, we introduce the operator-induced quadratic 
    	response pair
    \begin{align}
    	\mathcal{R}^{\alpha,k}(\mathbf{x})
    	&:= \|\mathbf{P}^{\alpha,k}\mathbf{x}\|_2^2
    	= \mathbf{x}^T (\mathbf{P}^{\alpha,k})^T \mathbf{P}^{\alpha,k} \mathbf{x},
    	\label{21}\\
    	\widetilde{\mathcal{R}}^{\alpha,k}(\mathbf{x})
    	&:= \|(\mathbf{P}^{\alpha,k})^T\mathbf{x}\|_2^2
    	= \mathbf{x}^T \mathbf{P}^{\alpha,k} (\mathbf{P}^{\alpha,k})^T \mathbf{x},
    	\label{22}
    \end{align}
    for $\mathbf{x} \in \mathbb{R}^N$, which jointly measure the energy of 
    the signal response under the forward and adjoint actions of the operator, 
    respectively. Although $\mathbf{P}^{\alpha,k}$ is generally asymmetric 
    for directed graphs, both Gram matrices 
    $(\mathbf{P}^{\alpha,k})^T\mathbf{P}^{\alpha,k}$ and 
    $\mathbf{P}^{\alpha,k}(\mathbf{P}^{\alpha,k})^T$ are symmetric positive 
    semidefinite by construction, so $\mathcal{R}^{\alpha,k}(\mathbf{x})$ and 
    $\widetilde{\mathcal{R}}^{\alpha,k}(\mathbf{x})$ constitute well-defined 
    quadratic forms on $\mathbb{R}^N$ and together serve as surrogates for 
    graph variation under the standard Euclidean inner product \cite{ref32}. 
    Substituting the SVD \eqref{16} and using the orthogonality of 
    $\mathbf{V}^{\alpha,k}$ and $\mathbf{U}^{\alpha,k}$ respectively, one obtains
    \begin{align}
    	\mathcal{R}^{\alpha,k}(\mathbf{x}) 
    	&= \|\mathbf{P}^{\alpha,k}\mathbf{x}\|_2^2 
    	= \mathbf{x}^T \mathbf{V}^{\alpha,k} (\boldsymbol{\Sigma}^{\alpha,k})^2 
    	(\mathbf{V}^{\alpha,k})^T \mathbf{x} \notag\\
    	&=\sum_{t=0}^{N-1} (\sigma_t^{\alpha,k})^2 ((\mathbf{v}_t^{\alpha,k})^T \mathbf{x})^2,
    	\label{23}\\
    	\widetilde{\mathcal{R}}^{\alpha,k}(\mathbf{x}) 
    	&= \|{\mathbf{P}^{\alpha,k}}^T\mathbf{x}\|_2^2 
    	= \mathbf{x}^T \mathbf{U}^{\alpha,k} (\boldsymbol{\Sigma}^{\alpha,k})^2 
    	(\mathbf{U}^{\alpha,k})^T \mathbf{x} \notag\\
    	&=\sum_{t=0}^{N-1} (\sigma_t^{\alpha,k})^2 ((\mathbf{u}_t^{\alpha,k})^T \mathbf{x})^2.
    	\label{24}
    \end{align}
    This plays a key role in spectral truncation: as formalized in Theorem~\ref{thm1}, the approximation residual is controlled by both the forward and adjoint operator responses, reflecting the involvement of both sets of singular vectors. This follows the approach of~\cite{ref44}. Signals concentrated on modes with smaller singular values yield smaller responses. In the smoothness regime ($\beta<0$), these correspond to low-variation graph signals, and are naturally treated as low-frequency modes, consistent with the ordering in~\eqref{12}.
    The qualitative role of the parameters is further clarified by defining
    \begin{equation}
    	\beta := (2k-1)(\alpha-1),
    	\label{25}
    \end{equation}
    and employing the identity $\mathbf{A} = \mathbf{D} - \mathbf{L}$ to rewrite the UGRM as
    \begin{equation}
    	\mathbf{P}^{\alpha,k} = \alpha\mathbf{D} + \beta\mathbf{A}
    	= (\alpha+\beta)\mathbf{D} - \beta\mathbf{L}.
    	\label{26}
    \end{equation}
    
    This decomposition reveals a fundamental dichotomy in spectral behavior governed by the sign of $\beta$. When $\beta < 0$, the Laplacian term $\mathbf{L}$ appears with a positive coefficient in \eqref{26}, and the operator behaves in a difference-like manner analogous to $\mathbf{L}$. In this regime, $\mathcal{R}^{\alpha,k}(\mathbf{x})$ is consistent with a smoothness-type measure, and signals with smaller response are better aligned with lower-variation modes of the operator; it is therefore natural to treat smaller singular values as lower-order spectral components and to retain those components for approximation. Conversely, when $\beta \geq 0$, the operator becomes more propagation-oriented and adjacency-like. In this regime, $\mathcal{R}^{\alpha,k}(\mathbf{x})$ measures the alignment of the signal with the dominant modes of the operator, and larger singular values correspond to components with greater energy compaction capability, so that retaining the leading singular modes provides an effective truncation strategy.
    
    Motivated by this analysis, we adopt the following operator-dependent spectral ordering:
    \begin{equation}
    	\begin{cases}
    		\sigma_0^{\alpha,k} \leq \sigma_1^{\alpha,k} \leq \cdots
    		\leq \sigma_{N-1}^{\alpha,k}, & \text{if } \beta < 0, \\[2mm]
    		\sigma_0^{\alpha,k} \geq \sigma_1^{\alpha,k} \geq \cdots
    		\geq \sigma_{N-1}^{\alpha,k} \geq 0, & \text{if } \beta \geq 0.
    	\end{cases}
    	\label{27}
    \end{equation}
    Under either ordering, the first $M$ retained components $\{\mathbf{u}_t^{\alpha,k}, \mathbf{v}_t^{\alpha,k}\}_{i=0}^{M-1}$ are always those most relevant for approximation in the respective regime. In particular, when $\beta \geq 0$, we refer to them as \emph{leading spectral components} rather than low-frequency components, to avoid conflating the two qualitatively distinct regimes. We emphasize that \eqref{27} constitutes an operator-dependent truncation convention tailored to the parameter regime, rather than a universal definition of graph frequency valid for all directed graphs.
    
    With this ordering in place, we now establish the approximation guarantees associated with the two parameter regimes. The following theorem quantifies the approximation error incurred when retaining the $M$ components with the smallest singular values in the regime $\beta < 0$, showing that the error is controlled by the operator response of the signal.
    
    \begin{theorem}\label{thm1}
    	Let $\mathcal{G}$ be a directed graph with $N$ nodes, and let $\mathbf{P}^{\alpha,k}$ be the UGRM. Let the SVD of $\mathbf{P}^{\alpha,k}$ be as in \eqref{16}, and let $\beta = (2k-1)(\alpha-1)$. For a bandwidth $M \in \{1, \ldots, N\}$, define the spectral approximation $\mathbf{x}_{M}^{\alpha,k}$ of a signal $\mathbf{x} \in \mathbb{R}^N$ by
    \begin{equation}
    	\begin{aligned}
    		\mathbf{x}_{M}^{\alpha,k} 
    		&= \frac{1}{2}\sum_{t=0}^{M-1} \left[ \left(\mathbf{z}_{1,t}^{\alpha,k}+\mathbf{z}_{2,t}^{\alpha,k}\right)\mathbf{u}_t + \left(\mathbf{z}_{1,t}^{\alpha,k}-\mathbf{z}_{2,t}^{\alpha,k}\right)\mathbf{v}_t \right] \\
    		&= \frac{1}{2}\sum_{t=0}^{M-1} \left( \mathbf{u}_t^{\alpha,k}(\mathbf{u}_t^{\alpha,k})^T + \mathbf{v}_t^{\alpha,k}(\mathbf{v}_t^{\alpha,k})^T \right)\mathbf{x},
    	\end{aligned}
    	\label{28}
    \end{equation}
    	where $\mathbf{z}_{1,t}^{\alpha,k} = (\mathbf{u}_t^{\alpha,k}+\mathbf{v}_t^{\alpha,k})^T\mathbf{x}/2$ and $\mathbf{z}_{2,t}^{\alpha,k} = (\mathbf{u}_t^{\alpha,k}-\mathbf{v}_t^{\alpha,k})^T\mathbf{x}/2$, \; $0 \le t \le M-1$.
    	
    	\begin{enumerate}
    		\item[\textbf{(1)}] \textbf{Case $\boldsymbol{\beta < 0}$.} Let the singular values be ordered non-decreasingly as
    		\[
    		0 \le \sigma_0^{\alpha,k} \le \sigma_1^{\alpha,k} \le \cdots \le \sigma_{N-1}^{\alpha,k}.
    		\]
    		If $\sigma_{M-1}^{\alpha,k} > 0$, then
    		\begin{equation}
    			\|\mathbf{x} - \mathbf{x}_{M}^{\alpha,k}\|_2 \le \frac{1}{2\sigma_{M-1}^{\alpha,k}}\left(\|\mathbf{P}^{\alpha,k}\mathbf{x}\|_2 + \|(\mathbf{P}^{\alpha,k})^T\mathbf{x}\|_2\right).
    			\label{29}
    		\end{equation}
    		
    		\item[\textbf{(2)}]\textbf{Case $\boldsymbol{\beta \ge 0}$.}
    		Let the singular values be ordered non-increasingly as
    		$\sigma^{\alpha,k}_{0}\geq\sigma^{\alpha,k}_{1}\geq\cdots
    		\geq\sigma^{\alpha,k}_{N-1}\geq 0$.
    		Suppose $\mathbf{x}$ satisfies the operator consistency condition:
    		there exists a constant $C>0$ such that
    		\begin{equation*}
    			(\mathbf{u}^{\alpha,k}_{t})^{T}\mathbf{x}\
    			\leq C\sigma^{\alpha,k}_{t}
    			\quad\text{and}\quad
    			(\mathbf{v}^{\alpha,k}_{t})^{T}\mathbf{x}
    			\leq C\sigma^{\alpha,k}_{t}
    		\end{equation*}
    		for all $t\geq M$. Then
    		\begin{equation}
    			\|\mathbf{x} - \mathbf{x}^{\alpha,k}_{M}\|_{2}
    			\leq
    			C\left(\sum_{t=M}^{N-1}(\sigma^{\alpha,k}_{t})^{2}\right)^{1/2}.
    			\label{30}
    		\end{equation}
    	\end{enumerate}
    \end{theorem}
    
    \begin{proof}
    	See Appendix A.
    \end{proof}

 The approximation error bounds in Theorem~\ref{thm1} provide a 
 structural explanation for why the UGRM spectral basis 
 achieves better energy compaction than any fixed operator. 
 Recall from~\eqref{16} that the UGRM admits 
 the decomposition~\eqref{26}
 revealing that the UGRM continuously interpolates between 
 the degree matrix (encoding local connectivity) and the 
 Laplacian (encoding global variation). For a signal 
 $\mathbf{x}$ concentrated on the leading $M$ singular 
 modes, the bound in~\eqref{29} shows that 
 the residual energy is controlled by 
 $\sigma_{M-1}^{\alpha,k}$: this bound is minimized when 
 $\sigma_{M-1}^{\alpha,k}$ is maximized, which occurs 
 precisely when $\mathbf{P}^{\alpha,k}$ is spectrally 
 aligned with the signal's correlation structure. A fixed 
 operator such as $\mathbf{L}$ or $\mathbf{A}$ has no 
 freedom to adapt its spectral basis; by contrast, the 
 parameters $(\alpha, k)$ select the combination of local 
 and global graph properties that maximizes 
 $\sigma_{M-1}^{\alpha,k}$ for the specific signal at 
 hand, thereby achieving tighter spectral concentration. 
 This same principle extends to the directed Cartesian 
 product graph setting through Theorems~\ref{thm2} and~\ref{thm3}, where 
 the approximation error is similarly controlled by the 
 singular values of the factor graph UGRMs. The specific 
 parameter values $(\alpha^*, k^*)$ that achieve optimal 
 spectral alignment depend on the intrinsic correlation 
 structure of the signal and the underlying graph 
 topology, and are identified in practice via Bayesian 
 optimization as described in Section~\ref{sec:chapter6}.
    
    \section{SVD-BASED UGRM-GFT-I ON DIRECTED CARTESIAN PRODUCT  GRAPHS}
       \label{sec:chapter4}
    In this section, following the approach in Section \ref{sec:chapter3}
    , we introduce a UGRM-GFT-I $\mathcal{F}_\boxtimes^{\alpha,k}$ on the directed Cartesian product graph $\mathcal{G}$.
    
    Signals in the Cartesian product graph structure $\mathcal{G}$ can be expressed in two matrix forms. The first form is $\mathbf{X} = [\mathbf{x}_i],{i \in \mathcal{V}_1} \in \mathbb{R}^{N_2 \times N_1}$, where each column $\mathbf{x}_i$ represents a graph signal defined on $\mathcal{G}_2$. and the overall signal can be converted to vector form through vectorization operation $\mathbf{x} = \operatorname{vec}(\mathbf{X})$. The second form uses $\mathbf{Y} = [\mathbf{y}_j^\top],{j \in \mathcal{V}_2}$, where each row represents a signal $\mathbf{y}_j$ defined on $\mathcal{G}_1$, which can also be vectorized as $\mathbf{y} = \operatorname{vec}(\mathbf{Y})$. In spatio-temporal data processing, $\mathbf{x}_i$ can be understood as the observation values at all spatial positions at the $i$th time point, while $\mathbf{y}_j$ represents the observation sequence over a period of time at the $j$th spatial position\cite{ref44}.
    
    For the UGRM $\mathbf{P}_{\boxtimes}^{\alpha,k}$ on a directed Cartesian product graph $\mathcal{G}$, given by (\ref{7}), we take its SVD as follows,
    \begin{equation}
    	\mathbf{P}_{\boxtimes}^{\alpha,k} = \mathbf{U}_{\boxtimes}^{\alpha,k} \boldsymbol{\Sigma}^{\alpha,k} (\mathbf{V}_{\boxtimes}^{\alpha,k})^T = \sum_{t=0}^{N - 1} \sigma_{t}^{\alpha,k}\mathbf{u}_{t}^{\alpha,k} (\mathbf{v}_{t}^{\alpha,k})^T, 
    	\label{31}
    \end{equation}
    where \( N = N_1 N_2 \), \( \mathbf{U}_{\boxtimes}^{\alpha,k} = [\mathbf{u}_{0}^{\alpha,k}, \ldots, \mathbf{u}_{N-1}^{\alpha,k}] \) and \( \mathbf{V}_{\boxtimes}^{\alpha,k} = [\mathbf{v}_{0}^{\alpha,k}, \ldots, \mathbf{v}_{N-1}^{\alpha,k}] \) are orthogonal matrices, and the diagonal matrix \( \boldsymbol{\Sigma}^{\alpha,k} = \operatorname{diag}(\sigma_{0}^{\alpha,k}, \ldots, \sigma_{N-1}^{\alpha,k}) \) has singular values of the UGRM \( \mathbf{P}_{\boxtimes}^{\alpha,k} \) deployed on the diagonal in a nondecreasing  or non-increasing order. The computational complexity to perform the SVD \eqref{31} is $O(N^{3})$.
    
    When $\alpha = 0.5$ and $k = 1$, the matrix $\mathbf{P}_{\boxtimes}^{\alpha,k}$ degenerates into the form of SVD based on the Laplacian in (\ref9). Specifically, at this point, $\mathbf{P}_{\boxtimes}^{0.5,1}$ is equivalent to the SVD decomposition of the graph Laplacian matrix $\mathbf{L}_\boxtimes$.
    
    For the undirected graph setting, i.e., $\mathcal{G}_1$ and $\mathcal{G}_2$ are undirected graphs, the UGRMs $\mathbf{P}_{l}^{\alpha,k},l =1, 2$, are positive semidefinite and they have the following eigendecompositions
    \begin{equation}
    \mathbf{P}_{l}^{\alpha,k} = \sum_{i=0}^{N_l - 1} \lambda_{l,i}^{\alpha,k} \mathbf{w}_{l,i}^{\alpha,k} (\mathbf{w}_{l,i}^{\alpha,k})^T, \, l = 1, 2,
    \label{32}
    \end{equation}
    where \(  \lambda_{l,i}^{\alpha,k}\) are eigenvalues of \( \mathbf{P}_{l}^{\alpha,k} \), and \( \mathbf{w}_{l,i}^{\alpha,k}, 0 \leq i \leq N_l - 1 \), form an orthonormal basis of \( \mathbb{R}^{N_l} \). Therefore, eigenvalues (singular values) of the UGRM \( \mathbf{P}_{\boxtimes}^{\alpha,k} \) on the undirected Cartesian product graph \( \mathcal{G} \) are the sum of eigenvalues of \( \mathbf{P}_{1}^{\alpha,k} \) and \( \mathbf{P}_{2}^{\alpha,k} \), and orthogonal matrices \( \mathbf{U}_{\boxtimes}^{\alpha,k} \) and \( \mathbf{V}_{\boxtimes}^{\alpha,k} \) are the same and consist of Kronecker products of eigenvectors of UGRMs \( \mathbf{P}_{1}^{\alpha,k} \) and \( \mathbf{P}_{2}^{\alpha,k} \)\cite{ref40,ref48,ref49,ref50}, i.e.,
    \begin{equation}
    	\begin{split}
    		\mathbf{P}_{\boxtimes}^{\alpha,k} &= \sum_{i=0}^{N_1 - 1} \sum_{j=0}^{N_2 - 1} (\lambda_{1,i}^{\alpha,k} + \lambda_{2,j}^{\alpha,k}) \\
    		&\quad (\mathbf{w}_{1,i}^{\alpha,k} \otimes \mathbf{w}_{2,j}^{\alpha,k}) (\mathbf{w}_{1,i}^{\alpha,k} \otimes \mathbf{w}_{2,j}^{\alpha,k})^T.
    	\end{split}
    	\label{33}
    \end{equation}
    The SVD of the full $N_1 N_2 \times N_1 N_2$ matrix $\mathbf{P}_{\boxtimes}^{\alpha,k}$ has a per-evaluation computational complexity of $O(N_1^3 N_2^3)$\cite{ref47},\cite{ref49}. To identify the optimal parameters $(\alpha^*, k^*)$, we employ Bayesian optimization over the continuous domain $[0, 1] \times [0, 1]$, which efficiently explores the parameter space by maximizing the expected improvement acquisition function. Denoting the total number of objective function evaluations by $T_{\text{eval}}$, the overall computational complexity is $O(T_{\text{eval}} \cdot (N_1 N_2)^3)$. For the undirected graph scenario, the Kronecker eigendecomposition structure~\eqref{33} reduces the per-evaluation complexity to $O(N_1^3 + N_2^3)$, yielding an overall complexity of $O(T_{\text{eval}} \cdot (N_1^3 + N_2^3))$.
    
    Based on the SVD of the UGRM $\mathbf{P}_{\boxtimes}^{\alpha,k}$, given by (\ref{31}), we can follow the approach in Section \ref{sec:chapter3} to define the GFT on the directed Cartesian product graph $\mathcal{G}$.
    
    \textit{Definition 2}: Let $\mathcal{G}$ be the directed Cartesian product graph with UGRM $\mathbf{P}_{\boxtimes}^{\alpha,k}$ given by (\ref{7}), and let $\mathbf{U}_{\boxtimes}^{\alpha,k}, \mathbf{V}_{\boxtimes}^{\alpha,k}$ be the $N \times N$ orthogonal matrices found in (\ref{31}). The UGRM-GFT-I $\mathcal{F}_{\boxtimes}^{\alpha,k}: \mathbb{R}^N \rightarrow \mathbb{R}^{2N}$ is defined as
    \begin{equation}
    	\begin{split}
    		\mathcal{F}_{\boxtimes}^{\alpha,k} \mathbf{x}: &= \frac{1}{2} \begin{pmatrix}
    			(\mathbf{U}_{\boxtimes}^{\alpha,k} + \mathbf{V}_{\boxtimes}^{\alpha,k})^T \mathbf{x} \\
    			(\mathbf{U}_{\boxtimes}^{\alpha,k} - \mathbf{V}_{\boxtimes}^{\alpha,k})^T \mathbf{x}
    		\end{pmatrix} \\
    		&= \frac{1}{2} \begin{pmatrix}
    			(\mathbf{u}_{0}^{\alpha,k} + \mathbf{v}_{0}^{\alpha,k})^T \mathbf{x} \\
    			\vdots \\
    			(\mathbf{u}_{N-1}^{\alpha,k} + \mathbf{v}_{N-1}^{\alpha,k})^T \mathbf{x} \\
    			(\mathbf{u}_{0}^{\alpha,k} - \mathbf{v}_{0}^{\alpha,k})^T \mathbf{x} \\
    			\vdots \\
    			(\mathbf{u}_{N-1}^{\alpha,k} - \mathbf{v}_{N-1}^{\alpha,k})^T \mathbf{x}
    		\end{pmatrix},
    	\end{split}
    	\label{34}
    \end{equation}
    and its inverse UGRM-GFT-I $(\mathcal{F}_{\boxtimes}^{\alpha,k})^{-1}: \mathbb{R}^{2N} \mapsto \mathbb{R}^N$ is given by
    \begin{equation}
    	\begin{split}
    		(\mathcal{F}_{\boxtimes}^{\alpha,k})^{-1} \begin{pmatrix} \mathbf{z}_{1}^{\alpha,k} \\ \mathbf{z}_{2}^{\alpha,k} \end{pmatrix} 
    		:&=\frac{1}{2} \bigl( \mathbf{U}_{\boxtimes}^{\alpha,k} (\mathbf{z}_{1}^{\alpha,k} + \mathbf{z}_{2}^{\alpha,k}) \\ 
    		& \quad + \mathbf{V}_{\boxtimes}^{\alpha,k} (\mathbf{z}_{1}^{\alpha,k} - \mathbf{z}_{2}^{\alpha,k}) \bigr) \\
    		&=\frac{1}{2} \sum_{t=0}^{N - 1} \bigl( z_{1,t}^{\alpha,k} + z_{2,t}^{\alpha,k} \bigr) \mathbf{u}_t^{\alpha,k} \\
    		& \quad + \bigl( z_{1,t}^{\alpha,k} - z_{2,t}^{\alpha,k} \bigr) \mathbf{v}_t^{\alpha,k}
    	\end{split}
    	\label{35}
    \end{equation}
    for all $\mathbf{z}_{l}^{\alpha,k} = [z_{l,0}^{\alpha,k}, \ldots, z_{l,N-1}^{\alpha,k}]^T \in \mathbb{R}^N$, $l = 1, 2$. 
    
    By the orthogonality of the matrices $\mathbf{U}_\boxtimes^{\alpha,k}$, $\mathbf{V}_\boxtimes^{\alpha,k}$, one may verify that
    \begin{equation}
    	\|\mathcal{F}_{\boxtimes}^{\alpha,k} \mathbf{x}\|_2 = \|\mathbf{x}\|_2
    	\label{36}
    \end{equation}
    and
    \begin{equation}
    	(\mathcal{F}_{\boxtimes}^{\alpha,k})^{-1} \mathcal{F}_{\boxtimes}^{\alpha,k} \mathbf{x} = \mathbf{x}
    	\label{37}
    \end{equation}
    hold for all signals $\mathbf{x}$ on the directed Cartesian product graph $\mathcal{G}$.
   
  \begin{theorem} \label{thm2}
  Let $\mathcal{G}$ be the Cartesian product of directed graphs $\mathcal{G}_1$ and $\mathcal{G}_2$, $\mathbf{P}_{\boxtimes}^{\alpha,k}$ be the UGRM \eqref{7} on $\mathcal{G}$, and $\mathbf{u}_t, \mathbf{v}_t, \sigma_t$, $0 \leq t \leq N - 1$, be as in \eqref{31}, where $N = N_1N_2$, $N_1$ and $N_2$ are the orders of graphs $\mathcal{G}_1$ and $\mathcal{G}_2$ respectively.  For a bandwidth $M \in \{1, \dots, N\}$, define the spectral approximation $\mathbf{x}_{M}^{\alpha,k}$ of a signal $\mathbf{x} \in \mathbb{R}^N$ by
  	\begin{equation} 
  		\mathbf{x}_{M}^{\alpha,k} = \frac{1}{2} \sum_{t=0}^{M-1} \left[ \mathbf{u}_t^{\alpha,k}(\mathbf{u}_t^{\alpha,k})^T + \mathbf{v}_t^{\alpha,k}(\mathbf{v}_t^{\alpha,k})^T \right]\mathbf{x}.
  		\label{38}
  	\end{equation}
  	
  	\begin{enumerate}
  		\item[\textbf{(1)}] \textbf{Case $\boldsymbol{\beta < 0}$.} Let the singular values be ordered non-decreasingly as
  		\[
  		0 \le \sigma_0^{\alpha,k} \le \sigma_1^{\alpha,k} \le \cdots \le \sigma_{N-1}^{\alpha,k}.
  		\]
  		If $\sigma_{M-1}^{\alpha,k} > 0$, then
  		\begin{equation}
  			\begin{aligned}
  				\|\mathbf{x} - \mathbf{x}_{M,\boxtimes}^{\alpha,k}\|_2 
  				&\le \frac{1}{2\sigma_{M-1}^{\alpha,k}} \left(\|\mathbf{P}_{\boxtimes}^{\alpha,k}\mathbf{x}\|_2 \right. \left. + \|(\mathbf{P}_{\boxtimes}^{\alpha,k})^T\mathbf{x}\|_2\right) \\
  				&\le \frac{1}{2\sigma_{M-1}^{\alpha,k}} \Bigl( \|(\mathbf{P}_1^{\alpha,k} \otimes \mathbf{I}_{N_2})\mathbf{x}\|_2 \\
  				&\qquad + \|((\mathbf{P}_1^{\alpha,k})^T \otimes \mathbf{I}_{N_2})\mathbf{x}\|_2 \\
  				&\qquad + \|(\mathbf{I}_{N_1} \otimes \mathbf{P}_2^{\alpha,k})\mathbf{x}\|_2 \\
  				&\qquad + \|(\mathbf{I}_{N_1} \otimes (\mathbf{P}_2^{\alpha,k})^T)\mathbf{x}\|_2 \Bigr).
  			\end{aligned}
  		    \label{39}
  		\end{equation}

  		\item[\textbf{(2)}] \textbf{Case $\boldsymbol{\beta \ge 0}$.} Let the singular values be ordered nonincreasingly as
  		\[
  		\sigma_0^{\alpha,k} \ge \sigma_1^{\alpha,k} \ge \cdots \ge \sigma_{N-1}^{\alpha,k} \ge 0.
  		\]
  		Suppose $\mathbf{x}$ satisfies the \emph{operator consistency} condition: there exists a constant $C>0$ such that
  		\[
  		 (\mathbf{u}_t^{\alpha,k})^T \mathbf{x}  \le C \sigma_t^{\alpha,k}
  		\quad\text{and}\quad
  		 (\mathbf{v}_t^{\alpha,k})^T \mathbf{x}  \le C \sigma_t^{\alpha,k}
  		\]
  		for all $t \ge M$. Then
  		\begin{equation} \label{40}
  			\|\mathbf{x} - \mathbf{x}_{M,\boxtimes}^{\alpha,k}\|_2 \le C \sqrt{\sum_{t=M}^{N-1} (\sigma_t^{\alpha,k})^2}.
  		\end{equation}
  	\end{enumerate}
  \end{theorem}

   \begin{proof}
   	See Appendix B.
   \end{proof}

    \section{SVD-BASED UGRM-GFT-II ON DIRECTED  PRODUCT GRAPHS}
       \label{sec:chapter5}
    In some application scenarios (e.g., spatio-temporal signal processing), graph signals often exhibit significantly different correlation characteristics in different directions. To address this characteristic, the corresponding GFT design should be able to effectively capture and characterize the differentiated spectral properties of the signal in the direction dimension. In this section, based on the SVD of the UGRM on $\mathcal{G}_1$ and $\mathcal{G}_2$, we introduce another UGRM-GFT-II $\mathcal{F}_\otimes^{\alpha,k}$, which is defined on the directed Cartesian product graph $\mathcal{G}$.  Compared to $\mathcal{F}_\boxtimes^{\alpha,k}$, the computational complexity of the new GFT $\mathcal{F}_\otimes^{\alpha,k}$ is lower. On the other hand, they maintain similar performance, an advantage verified in the numerical experiments in Section \ref{sec:chapter6}.
    
    Let \( \mathcal{G}_1 = (\mathcal{V}_1, \mathcal{E}_1) \) and \( \mathcal{G}_2 = (\mathcal{V}_2, \mathcal{E}_2) \) be two directed graphs, and denote their UGRMs and orders by \( \mathbf{P}_{l}^{\alpha,k} \) and \( N_l, l = 1, 2 \) respectively. For the UGRMs \( \mathbf{P}_{l}^{\alpha,k}, l = 1, 2 \), we take their SVDs
    \begin{equation}
    	\mathbf{P}_{l}^{\alpha,k} = \mathbf{U}_{l}^{\alpha,k} \mathbf{\Sigma}_{l}^{\alpha,k} (\mathbf{V}_{l}^{\alpha,k})^T = \sum_{i=0}^{N_l - 1} \sigma_{l,i}^{\alpha,k} u_{l,i}^{\alpha,k} (v_{l,i}^{\alpha,k})^T, 
    	\label{41}
    \end{equation}
    where \( \sigma_{l,i}^{\alpha,k}, 0 \leq i \leq N_l - 1 \), are singular values of the UGRM \( \mathbf{P}_{l}^{\alpha,k} \) with a nondecreasing order. \( \mathbf{U}_{l}^{\alpha,k} = [\mathbf{u}_{l,0}^{\alpha,k}, \ldots, \mathbf{u}_{l,N_l - 1}^{\alpha,k}] \) and \( \mathbf{V}_{l}^{\alpha,k} = [\mathbf{v}_{l,0}^{\alpha,k}, \ldots, \mathbf{v}_{l,N_l - 1}^{\alpha,k}] \) are orthonormal matrices. Set
    \begin{equation}
    	\mathbf{U}_{\otimes}^{\alpha,k} = \mathbf{U}_{1}^{\alpha,k} \otimes \mathbf{U}_{2}^{\alpha,k} \quad \text{and} \quad \mathbf{V}_{\otimes}^{\alpha,k} = \mathbf{V}_{1}^{\alpha,k} \otimes \mathbf{V}_{2}^{\alpha,k}. 
    	\label{42}
    \end{equation}
    With the help of SVDs of UGRMs \( \mathbf{P}_{l}^{\alpha,k}, l = 1, 2 \), we propose the second GFT on the directed product graph \( \mathcal{G} \) as follows.
    
    \textit{Definition 3}: Let directed graphs \( \mathcal{G}_l, l \in \{1, 2\} \), have orders \( N_l \) with UGRMs \( \mathbf{P}_{l}^{\alpha,k} \), \( \mathbf{U}_{l}^{\alpha,k}\) and \(\mathbf{V}_{l}^{\alpha,k} \) be the orthogonal matrices found in (\ref{41}), \( \mathbf{U}_{\otimes}^{\alpha,k} \) and \( \mathbf{V}_{\otimes}^{\alpha,k} \) be the orthogonal matrices found in (\ref{42}), and set \( N = N_1 N_2 \). The UGRM-GFT-II 
    $\mathcal{F}_\otimes^{\alpha,k}: \mathbb{R}^N \mapsto \mathbb{R}^{2N}$ on $\mathcal{G}$  is defined as
    \begin{equation}
    	\mathcal{F}_{\otimes}^{\alpha,k} \mathbf{x} := \frac{1}{2} \begin{pmatrix} (\mathbf{U}_{\otimes}^{\alpha,k} + \mathbf{V}_{\otimes}^{\alpha,k})^T \mathbf{x} \\ (\mathbf{U}_{\otimes}^{\alpha,k} - \mathbf{V}_{\otimes}^{\alpha,k})^T \mathbf{x} \end{pmatrix}, 
    	\label{43}
    \end{equation}
    and its inverse UGRM-GFT-II $(\mathcal{F}_{\otimes}^{\alpha,k})^{-1}: \mathbb{R}^{2N} \mapsto \mathbb{R}^N$ is given by
   \begin{equation}
   	\begin{split}
   		(\mathcal{F}_{\otimes}^{\alpha,k})^{-1} \begin{pmatrix} \mathbf{z}_{1}^{\alpha,k} \\ \mathbf{z}_{2}^{\alpha,k} \end{pmatrix} :&= \frac{1}{2} \bigl( \mathbf{U}_{\otimes}^{\alpha,k} (\mathbf{z}_{1}^{\alpha,k} + \mathbf{z}_{2}^{\alpha,k}) \\
   		&\quad + \mathbf{V}_{\otimes}^{\alpha,k} (\mathbf{z}_{1}^{\alpha,k} - \mathbf{z}_{2}^{\alpha,k}) \bigr).
   	\end{split}
   	\label{44}
   \end{equation}

    The computational complexity to evaluate the GFT $\mathcal{F}_{\otimes}^{\alpha,k}$ is $O(N_1^3+N_2^3)$ \cite{ref47,ref49}. Following the parameter selection procedure in Section~\ref{sec:chapter4}, we use the same Bayesian optimization strategy to determine the optimal parameters \((\alpha^*, k^*)\). Denoting the total number of objective function evaluations by $T_{\mathrm{eval}}$, the overall computational complexity is $O(T_{\mathrm{eval}} (N_1^3 +N_2^3))$. For the undirected graph scenario, the Kronecker eigendecomposition structure \eqref{33} reduces the per-evaluation complexity to $O(N_1^3 + N_2^3)$, yielding an overall complexity of $O(T_{\mathrm{eval}} (N_1^3 + N_2^3))$.
    
    By the orthogonality of the matrices $\mathbf{U}_l^{\alpha,k}, \mathbf{V}_l^{\alpha,k}, l = 1,2$, one may verify that
    \begin{equation}
    	\|\mathcal{F}_{\otimes}^{\alpha,k} \mathbf{x}\|_2 = \|\mathbf{x}\|_2
    	\label{45}
    \end{equation}
    and
    \begin{equation}
    	(\mathcal{F}_{\otimes}^{\alpha,k})^{-1} \mathcal{F}_{\otimes}^{\alpha,k} \mathbf{x} = \mathbf{x}
    	\label{46}
    \end{equation}
    hold for all signals $\mathbf{x}$ on the directed Cartesian product graph $\mathcal{G}$. 
    
    \begin{theorem} \label{thm3}
    	Let $\mathcal{G}$ be the Cartesian product of directed graphs $\mathcal{G}_1$ and $\mathcal{G}_2$, $\sigma_{l,i}^{\alpha,k}, \mathbf{u}_{l,i}^{\alpha,k}, \mathbf{v}_{l,i}^{\alpha,k}$, $0 \leq i \leq N_l - 1$, $l = 1, 2$, be as in \eqref{41}, and $\mu_t^{\alpha,k}$, $0 \leq t \leq N - 1$, be the ascending order of $\sigma_{1,i}^{\alpha,k} + \sigma_{2,j}^{\alpha,k}$, $0 \leq i \leq N_1 - 1$, $0 \leq j \leq N_2 - 1$, where $N = N_1 N_2$. For a bandwidth $M$, define the spectral approximation $\mathbf{x}_{M,\otimes}^{\alpha,k}$ of a signal $\mathbf{x}$ using a selected index set $\mathcal{S}_M$ by
    	\begin{equation}
    		\begin{aligned}
    			\mathbf{x}_{M, \otimes}^{\alpha,k} = \frac{1}{2} \sum_{(i,j) \in S_M} \Biggl( & (\mathbf{u}_{1,i}^{\alpha,k} \otimes \mathbf{u}_{2,j}^{\alpha,k})(\mathbf{u}_{1,i}^{\alpha,k} \otimes \mathbf{u}_{2,j}^{\alpha,k})^T \mathbf{x} \\
    			& + (\mathbf{v}_{1,i}^{\alpha,k} \otimes \mathbf{v}_{2,j}^{\alpha,k})(\mathbf{v}_{1,i}^{\alpha,k} \otimes \mathbf{v}_{2,j}^{\alpha,k})^T \mathbf{x} \Biggr),
    		\end{aligned}
    		\label{47}
    	\end{equation}
    	\begin{enumerate}
    		\item[\textbf{(1)}] \textbf{Case $\boldsymbol{\beta < 0}$.} Let $\mu_t$ denote the values of sums $\sigma_{1,i}^{\alpha,k} + \sigma_{2,j}^{\alpha,k}$ ordered non-decreasingly as
    		\[
    		0 \le \mu_0^{\alpha,k} \le \mu_1^{\alpha,k} \le \cdots \le \mu_{N-1}^{\alpha,k}.
    		\]
    		Let $\mathcal{S}_M$ correspond to the first $M$ indices. If $\mu_{M-1}^{\alpha,k} > 0$, then
    		\begin{equation} 
    			\begin{aligned}
    				\|\mathbf{x} - \mathbf{x}_{M,\otimes}^{\alpha,k}\|_2
    				&\le \frac{1}{2\mu_{M-1}^{\alpha,k}}
    				\Bigl(
    				\|(\mathbf{P}^{\alpha,k}_1 \otimes \mathbf{I}_{N_2})\mathbf{x}\|_2  \\
    				&\qquad
    				+ \|(\mathbf{I}_{N_1} \otimes \mathbf{P}^{\alpha,k}_2)\mathbf{x}\|_2  \\
    				&\qquad
    				+ \|((\mathbf{P}^{\alpha,k}_1)^T \otimes \mathbf{I}_{N_2})\mathbf{x}\|_2  \\
    				&\qquad
    				+ \|(\mathbf{I}_{N_1} \otimes (\mathbf{P}^{\alpha,k}_2)^T)\mathbf{x}\|_2
    				\Bigr).
    			\end{aligned}
    			\label{48}
    		\end{equation}
    		
    		\item[\textbf{(2)}] \textbf{Case $\boldsymbol{\beta \ge 0}$.} Let $\mu_t^{\alpha,k}$ denote the values of sums $\sigma_{1,i}^{\alpha,k} + \sigma_{2,j}^{\alpha,k}$ ordered nonincreasingly as
    		\[
    		\mu_0^{\alpha,k} \ge \mu_1^{\alpha,k} \ge \cdots \ge \mu_{N-1}^{\alpha,k} \ge 0.
    		\]
    		Let $\mathcal{S}_M$ correspond to the first $M$ indices. Suppose $\mathbf{x}$ satisfies the \emph{operator consistency} condition for the tensor basis: there exists $C_1,C_2,C>0$ such that
    		\[
    		(\mathbf{u}_{i,j}^{\alpha,k})^T \mathbf{x} \le C_1 \mu_t^{\alpha,k}
    		\quad\text{and}\quad
    		(\mathbf{v}_{i,j}^{\alpha,k})^T \mathbf{x} \le C_2 \mu_t^{\alpha,k}
    		\]
    		for all $(i,j)$ corresponding to $t \ge M$. Then
    		\begin{equation} \label{49}
    			\|\mathbf{x} - \mathbf{x}_{M,\otimes}^{\alpha,k}\|_2
    			\le C \sqrt{\sum_{t=M}^{N-1} (\mu_t^{\alpha,k})^2}.
    		\end{equation}
    	\end{enumerate}
    \end{theorem}
  
    \begin{proof}
    	See Appendix C.
    \end{proof}
    
    \begin{algorithm}[htbp]
    	\caption{Algorithm to Implement the GFT $\mathcal{F}_{\otimes}^{\alpha,k}$}
    	\textbf{Require:} Graph signal $\tilde{\mathbf{X}}$.
    	
    	\textbf{Steps:}
    	\begin{algorithmic}[1]
    		\State Do $\mathbf{Y}_1=\tilde{\mathbf{X}}\mathbf{U}_1^{\alpha,k}$ and 
    		$\tilde{\mathbf{Y}}_1=\tilde{\mathbf{X}}\mathbf{V}_1^{\alpha,k}$;
    		\State Do $\mathbf{Y}_2=(\mathbf{U}_2^{\alpha,k})^{\top}\mathbf{Y}_1$ and 
    		$\tilde{\mathbf{Y}}_2=(\mathbf{V}_2^{\alpha,k})^{\top}\tilde{\mathbf{Y}}_1$;
    		\State Do $\hat{\mathbf{X}}_1=(\mathbf{Y}_2+\tilde{\mathbf{Y}}_2)/2$ and 
    		$\hat{\mathbf{X}}_2=(\mathbf{Y}_2-\tilde{\mathbf{Y}}_2)/2$.
    	\end{algorithmic}
    	
    	\textbf{Ensure:} The first component $\hat{\mathbf{X}}_1$ and the second component
    	$\hat{\mathbf{X}}_2$ of the GFT $\mathcal{F}_{\otimes}^{\alpha,k}\,\mathrm{vec}(\mathbf{X})$.
    \end{algorithm}
    
    \begin{algorithm}[htbp]
    	\caption{Algorithm to Implement the Inverse GFT $(\mathcal{F}_{\otimes}^{\alpha,k})^{-1}$}
    	\textbf{Require:} $\mathbf{z}_1^{\alpha,k},\mathbf{z}_2^{\alpha,k}\in\mathbb{R}^{N}$.
    	
    	\textbf{Inverse vectorization:} 
    	$\mathbf{Z}_1^{\alpha,k}=\mathrm{vec}^{-1}(\mathbf{z}_1^{\alpha,k})$ and 
    	$\mathbf{Z}_2^{\alpha,k}=\mathrm{vec}^{-1}(\mathbf{z}_2^{\alpha,k})$.
    	
    	\textbf{Steps:}
    	\begin{algorithmic}[1]
    		\State Do $\mathbf{W}_1=(\mathbf{Z}_1^{\alpha,k}+\mathbf{Z}_2^{\alpha,k})(\mathbf{U}_1^{\alpha,k})^{\top}$ and 
    		$\tilde{\mathbf{W}}_1=(\mathbf{Z}_1^{\alpha,k}-\mathbf{Z}_2^{\alpha,k})(\mathbf{V}_1^{\alpha,k})^{\top}$;
    		\State Do $\mathbf{W}_2=\mathbf{U}_2^{\alpha,k}\mathbf{W}_1$ and 
    		$\tilde{\mathbf{W}}_2=\mathbf{V}_2^{\alpha,k}\tilde{\mathbf{W}}_1$;
    		\State Do $\mathbf{X}=(\mathbf{W}_2+\tilde{\mathbf{W}}_2)/2$.
    	\end{algorithmic}
    	
    	\textbf{Ensure:} $\mathbf{x}=\mathrm{vec}(\mathbf{X})=
    	(\mathcal{F}_{\otimes}^{\alpha,k})^{-1}
    	\begin{pmatrix}
    		\mathbf{z}_1^{\alpha,k}\\
    		\mathbf{z}_2^{\alpha,k}
    	\end{pmatrix}$.
    	\end{algorithm}

\begin{figure}
	\centering
	\begin{minipage}{0.49\linewidth}
		\centering
		{\footnotesize The first component of the UGRM-GFT-I $\mathcal{F}_{\boxtimes}^{\alpha,k}\mathbf{x}$.}\\[2pt]
		\includegraphics[width=\linewidth,height=4cm]{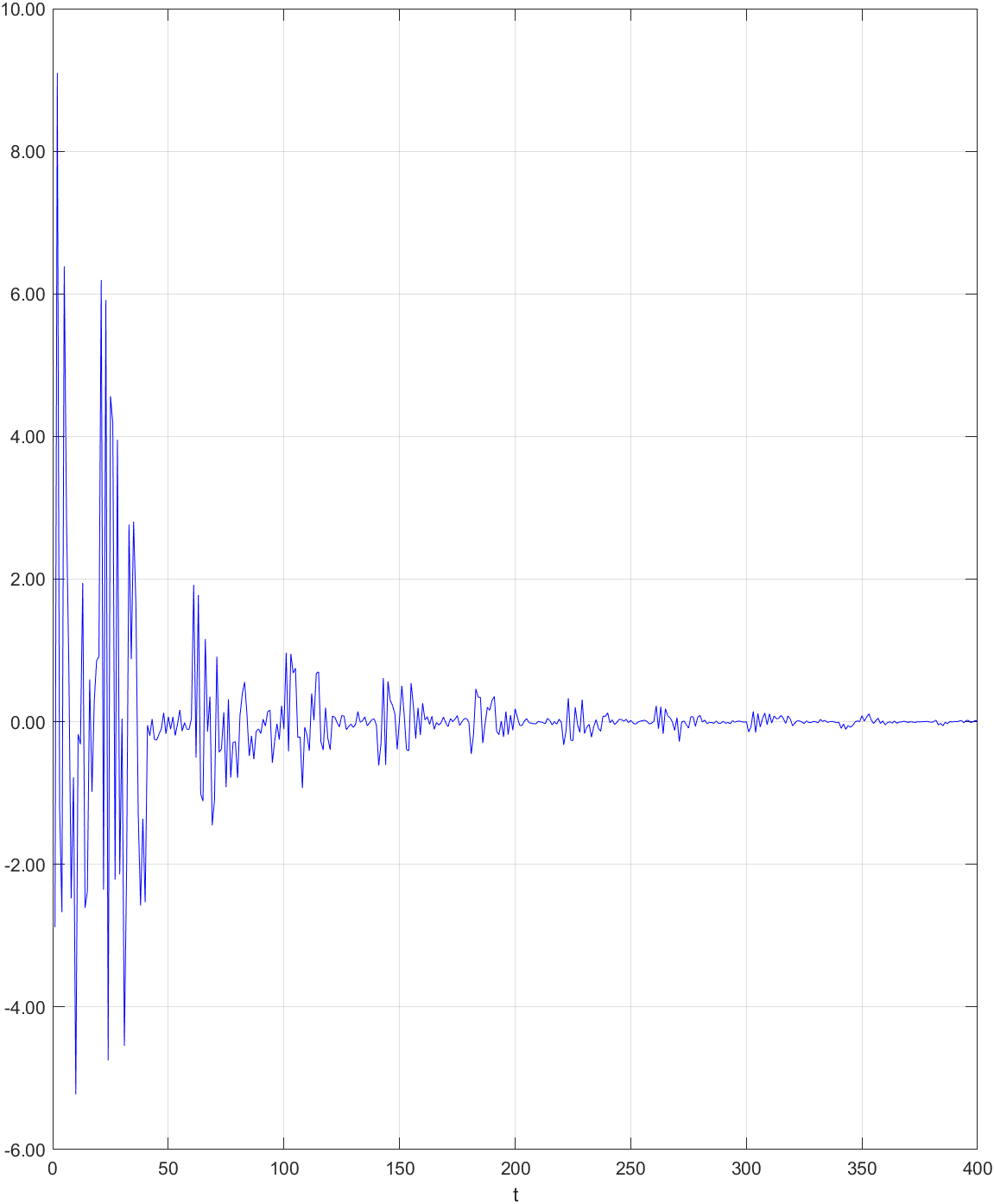}
	\end{minipage}
	\hfill
	\begin{minipage}{0.49\linewidth}
		\centering
		{\footnotesize The second component of the UGRM-GFT-I $\mathcal{F}_{\boxtimes}^{\alpha,k}\mathbf{x}$.}\\[2pt]
		\includegraphics[width=\linewidth,height=4cm]{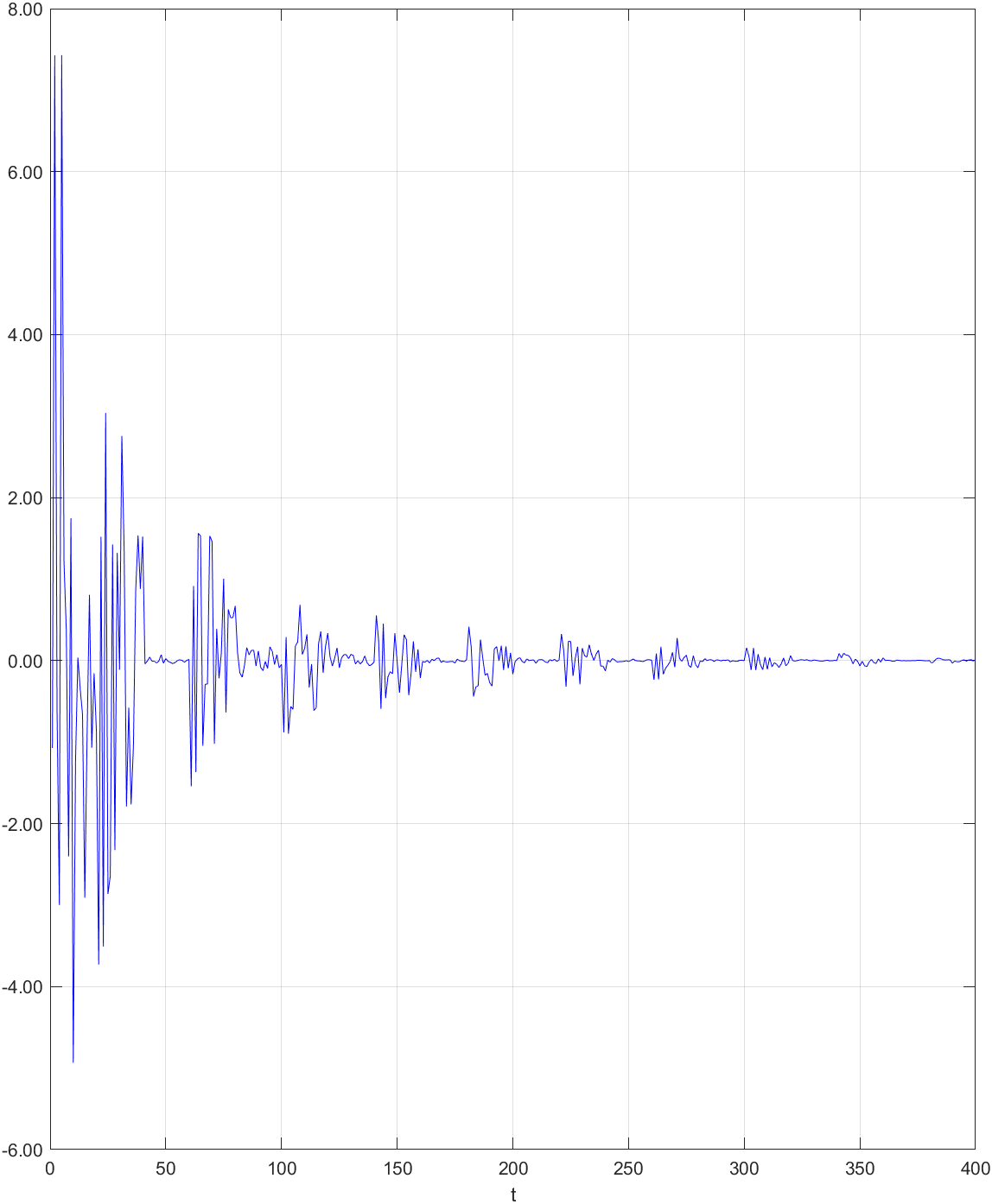}
	\end{minipage}
	
	\begin{minipage}{0.49\linewidth}
		\centering
		{\footnotesize The first component of the Lap-GFT-I $\mathcal{F}_{\boxtimes}^{\alpha,k}\mathbf{x}$.}\\[2pt]
		\includegraphics[width=\linewidth,height=4cm]{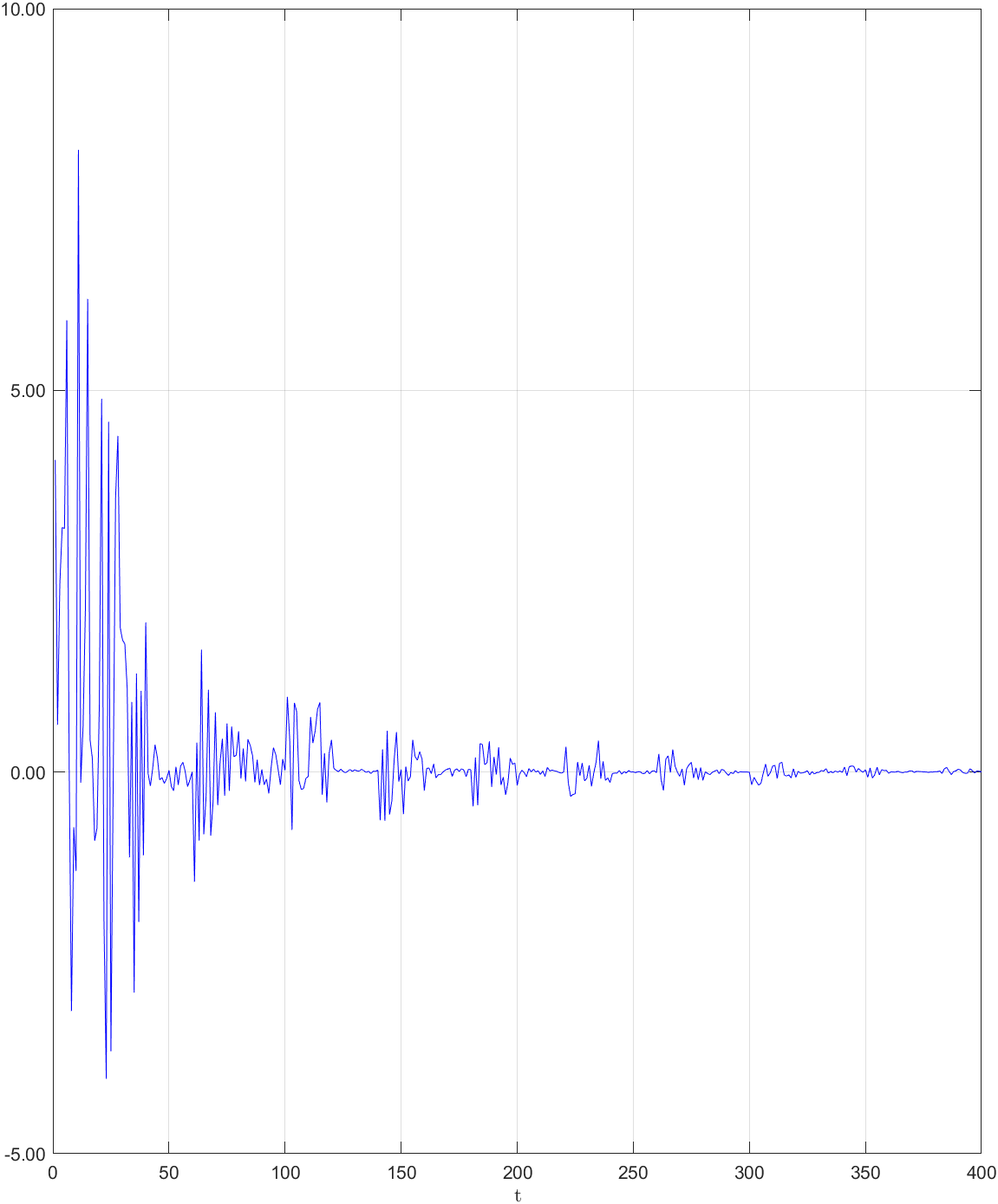}
	\end{minipage}
	\hfill
	\begin{minipage}{0.49\linewidth}
		\centering
		{\footnotesize The second component of the Lap-GFT-I $\mathcal{F}_{\boxtimes}^{\alpha,k}\mathbf{x}$.}\\[2pt]
		\includegraphics[width=\linewidth,height=4cm]{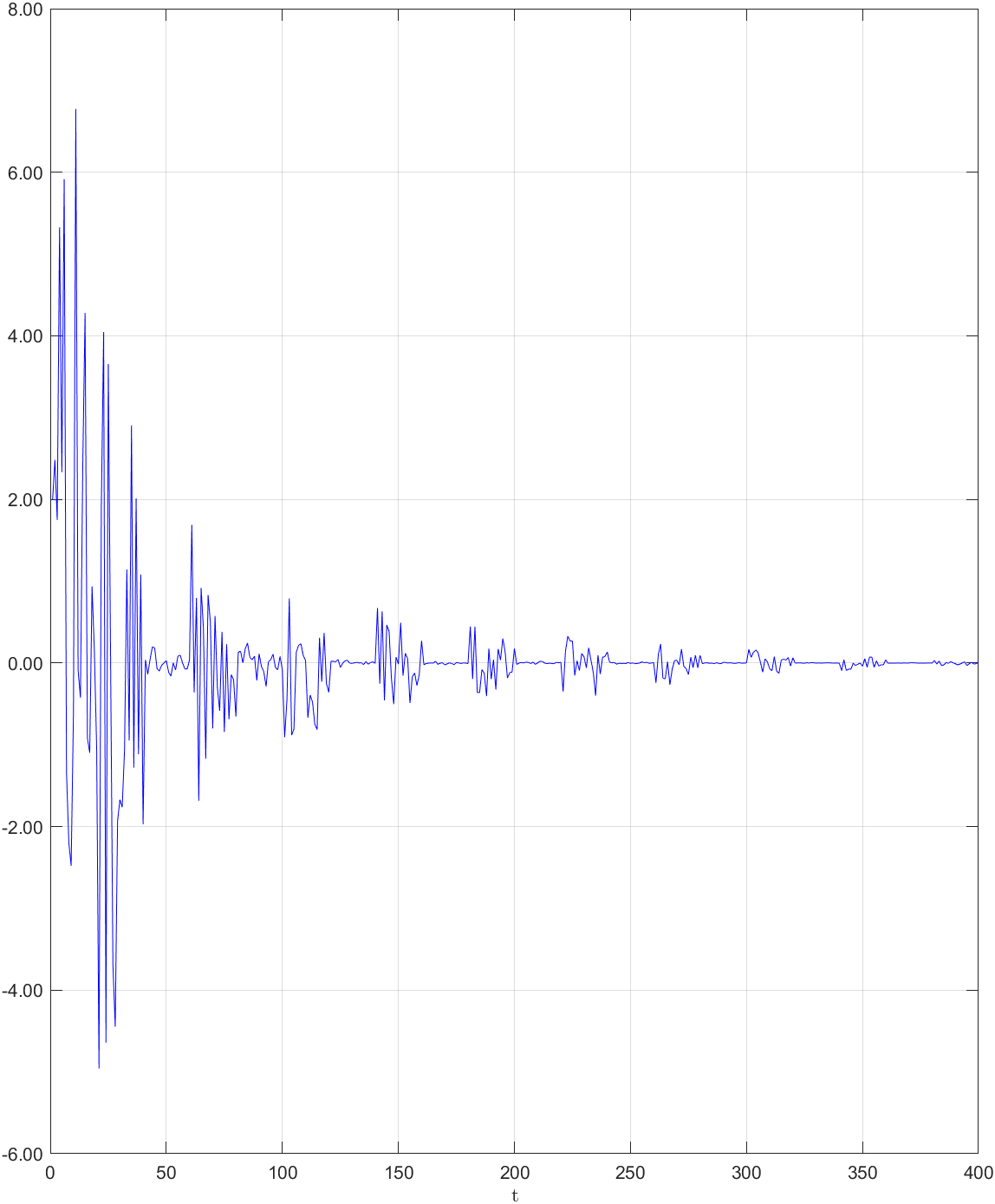}
	\end{minipage}
	
	\begin{minipage}{0.49\linewidth}
		\centering
		{\footnotesize The first component of the Adj-GFT-I $\mathcal{F}_{\boxtimes}^{\alpha,k}\mathbf{x}$.}\\[2pt]
		\includegraphics[width=\linewidth,height=4cm]{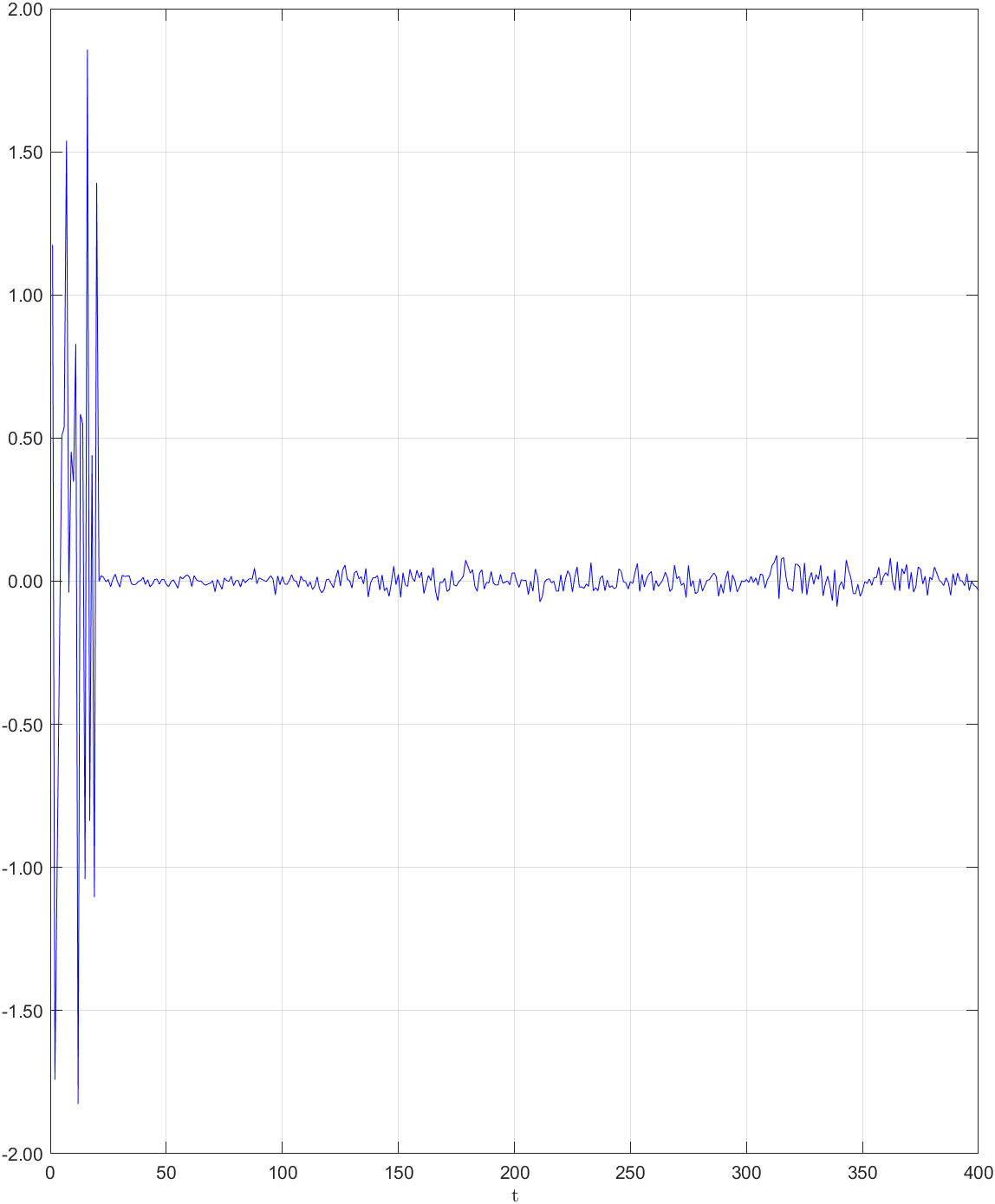}
	\end{minipage}
	\hfill
	\begin{minipage}{0.49\linewidth}
		\centering
		{\footnotesize The second component of the Adj-GFT-I $\mathcal{F}_{\boxtimes}^{\alpha,k}\mathbf{x}$.}\\[2pt]
		\includegraphics[width=\linewidth,height=4cm]{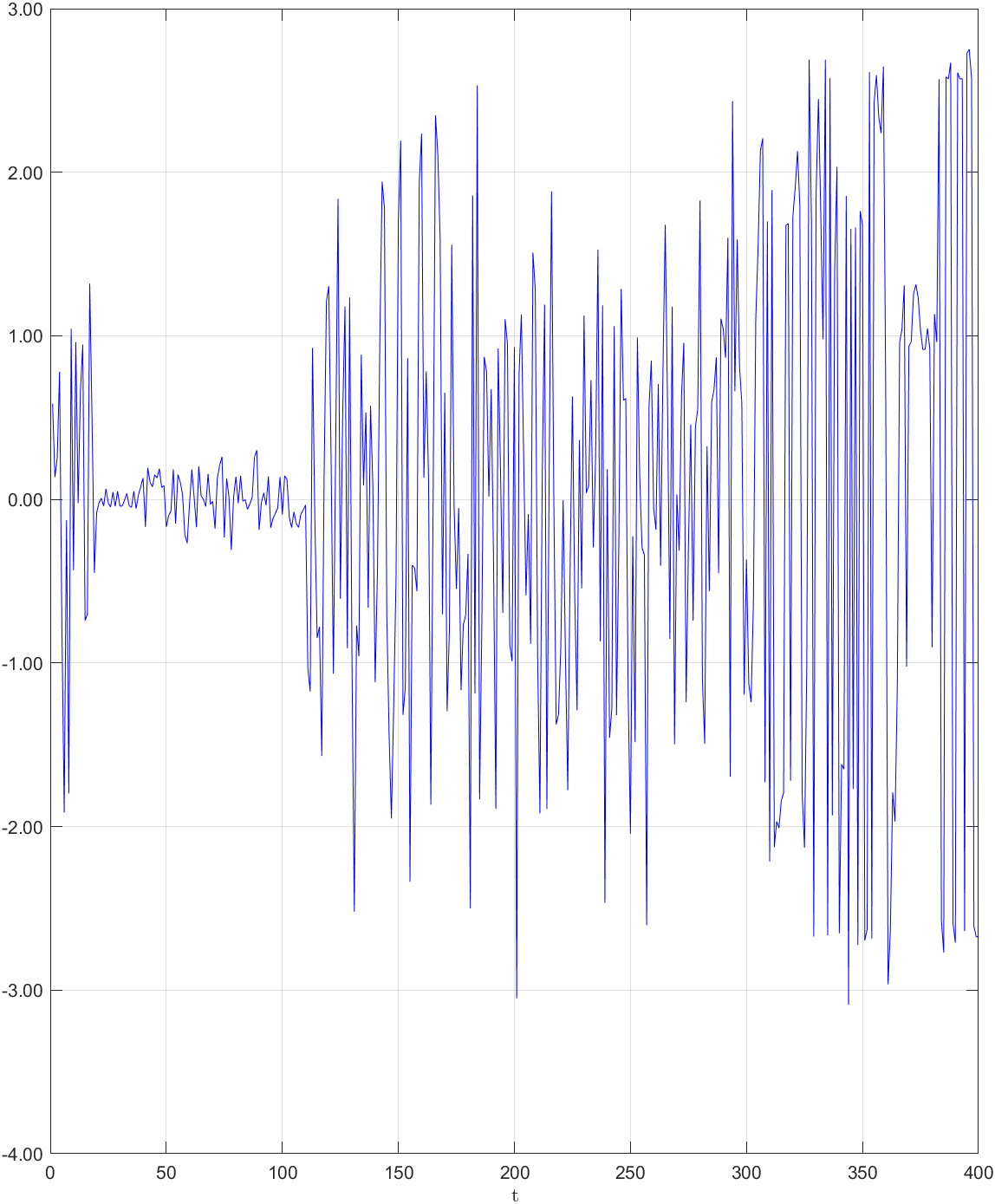}
	\end{minipage}
	
	\begin{minipage}{0.49\linewidth}
		\centering
		{\footnotesize The first component of the Id-GFT-I $\mathcal{F}_{\boxtimes}^{\alpha,k}\mathbf{x}$.}\\[2pt]
		\includegraphics[width=\linewidth,height=4cm]{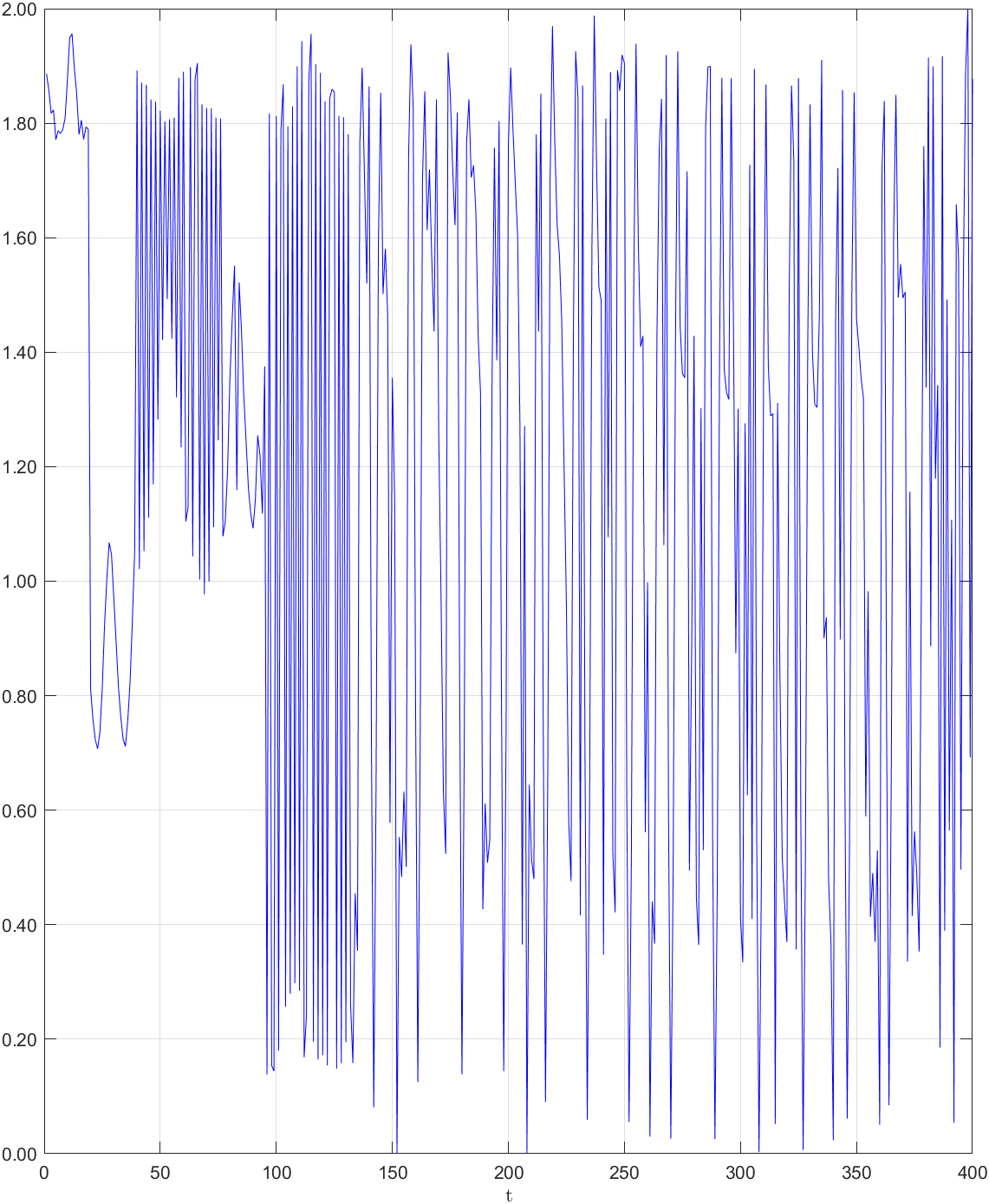}
	\end{minipage}
	\hfill
	\begin{minipage}{0.49\linewidth}
		\centering
		{\footnotesize The second component of the Id-GFT-I $\mathcal{F}_{\boxtimes}^{\alpha,k}\mathbf{x}$.}\\[2pt]
		\includegraphics[width=\linewidth,height=4cm]{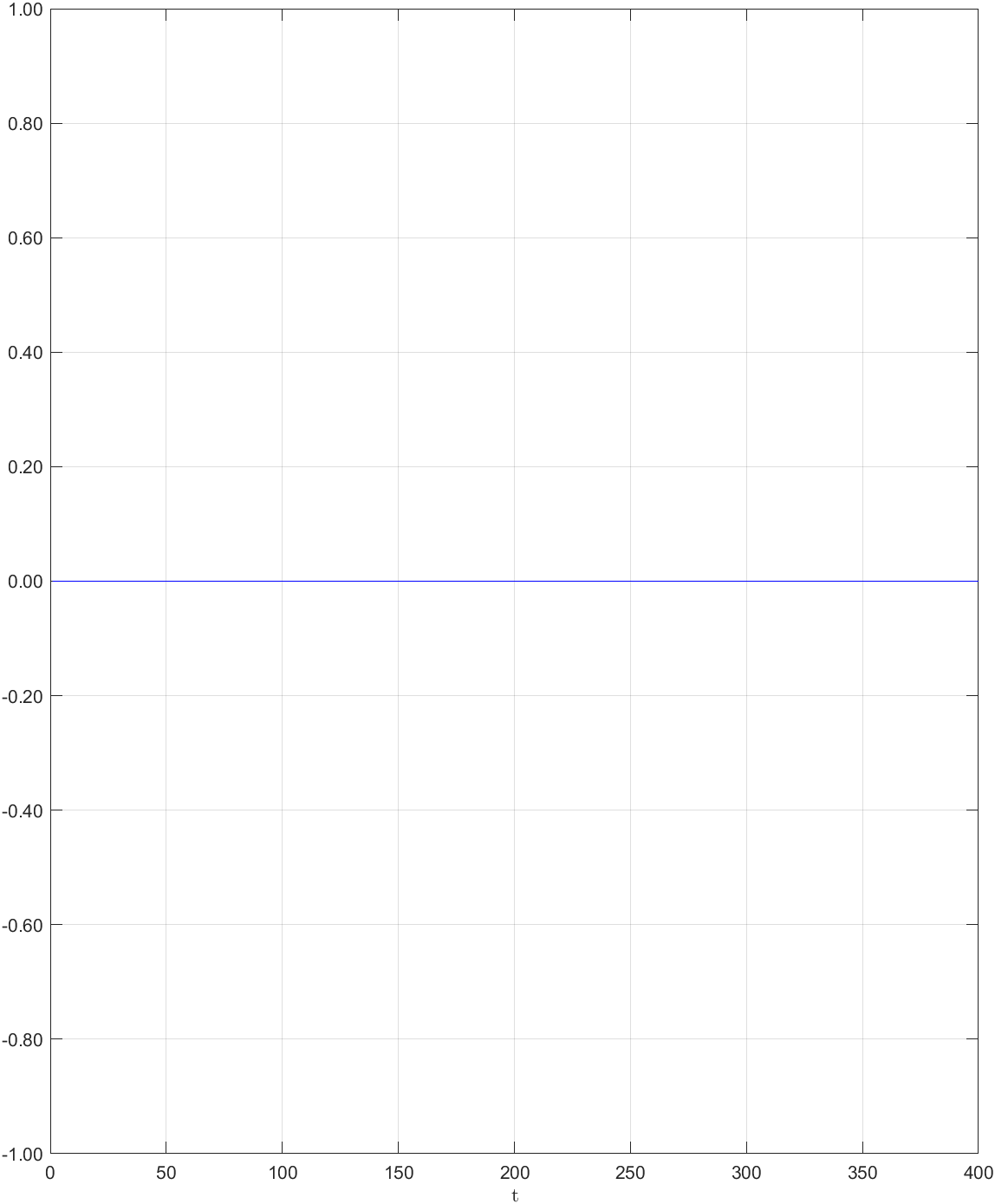}
	\end{minipage}
	
	\begin{minipage}{0.49\linewidth}
		\centering
		{\footnotesize The first component of the SLap-GFT-I $\mathcal{F}_{\boxtimes}^{\alpha,k}\mathbf{x}$.}\\[2pt]
		\includegraphics[width=\linewidth,height=4cm]{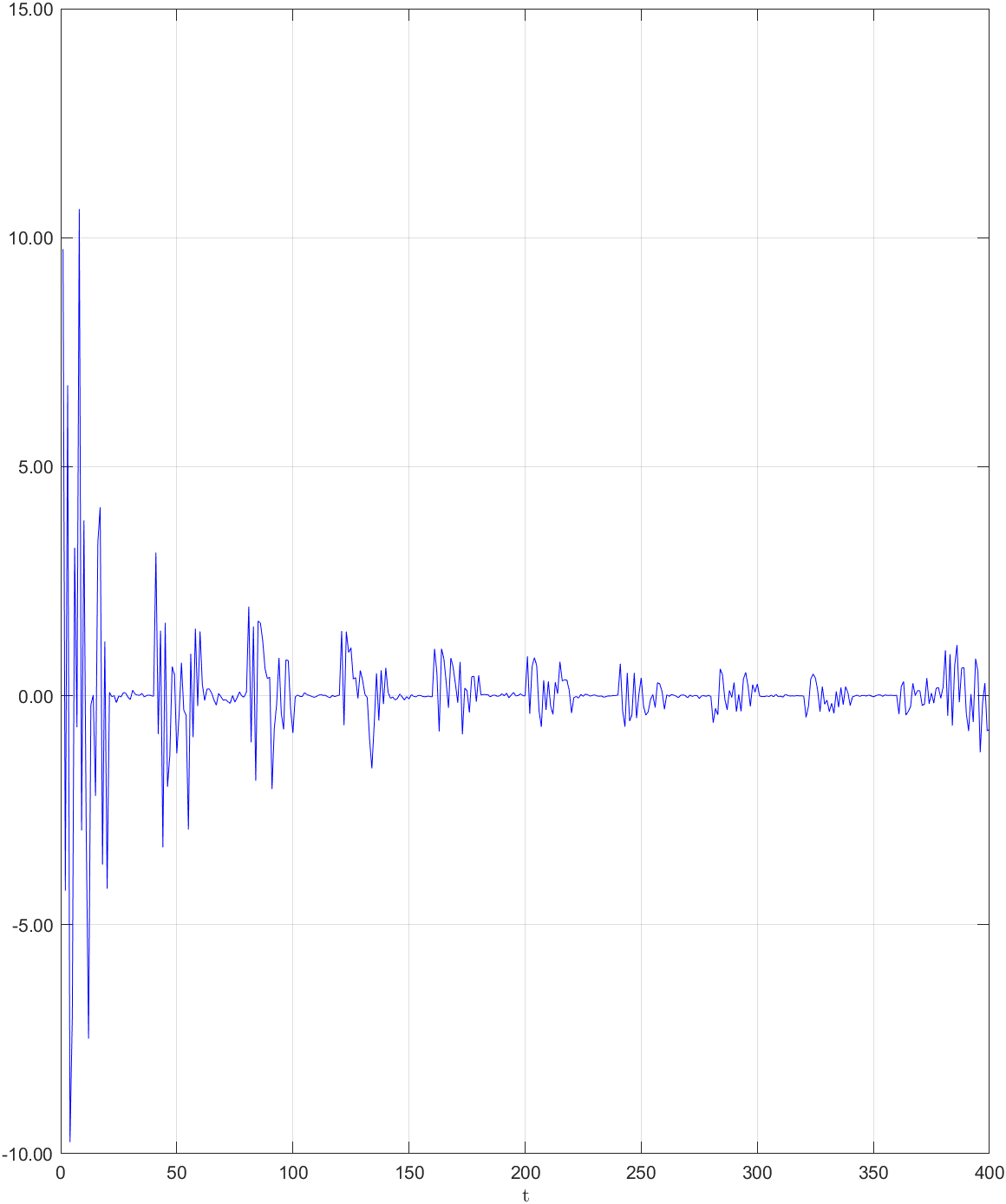}
	\end{minipage}
	\hfill
	\begin{minipage}{0.49\linewidth}
		\centering
		{\footnotesize The second component of the SLap-GFT-I $\mathcal{F}_{\boxtimes}^{\alpha,k}\mathbf{x}$.}\\[2pt]
		\includegraphics[width=\linewidth,height=4cm]{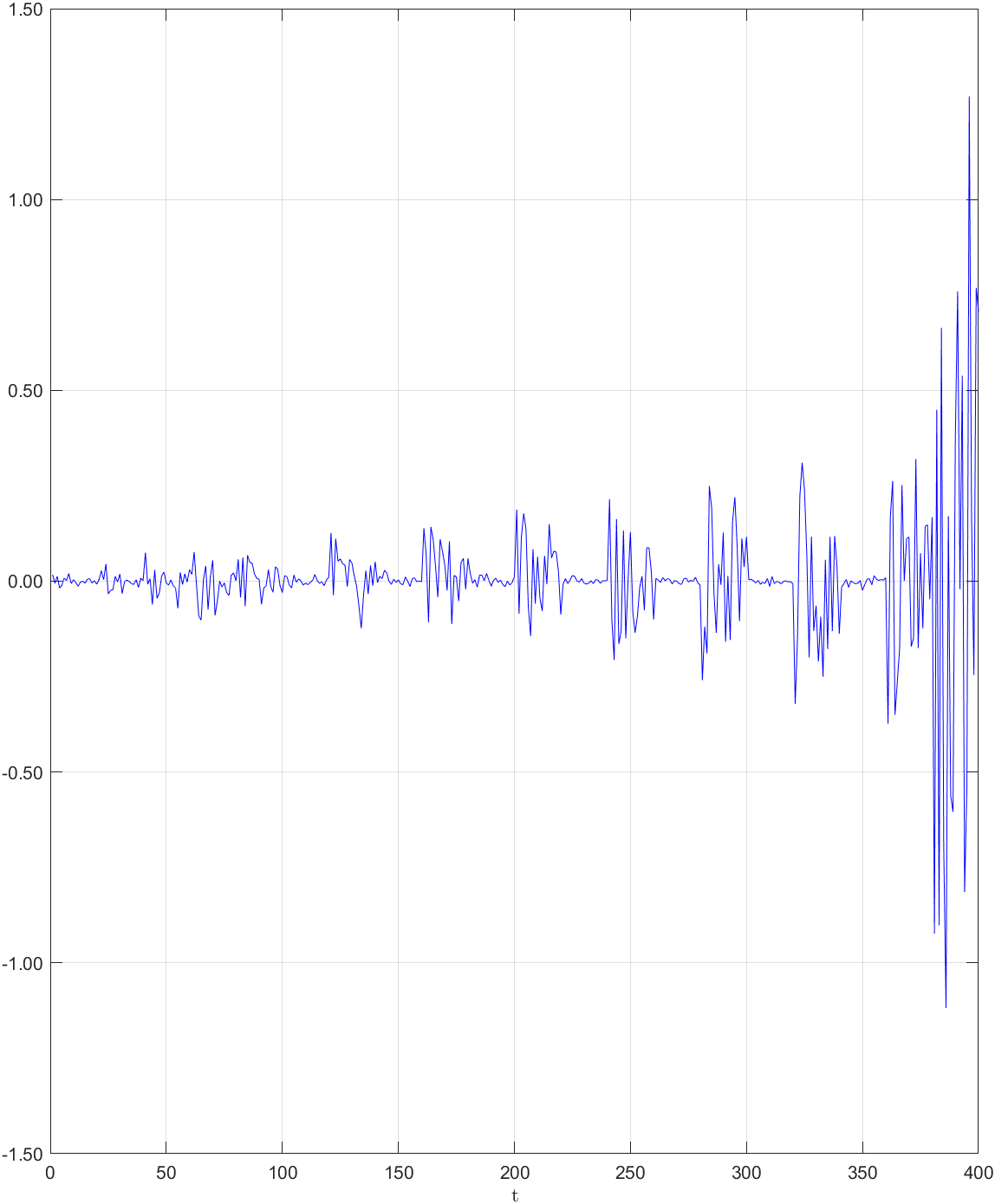}
	\end{minipage}
	
	\caption{Comparison of the first and second components produced by UGRM-GFT-I, Lap-GFT-I, Adj-GFT-I, Id-GFT-I and SLap-GFT-I under $\mathcal{F}_{\boxtimes}^{\alpha,k}$ for the signal $\mathbf{x}$.}
	\label{fig3}
\end{figure}

\begin{figure}
	\centering
	\begin{minipage}{0.49\linewidth}
		\centering
		{\footnotesize The first component of the UGRM-GFT-II $\mathcal{F}_{\otimes}^{\alpha,k}\mathbf{x}$.}\\[2pt]
		\includegraphics[width=\linewidth,height=4cm]{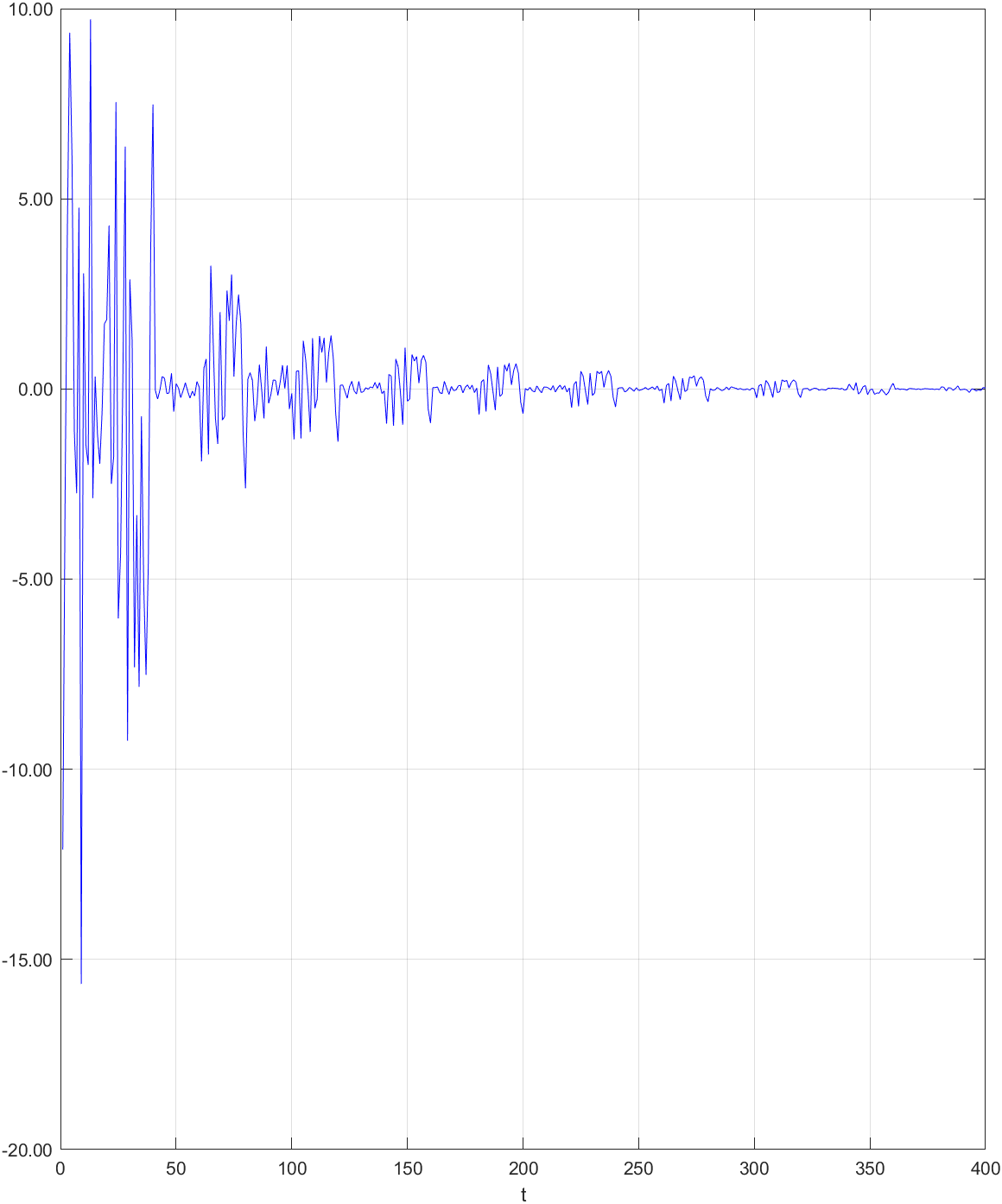}
	\end{minipage}
	\hfill
	\begin{minipage}{0.49\linewidth}
		\centering
		{\footnotesize The second component of the UGRM-GFT-II $\mathcal{F}_{\otimes}^{\alpha,k}\mathbf{x}$.}\\[2pt]
		\includegraphics[width=\linewidth,height=4cm]{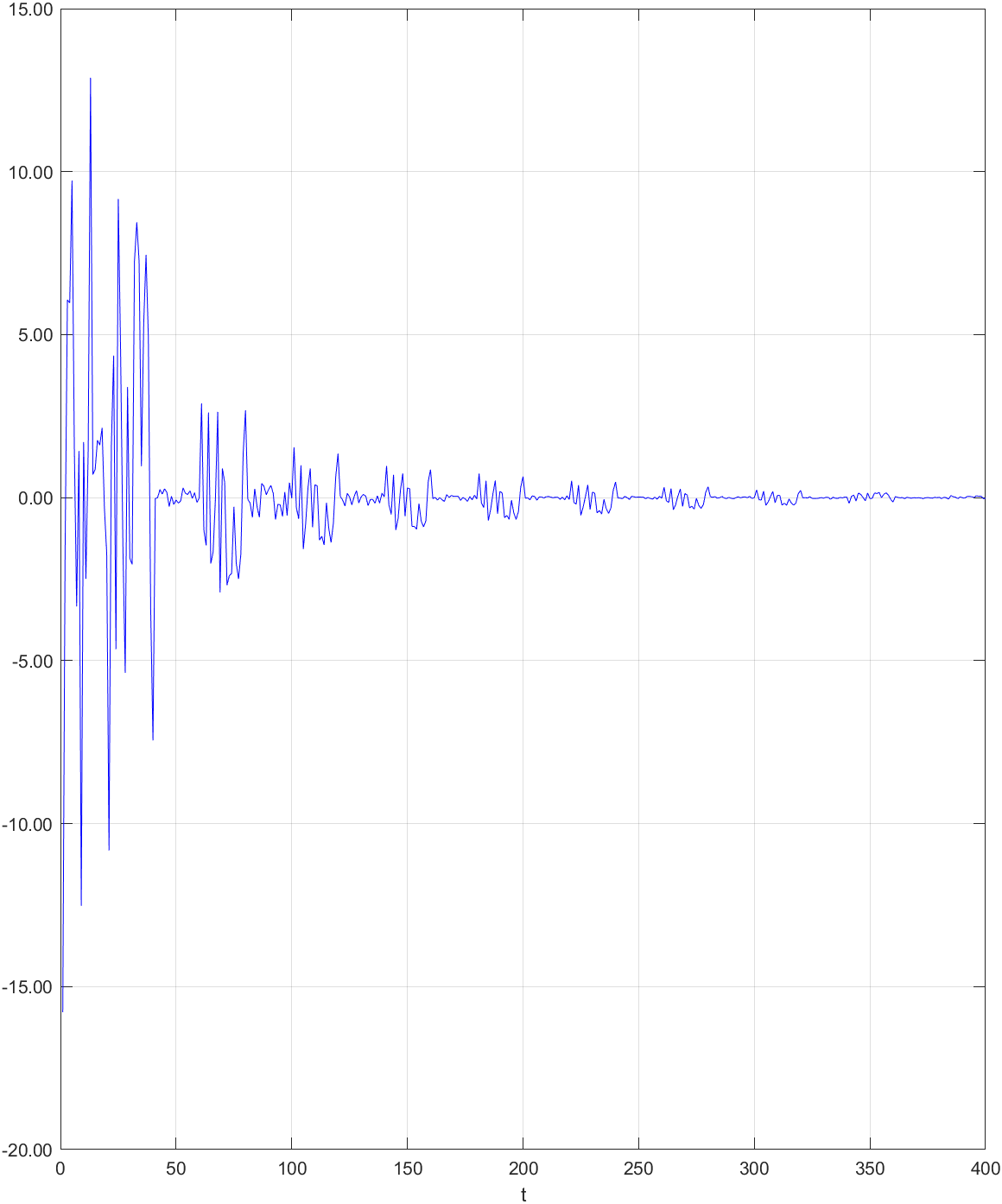}
	\end{minipage}
	
	\begin{minipage}{0.49\linewidth}
		\centering
		{\footnotesize The first component of the Lap-GFT-II $\mathcal{F}_{\otimes}^{\alpha,k}\mathbf{x}$.}\\[2pt]
		\includegraphics[width=\linewidth,height=4cm]{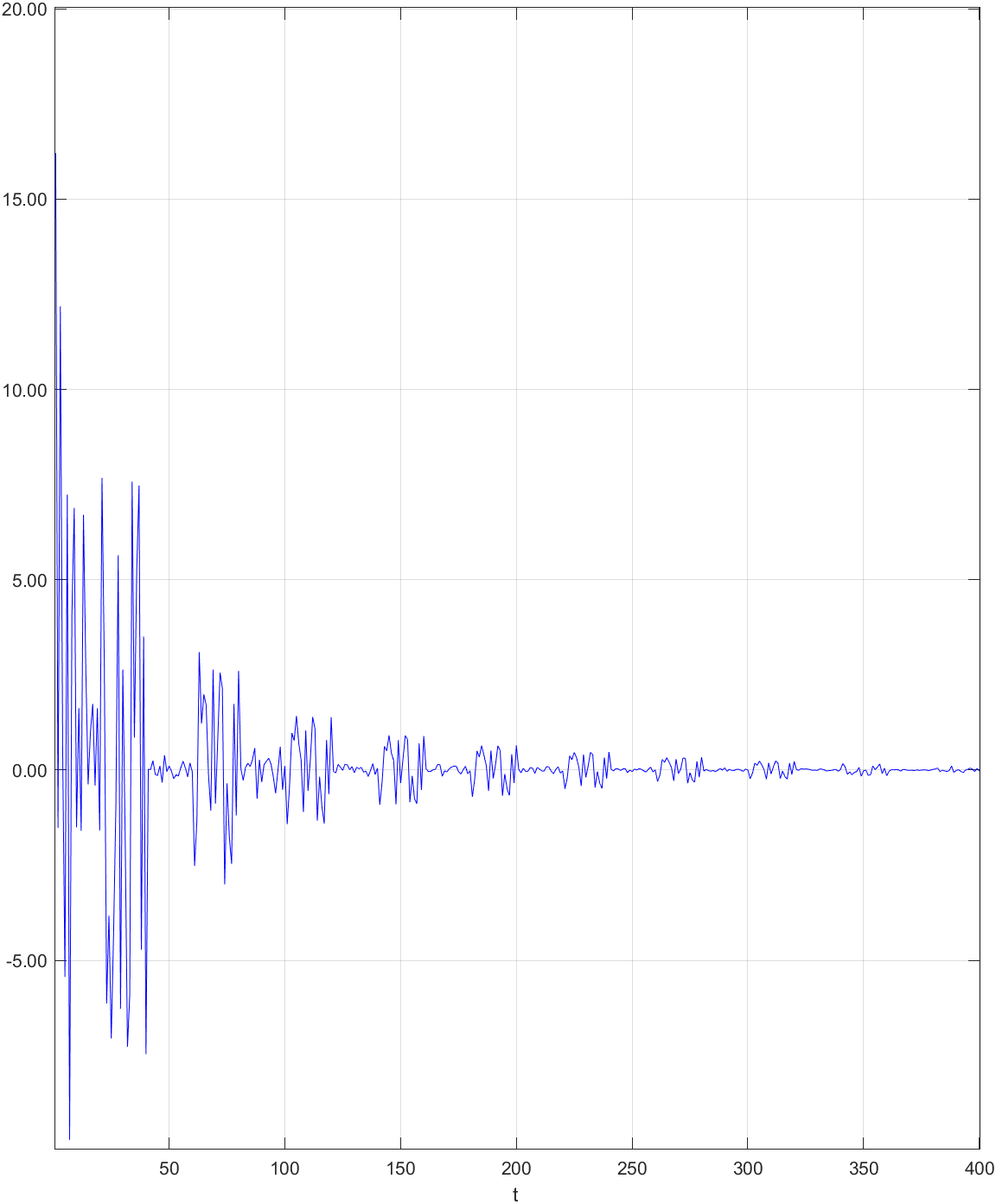}
	\end{minipage}
	\hfill
	\begin{minipage}{0.49\linewidth}
		\centering
		{\footnotesize The second component of the Lap-GFT-II $\mathcal{F}_{\otimes}^{\alpha,k}\mathbf{x}$.}\\[2pt]
		\includegraphics[width=\linewidth,height=4cm]{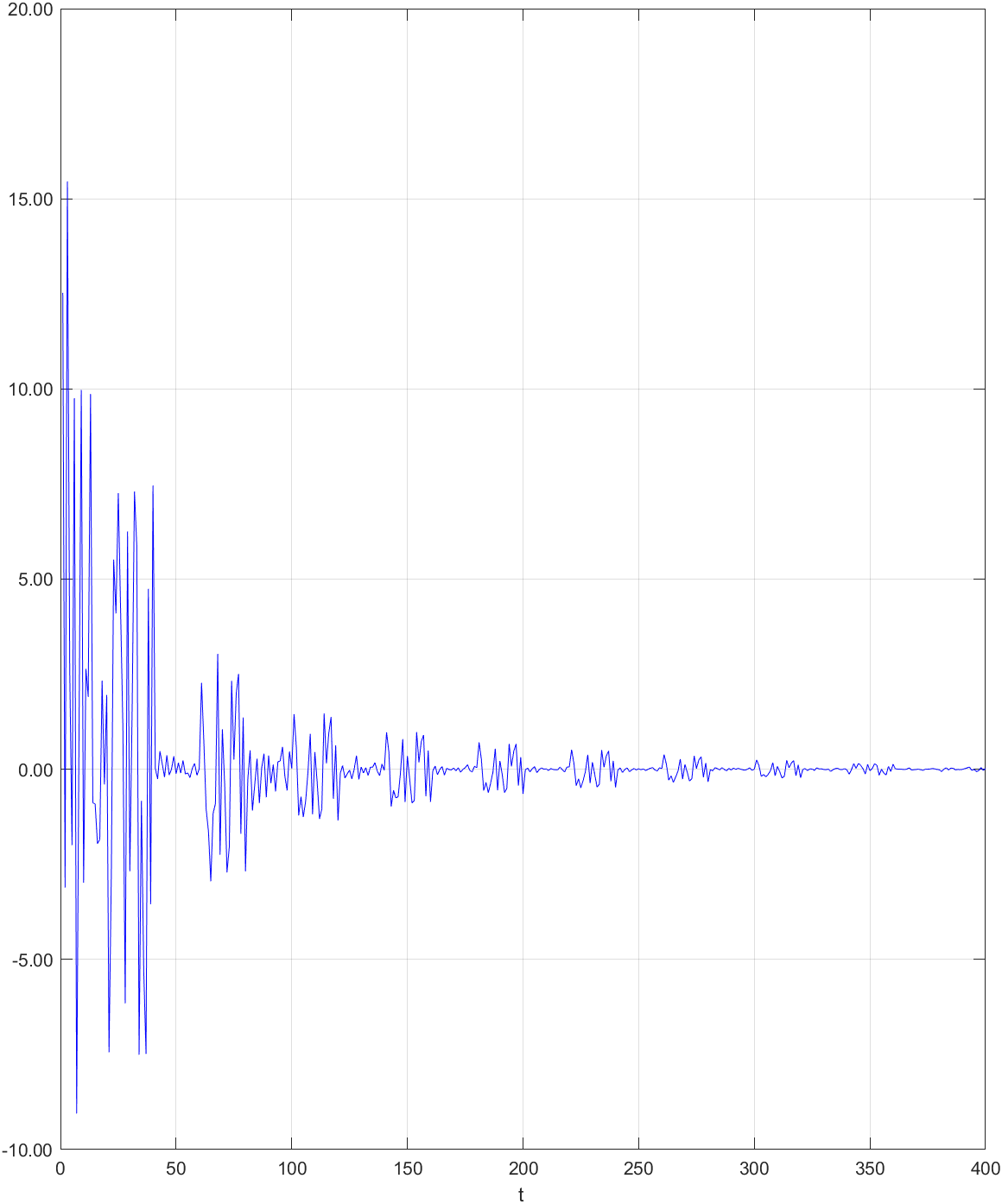}
	\end{minipage}
	
	\begin{minipage}{0.49\linewidth}
		\centering
		{\footnotesize The first component of the Adj-GFT-II $\mathcal{F}_{\otimes}^{\alpha,k}\mathbf{x}$.}\\[2pt]
		\includegraphics[width=\linewidth,height=4cm]{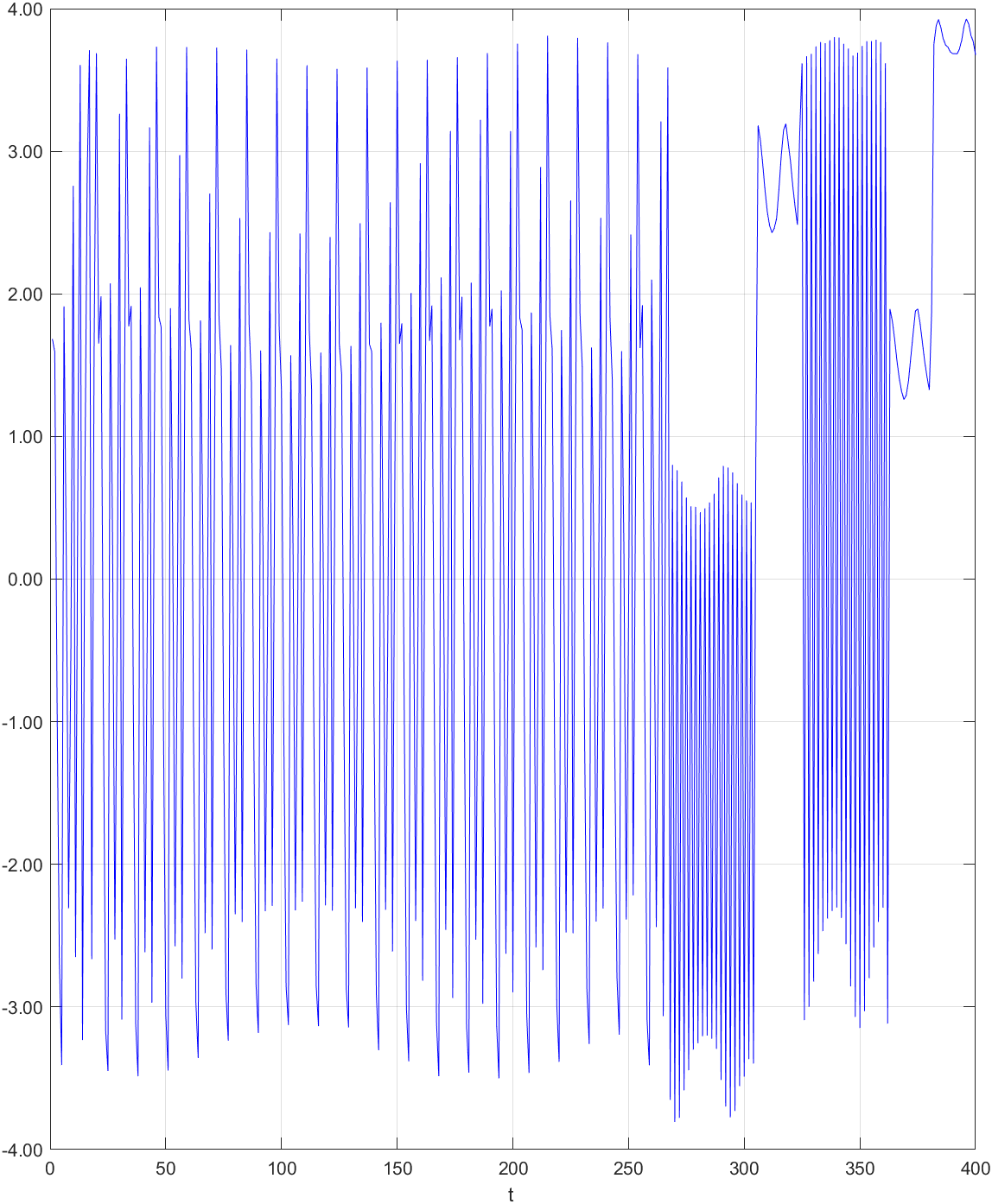}
	\end{minipage}
	\hfill
	\begin{minipage}{0.49\linewidth}
		\centering
		{\footnotesize The second component of the Adj-GFT-II $\mathcal{F}_{\otimes}^{\alpha,k}\mathbf{x}$.}\\[2pt]
		\includegraphics[width=\linewidth,height=4cm]{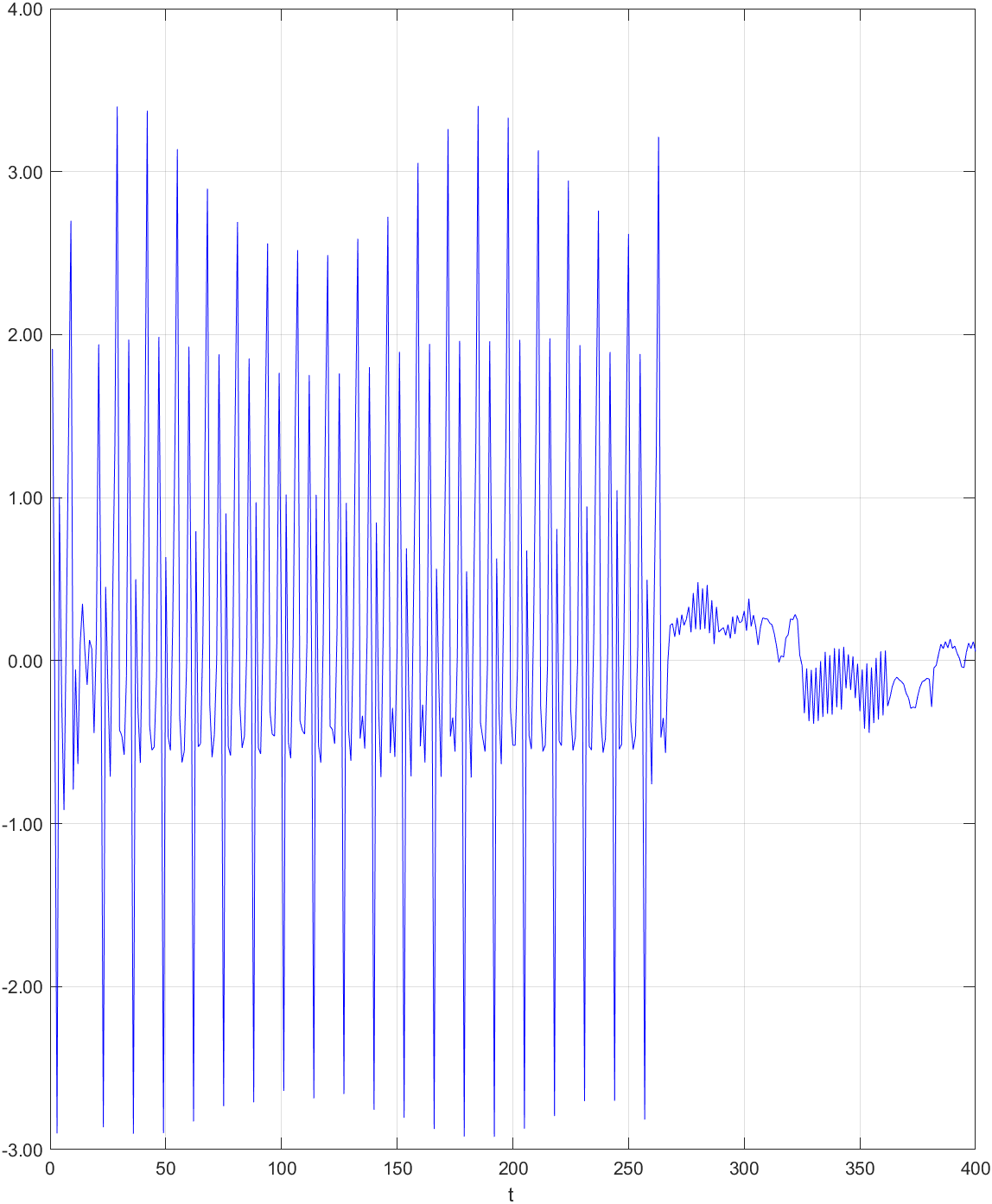}
	\end{minipage}
	
	\begin{minipage}{0.49\linewidth}
		\centering
		{\footnotesize The first component of the Id-GFT-II $\mathcal{F}_{\otimes}^{\alpha,k}\mathbf{x}$.}\\[2pt]
		\includegraphics[width=\linewidth,height=4cm]{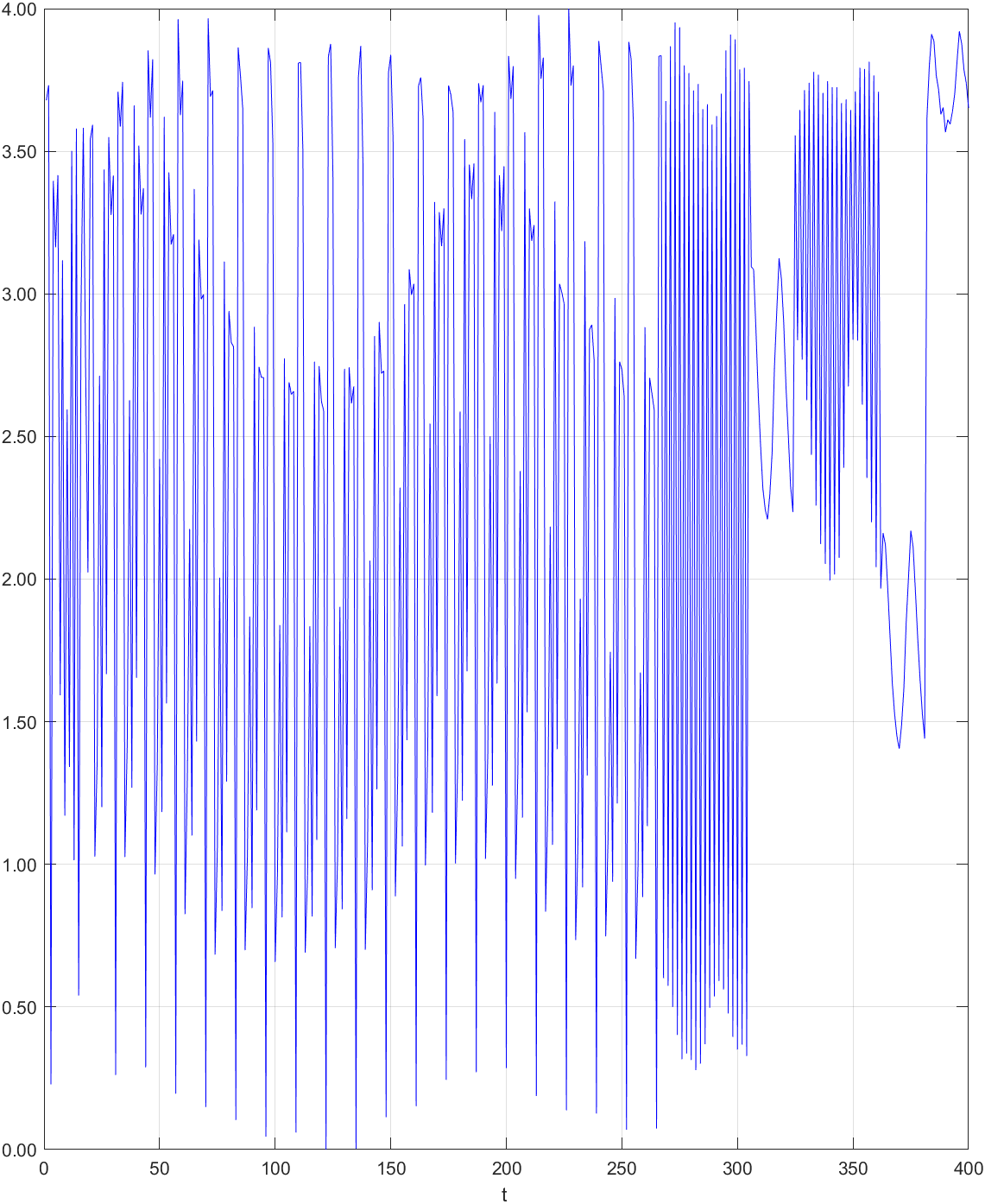}
	\end{minipage}
	\hfill
	\begin{minipage}{0.49\linewidth}
		\centering
		{\footnotesize The second component of the Id-GFT-II $\mathcal{F}_{\otimes}^{\alpha,k}\mathbf{x}$.}\\[2pt]
		\includegraphics[width=\linewidth,height=4cm]{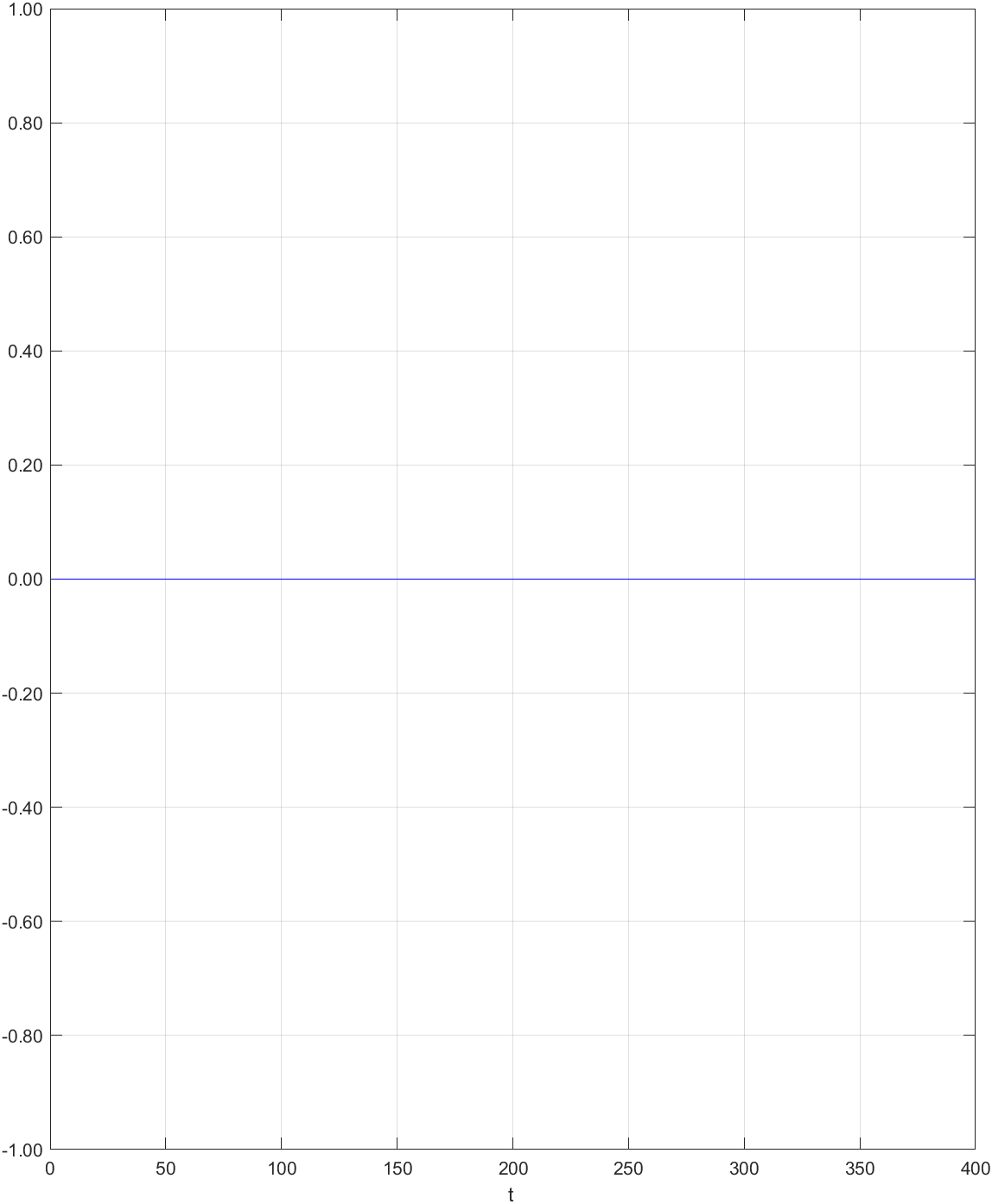}
	\end{minipage}
	
	\begin{minipage}{0.49\linewidth}
		\centering
		{\footnotesize The first component of the SLap-GFT-II $\mathcal{F}_{\otimes}^{\alpha,k}\mathbf{x}$.}\\[2pt]
		\includegraphics[width=\linewidth,height=4cm]{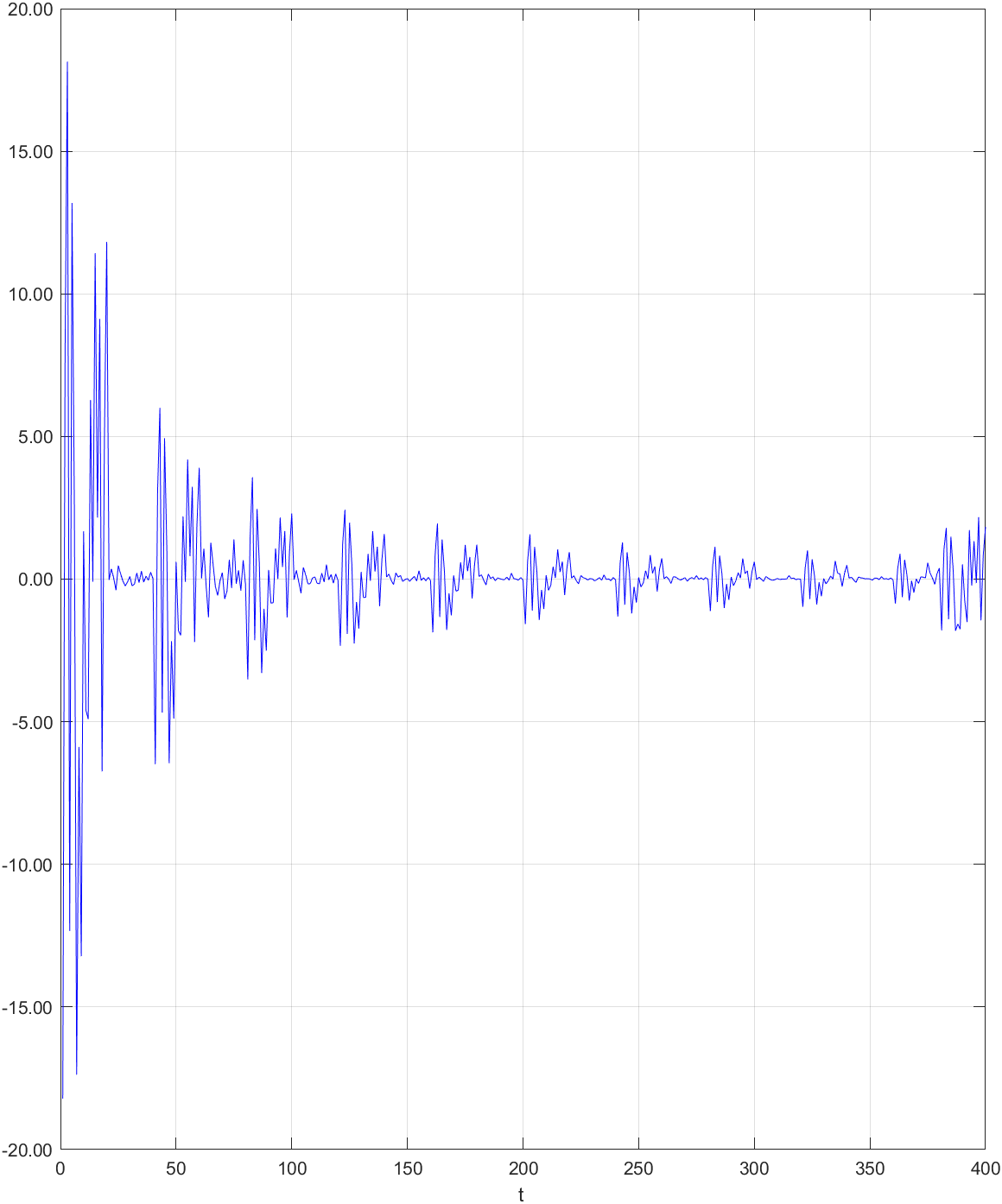}
	\end{minipage}
	\hfill
	\begin{minipage}{0.49\linewidth}
		\centering
		{\footnotesize The second component of the SLap-GFT-II $\mathcal{F}_{\otimes}^{\alpha,k}\mathbf{x}$.}\\[2pt]
		\includegraphics[width=\linewidth,height=4cm]{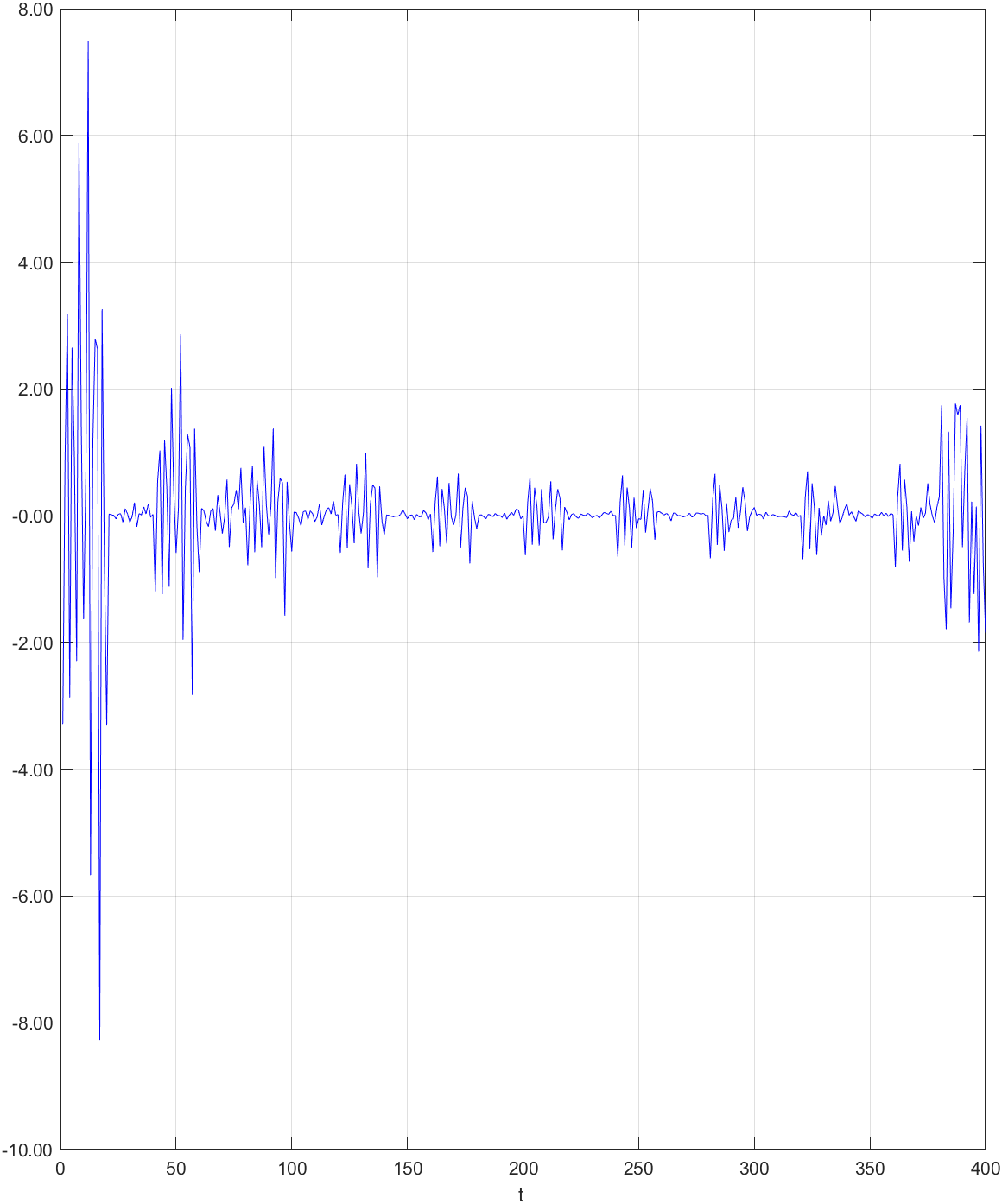}
	\end{minipage}
	
	\caption{Comparison of the first and second components produced by UGRM-GFT-II, Lap-GFT-II, Adj-GFT-II, Id-GFT-II and SLap-GFT-II under $\mathcal{F}_{\otimes}^{\alpha,k}$ for the signal $\mathbf{x}$.}
	\label{fig4}
\end{figure}

\section{Application}
\label{sec:chapter6} 
In this section,  we performed denoising experiments on graphs generated from the 
real SST, PM-25, and COVID datasets; \cite{ref51} 
provides the pre-processed versions.

The SST dataset consists of monthly captured SSTs. We use a 
subset of the first $N = 20$ vertices in the first $T = 20$-month time frame. The 
PM-25 dataset consists of daily-averaged Particulate Matter 2.5 concentration values 
in California in 2015, and we use a subset of the first $N = 20$ vertices in the 
first $T = 20$-month time frame. Johns Hopkins University provided the COVID 
dataset. This dataset consists of COVID daily confirmed cases from January 22, 
2020 to November 18, 2020 at 3232 locations in the United States. We use $N = 20$ 
vertices and the days are chosen from day 41 to day 60, i.e., $T = 20$. The SST and 
PM-25 datasets can be viewed as discretizations of the streaming form, whereas the 
COVID dataset is naturally discrete. The SST and PM-25 signals exhibit strong 
spatiotemporal smoothness characteristic of continuous geophysical fields, whereas the 
COVID signal is governed by contact-based propagation dynamics, producing 
pronounced directional and adjacency-type correlations.

Let the observation dataset be the matrix $\mathbf{X} \in \mathbb{R}^{N \times T}$, 
$\mathbf{X} = [\mathbf{x}(t_0), \cdots, \mathbf{x}(t_{19})]$, where the column 
vector $\mathbf{x}(t_i)$, $0 \leq i \leq 19$, is the zonal temperature for the $t$ 
day or $t$ month. We first normalize $\mathbf{X}$ to obtain $\mathbf{X}_{\text{norm}}$ 
via
\begin{equation}
	\mathbf{X}_{\text{norm}} = \frac{\mathbf{X} - \min(\mathbf{X})}
	{\max(\mathbf{X}) - \min(\mathbf{X})},
	\label{50}
\end{equation}
where $\min(\mathbf{X})$ and $\max(\mathbf{X})$ denote the global minimum and maximum 
of all entries of $\mathbf{X}$, respectively. To account for the different dynamic 
ranges across datasets, the normalized signals are rescaled by a dataset-specific 
factor $c$ and corrupted by additive noise to form the noisy observation matrix
\begin{equation}
	\tilde{\mathbf{X}} = c \cdot \mathbf{X}_{\text{norm}} + \eta,
	\label{51}
\end{equation}
where the scaling factor $c$ is set to $c = 1$ for the SST dataset, $c = 2$ for the 
PM-25 dataset, and $c = 4$ for the COVID dataset, and $\eta \in 
\mathbb{R}^{N \times T}$ is a noise matrix whose entries are independently and 
identically distributed (i.i.d.) following $\mathcal{N}(0, \sigma^2)$ with 
$\sigma \in \{0.1, 0.2, 0.3, 0.4\}$. This rescaling ensures that the noise levels 
induce comparable input signal-to-noise ratios (ISNRs) across the three datasets, 
thereby enabling fair performance comparisons among the different GFT methods.

We model the matrix $\mathbf{X}_{\text{scaled}}$ as a signal on a $400 = 20\times 20$ 
order Cartesian product graph $T \boxtimes N$, where $T$ is an unweighted directed 
line graph of 20 vertices, and $N$ is a directed graph of 20 meteorological 
observation stations as vertices. Edges are constructed from the $k$ nearest-neighbor 
stations with weights $W$ determined by the physical distances between node $i$ and 
its $k$ nearest neighbor nodes $j \in N_i$,
\begin{equation}
	W_{ij} = 
	\begin{cases} 
		\exp\!\left(-\dfrac{\|p_i - p_j\|_2^2}{\varepsilon^2}\right) 
		& \text{if } j \in N_i \\[6pt]
		0 & \text{otherwise}
	\end{cases}, 
	\label{52}
\end{equation}
where $p_i = (\theta_i, \phi_i)$, $p_j = (\theta_j, \phi_j)$ denote the geospatial 
coordinates of node $i$ and node $j$, respectively, $\theta$ is the longitude, and 
$\phi$ is the latitude. Experiments are conducted under three connectivity settings: 
$k \in \{3, 5, 7\}$.

A fundamental methodological consideration in comparing GFT-based 
denoising across different operators is that the Laplacian 
$\mathbf{L}$, adjacency $\mathbf{A}$, in-degree $\mathbf{D}$, 
signless Laplacian $\mathbf{Q}$, and the proposed UGRM 
$\mathbf{P}^{\alpha,k}$ induce qualitatively different spectral 
orderings. To ensure a fair comparison, we adopt the following 
unified convention: for each operator, the $M$ retained spectral 
components are always those that maximize the energy of the 
rank-$M$ signal approximation, i.e., those yielding the smallest 
residual $\|\mathbf{x} - \hat{\mathbf{x}}_M^{\alpha,k}\|_2$. 
Specifically, for operators in the Laplacian-like regime 
($\beta < 0$), the first $M$ components correspond to the $M$ 
smallest singular values; for operators in the adjacency-like 
regime ($\beta \geq 0$), they correspond to the $M$ largest 
singular values. The same convention is applied consistently to 
all baselines, ensuring that each method is evaluated at its 
optimal spectral truncation.

On the Cartesian product $\mathcal{G}_1 \boxtimes \mathcal{G}_2$, 
let $\sigma_t^{\alpha,k}$, $\mathbf{u}_t^{\alpha,k}$, 
$\mathbf{v}_t^{\alpha,k}$, $0 \leq t \leq 399$, be chosen such 
that the singular values 
$\sigma_0^{\alpha,k}, \sigma_1^{\alpha,k}, \ldots, 
\sigma_{399}^{\alpha,k}$ of the UGRM-GFT-I 
$\mathcal{F}_{\boxtimes}^{\alpha,k}$ are in nondecreasing or 
nonincreasing order, and arrange the singular value pairs 
$(\sigma_{1,i}^{\alpha,k}, \sigma_{2,j}^{\alpha,k})$ of the 
UGRM-GFT-II operator $\mathcal{F}_{\otimes}^{\alpha,k}$ in 
nondecreasing or nonincreasing order, denoted as 
$\mu_0^{\alpha,k}, \mu_1^{\alpha,k}, \ldots, \mu_{399}^{\alpha,k}$. 
We set the bandwidth to $M = 40$, corresponding to retaining the 
first 10\% of the 400 spectral components. This choice strikes a 
balance between capturing the dominant signal structure and 
suppressing noise: as confirmed by Table~\ref{tab1}, the leading 40 
components of the UGRM-GFTs already concentrate over 95.91\% 
(SST), 80.54\% (PM-25), and 77.52\% (COVID) of the total signal 
energy, indicating that the remaining 360 components are 
predominantly noise. We bandlimit the first $M = 40$ spectral 
components of various GFTs including UGRM-GFT-I, UGRM-GFT-II, 
Adj-GFT-I, Adj-GFT-II, Id-GFT-I, Id-GFT-II, SLap-GFT-I, 
SLap-GFT-II, Lap-GFT-I, and Lap-GFT-II and apply them to the 
noisy signal $\tilde{\mathbf{X}}$. The reconstructed signals are 
obtained as follows. For the UGRM-GFT-I type method,
\begin{equation}
	\begin{aligned}
		\hat{\mathbf{X}}_{M,\boxtimes}^{\alpha,k} 
		&= \mathrm{vec}^{-1}
		\left(\frac{1}{2}\sum_{t=0}^{M-1}\left(\mathbf{u}_t^{\alpha,k}
		(\mathbf{u}_t^{\alpha,k})^T \right.\right. \\
		&\quad \left.\left.+  \mathbf{v}_t^{\alpha,k}
		(\mathbf{v}_t^{\alpha,k})^T\right)\mathrm{vec}
		(\tilde{\mathbf{X}})\right),
	\end{aligned}
	\label{53}
\end{equation}
and for the UGRM-GFT-II type method,
\begin{equation}
	\begin{aligned}
		\hat{\mathbf{X}}_{M,\otimes}^{\alpha,k} 
		&= \frac{1}{2} \sum_{(i,j)\in\mathcal{S}_M} 
		\left( (\mathbf{u}_{2,j}^{\alpha,k})^T \tilde{\mathbf{X}} \mathbf{u}_{1,i}^{\alpha,k} \mathbf{u}_{2,j}^{\alpha,k} (\mathbf{u}_{1,i}^{\alpha,k})^T \right. \\
		&\quad \left. + ((\mathbf{v}_{2,j}^{\alpha,k})^T \tilde{\mathbf{X}} \mathbf{v}_{1,i}^{\alpha,k}) \mathbf{v}_{2,j}^{\alpha,k} (\mathbf{v}_{1,i}^{\alpha,k})^T \right),
	\end{aligned}
	\label{54}
\end{equation}
where $\mathcal{S}_M$ contains all pairs $(i,j)$ with 
$\sigma_{1,i} + \sigma_{2,j}$ being some $\mu_k$ 
($0 \leq k \leq M-1$), one of the first $M$ spectral components in the 
spectral domain of the UGRM-GFT-II 
$\mathcal{F}_{\otimes}^{\alpha,k}$. Define the input SNR (ISNR), 
the bandlimiting SNR, and the bandlimiting approximation error 
(BAE) by
\begin{equation}
	\mathrm{ISNR}(\sigma) = -20\log_{10}
	\frac{\|\tilde{\mathbf{X}} - \mathbf{X}\|_F}{\|\mathbf{X}\|_F},
\end{equation}
\begin{equation}
	\mathrm{SNR}(\sigma, M) = -20\log_{10}\left(
	\frac{\|\hat{\mathbf{X}}_M^{\alpha,k} - \mathbf{X}\|_F}
	{\|\mathbf{X}\|_F}\right),
\end{equation}
and
\begin{equation}
	\mathrm{BAE}(\sigma, M) = \|\hat{\mathbf{X}}_M^{\alpha,k} 
	- \mathbf{X}\|_\infty.
\end{equation}
To maximize the denoising performance, the optimal parameters 
$(\alpha^*, k^*)$ of the UGRM are selected via Bayesian 
optimization over the continuous domain $[0,1] \times [0,1]$, 
which efficiently identifies high-performing configurations by 
maximizing the expected improvement acquisition function.

\begin{table*}
	\caption{ENERGY CONCENTRATION RATIOS OF THE LEADING SPECTRAL COMPONENTS FOR DIFFERENT GFT METHODS ON THE SST, PM-25, AND COVID DATASETS WITH 5-NN GRAPH CONNECTIVITY}
	\label{tab1}
	\centering
	\begin{tabular}{lccc}
		\toprule
		Method & SST & PM-25 & COVID \\
		\bottomrule
		UGRM-GFT-I  & \textbf{95.91}\% & \textbf{80.54}\% & \textbf{77.52}\% \\
		& $(\alpha = 0.4303,\ k = 0.8630)$ 
		& $(\alpha = 0.4970,\ k = 0.8310)$ 
		& $(\alpha = 0.9552,\ k = 0.9731)$ \\
		Lap-GFT-I   & 95.89\% & 75.74\% & 74.96\% \\
		Adj-GFT-I   & 21.95\% & 22.44\% & 29.97\% \\
		Id-GFT-I    & 34.17\% & 32.11\% & 0.00\%  \\
		SLap-GFT-I  & 89.21\% & 55.79\% & 19.20\% \\
		\bottomrule
		\bottomrule
		UGRM-GFT-II & \textbf{95.91}\% & \textbf{77.98}\% & \textbf{76.54}\% \\
		& $(\alpha = 0.3860,\ k = 0.8022)$ 
		& $(\alpha = 0.2386,\ k = 0.6478)$ 
		& $(\alpha = 0.9195,\ k = 0.7569)$ \\
		Lap-GFT-II  & 95.89\% & 77.77\% & 74.37\% \\
		Adj-GFT-II  & 31.31\% & 37.13\% & 28.49\% \\
		Id-GFT-II   & 31.66\% & 50.29\% & 0.00\%  \\
		SLap-GFT-II & 89.21\% & 55.05\% & 19.81\% \\
		\bottomrule
	\end{tabular}
\end{table*}
Figs.~\ref{fig3} and~\ref{fig4} present the spectral coefficient distributions 
of the proposed UGRM-GFT-I and UGRM-GFT-II compared with 
traditional methods (Lap-GFT, Adj-GFT, Id-GFT, and SLap-GFT) 
on the SST dataset, with each subplot title indicating the 
specific operator used to construct the spectral basis. The 
figures demonstrate that the transformation coefficients of both 
UGRM-GFT-I and UGRM-GFT-II exhibit a markedly more pronounced 
energy concentration in the leading spectral components compared 
to all baseline methods. Specifically, the first components of 
UGRM-GFT-I and UGRM-GFT-II display a sharp initial peak 
followed by rapid decay, indicating that the dominant signal 
energy is compactly captured within the first $M = 40$ spectral 
modes. In contrast, the Adj-GFT and Id-GFT coefficients show 
relatively flat or dispersed distributions across the entire 
spectral domain, reflecting poor energy compaction. The Lap-GFT 
coefficients exhibit energy concentration comparable to but 
slightly less compact than the UGRM-GFTs, while SLap-GFT shows 
intermediate behavior. The second components, which capture the 
asymmetric part of the spectral decomposition arising from graph 
directionality, are substantially smaller in magnitude for the 
UGRM-GFTs, confirming that the optimized parameterization aligns 
the spectral basis more effectively with the underlying signal 
structure. This visual evidence of superior energy compaction is 
quantitatively confirmed by Table~\ref{tab1}: with the fixed bandwidth 
$M = 40$, both UGRM-GFT-I and UGRM-GFT-II concentrate 95.91\% 
of the total signal energy on the SST dataset within these 40 
leading components, surpassing all fixed-matrix methods. This 
enhanced energy compaction is a direct consequence of the adaptive 
interpolation mechanism of the UGRM, as established by the 
theoretical analysis following Theorem~\ref{thm1}.

Shown in Tables~\ref{tab1}--\ref{tab4} are the denoising performances of the 
proposed UGRM-GFTs at bandwidth $M = 40$ for different noise 
levels $\sigma \in \{0.1, 0.2, 0.3, 0.4\}$ and varying graph 
connectivities on the SST, PM-25, and COVID datasets. From these 
results, we observe the following:

\begin{itemize}
	\item On denoising the real-world datasets, the proposed 
	UGRM-GFT-I and UGRM-GFT-II have comparably good performances 
	and consistently outperform the traditional fixed-matrix methods 
	(Lap-GFT, Adj-GFT, Id-GFT, and SLap-GFT) at the same bandwidth 
	$M = 40$. For signals exhibiting spatiotemporal smoothness like 
	the SST and PM-25 datasets, UGRM-GFTs achieve higher SNRs and 
	lower BAEs than Lap-GFT. For instance, on the SST dataset with 
	5-NN connectivity and $\sigma = 0.1$, UGRM-GFT-I attains an SNR 
	of 9.1847~dB compared to 9.1466~dB for Lap-GFT-I, while 
	UGRM-GFT-II achieves 9.5254~dB versus 9.2665~dB for Lap-GFT-II. 
	For the highly irregular COVID epidemic dataset, traditional 
	methods such as Id-GFT and SLap-GFT yield negative or near-zero 
	SNRs (e.g., Id-GFT-I produces $-0.0206$~dB with 3-NN at 
	$\sigma = 0.1$, and SLap-GFT-I produces only 0.1621~dB), whereas 
	the proposed UGRM-GFTs maintain significant performance 
	advantages and successfully capture the complex signal structure 
	(e.g., UGRM-GFT-I achieves 3.9388~dB with 5-NN, and UGRM-GFT-II 
	achieves 5.8404~dB with 7-NN). This consistent superiority across 
	datasets with fundamentally different correlation structures 
	validates the theoretical prediction following Theorem~\ref{thm1} that the 
	parameterized UGRM can spectrally align with diverse signal types 
	by continuously adjusting the balance between degree and Laplacian 
	contributions.
	
	\item At the fixed bandwidth $M = 40$ (10\% of total spectral 
	components), the UGRM-GFTs achieve significantly higher energy 
	concentration ratios than all baselines, as reported in Table~\ref{tab1}. 
	For the SST dataset, both UGRM-GFT-I and UGRM-GFT-II concentrate 
	95.91\% of the signal energy in the first 40 components, compared 
	to 95.89\% for Lap-GFT, and dramatically lower ratios for Adj-GFT 
	(21.95\%--31.31\%), Id-GFT (31.66\%--34.17\%), and SLap-GFT 
	(89.21\%). For the PM-25 dataset, UGRM-GFT-I achieves 80.54\% 
	versus 75.74\% for Lap-GFT-I. This superior energy compaction 
	within only 40 modes directly explains the denoising advantage: 
	with the signal energy tightly concentrated in the retained band, 
	the truncation at $M = 40$ effectively separates signal from 
	noise. The theoretical basis for this advantage, as analyzed 
	following Theorem~\ref{thm1}, lies in the UGRM's ability to continuously 
	interpolate between degree and Laplacian information through 
	$(\alpha, k)$, adapting to the signal's intrinsic correlation 
	structure in a way that no fixed operator can match.
	
	\item For the different datasets, the optimal parameters 
	$(\alpha^*, k^*)$ identified by Bayesian optimization adaptively 
	adjust the spectral operator within the Laplacian-like regime 
	($\beta < 0$), providing direct empirical validation of the 
	theoretical analysis following Theorem~\ref{thm1}. For the SST dataset, 
	the parameters yield $\beta$ well below zero (e.g., 
	$\alpha = 0.4303$, $k = 0.8630$, giving $\beta \approx -0.41$), 
	strongly emphasizing the Laplacian component and aligning with the 
	highly smooth nature of temperature fields. For the PM-25 dataset, 
	the parameters produce similar Laplacian-dominated behavior 
	($\alpha = 0.4970$, $k = 0.8310$, giving $\beta \approx -0.33$). 
	For the COVID dataset, the optimal parameters shift markedly 
	toward the regime boundary (e.g., $\alpha = 0.9552$, 
	$k = 0.9731$, giving $\beta \approx -0.04$), substantially 
	increasing the weight of the degree component relative to the 
	Laplacian. This near-zero $\beta$ reflects the weaker spatial 
	smoothness and stronger local connectivity dependence 
	characteristic of contact-based epidemic propagation, in contrast 
	to the strongly smooth geophysical fields of SST and PM-25. The 
	fact that the optimal $\beta$ systematically varies across 
	datasets---from strongly negative for smooth signals to near-zero 
	for propagation-type signals---demonstrates the UGRM's ability to 
	continuously adapt its spectral basis to the signal's intrinsic 
	correlation structure, a flexibility that no fixed operator can 
	provide. Furthermore, these optimal parameters remain highly 
	stable across varying noise levels $\sigma$, indicating that the 
	parameter selection is driven primarily by the intrinsic graph 
	topology rather than specific noise realizations.
	
	\item When comparing the two proposed methods, UGRM-GFT-II 
	exhibits competitive or occasionally superior denoising 
	performance relative to UGRM-GFT-I at the same bandwidth 
	$M = 40$, while significantly reducing the fundamental 
	computational complexity from $\mathcal{O}(N_1^3 N_2^3)$ to 
	$\mathcal{O}(N_1^3 + N_2^3)$. As reported in Table~\ref{tab2}, the total 
	parameter optimization runtime of UGRM-GFT-II (26.49~s) is 
	approximately 22.7\% lower than that of UGRM-GFT-I (34.27~s) on 
	the $20 \times 20$ product graph, with the computational advantage 
	expected to grow substantially for larger graphs since the 
	complexity gap widens as $N_1$ and $N_2$ increase. On the SST 
	dataset, UGRM-GFT-II frequently achieves slightly higher SNRs 
	than UGRM-GFT-I (e.g., 9.5254~dB vs.\ 9.1847~dB with 5-NN at 
	$\sigma = 0.1$), suggesting that the factored decomposition may 
	better capture the separable spatiotemporal structure of 
	geophysical signals. This efficiency-performance tradeoff makes 
	UGRM-GFT-II particularly attractive for large-scale applications, 
	where the factored spectral decomposition effectively 
	characterizes the distinct variations along different dimensions 
	while maintaining the approximation guarantees established in 
	Theorem~\ref{thm3}.
\end{itemize}

\begin{table}
	\centering
	\caption{Total Run Time for Parameter Optimization }
	\label{tab2}
	\begin{tabular}{llc}
		\toprule
		\textbf{Method} & \textbf{Run Time(s)} \\
		\midrule
		UGRM-GFT-I       & 34.27 \\
		UGRM-GFT-II      & 26.49 \\
		\bottomrule
	\end{tabular}
\end{table}
\begin{table*}[htbp]
	\centering
	\tiny
	\caption{The Bandlimiting SNR and BAE for UGRM-GFT-I and Baseline Methods with Different Noise Levels $\sigma \in \{0.1, 0.2, 0.3, 0.4\}$ and Graph Connectivities (3-NN, 5-NN, 7-NN) on the SST, PM-25, and COVID Datasets}
	\label{tab3}
	\begin{tabular}{cccccccccccccc}
		\thickrule
		\multicolumn{2}{c}{}
		& \multicolumn{4}{c}{3-NN}
		& \multicolumn{4}{c}{5-NN}
		& \multicolumn{4}{c}{7-NN} \\
		\cmidrule(lr){3-6} \cmidrule(lr){7-10} \cmidrule(lr){11-14}
		\multicolumn{2}{c}{SST} & \multicolumn{12}{c}{} \\
		& & $\sigma=0.1$ & $\sigma=0.2$ & $\sigma=0.3$ & $\sigma=0.4$
		& $\sigma=0.1$ & $\sigma=0.2$ & $\sigma=0.3$ & $\sigma=0.4$
		& $\sigma=0.1$ & $\sigma=0.2$ & $\sigma=0.3$ & $\sigma=0.4$ \\
		\midrule
		ISNR &
		& 16.6280 & 10.6074 & 7.0856 & 4.5868
		& 16.6280 & 10.6074 & 7.0856 & 4.5868
		& 16.6280 & 10.6074 & 7.0856 & 4.5868 \\
		\multirow{5}{*}{SNR}
		& \makecell{UGRM-GFT-I\\$(\alpha,k)$}
		& \makecell{\textbf{9.1832}\\(0.3785,0.8134)}
		& \makecell{\textbf{9.0250}\\(0.2362,0.6590)}
		& \makecell{\textbf{8.7661}\\(0.2357,0.6583)}
		& \makecell{\textbf{8.4230}\\(0.3184,0.7396)}
		& \makecell{\textbf{9.1847}\\(0.4289,0.8854)}
		& \makecell{\textbf{9.0250}\\(0.2413,0.6633)}
		& \makecell{\textbf{8.7656}\\(0.1839,0.6154)}
		& \makecell{\textbf{8.4054}\\(0.1981,0.6287)}
		& \makecell{\textbf{9.1793}\\(0.4393,0.9040)}
		& \makecell{\textbf{9.0259}\\(0.4786,0.9795)}
		& \makecell{\textbf{8.7635}\\(0.2214,0.6463)}
		& \makecell{\textbf{8.4168}\\(0.2707,0.6919)} \\
		& Lap-GFT-I
		& 9.1482 & 8.9880 & 8.7263 & 8.3811
		& 9.1466 & 8.9864 & 8.7249 & 8.3798
		& 9.1464 & 8.9862 & 8.7247 & 8.3796 \\
		& Adj-GFT-I
		& 2.6824 & 2.6363 & 2.5603 & 2.4562
		& 2.6860 & 2.6399 & 2.5639 & 2.4597
		& 2.6869 & 2.6408 & 2.5648 & 2.4605 \\
		& Id-GFT-I
		& 0.5739 & 0.5420 & 0.4893 & 0.4167
		& 0.6812 & 0.6483 & 0.5940 & 0.5192
		& 0.6555 & 0.6227 & 0.5686 & 0.4940 \\
		& SLap-GFT-I
		& 4.4607 & 4.3828 & 4.2560 & 4.0846
		& 4.4597 & 4.3818 & 4.2550 & 4.0837
		& 4.4593 & 4.3814 & 4.2546 & 4.0833 \\
		\cmidrule(lr){1-14}
		\multirow{5}{*}{BAE}
		& \makecell{UGRM-GFT-I\\$(\alpha,k)$}
		& \makecell{\textbf{0.5580}\\(0.4613,0.9363)}
		& \makecell{\textbf{0.5765}\\(0.4442,0.9001)}
		& \makecell{\textbf{0.6062}\\(0.4503,0.9052)}
		& \makecell{\textbf{0.6598}\\(0.4696,0.9398)}
		& \makecell{\textbf{0.5580}\\(0.4613,0.9363)}
		& \makecell{\textbf{0.5766}\\(0.4442,0.9001)}
		& \makecell{\textbf{0.6065}\\(0.3664,0.7869)}
		& \makecell{\textbf{0.6602}\\(0.4711,0.9404)}
		& \makecell{\textbf{0.5580}\\(0.4613,0.9363)}
		& \makecell{\textbf{0.5766}\\(0.4442,0.9001)}
		& \makecell{\textbf{0.6065}\\(0.3664,0.7869)}
		& \makecell{\textbf{0.6602}\\(0.4466,0.8994)} \\
		& Lap-GFT-I
		& 0.5609 & 0.5765 & 0.6067 & 0.6605
		& 0.5609 & 0.5766 & 0.6070 & 0.6609
		& 0.5609 & 0.5766 & 0.6070 & 0.6609 \\
		& Adj-GFT-I
		& 0.9305 & 0.9305 & 0.9305 & 0.9331
		& 0.9305 & 0.9305 & 0.9305 & 0.9332
		& 0.9305 & 0.9305 & 0.9305 & 0.9332 \\
		& Id-GFT-I
		& 1.0000 & 1.0000 & 1.0011 & 1.0384
		& 1.0000 & 1.0000 & 1.0010 & 1.0442
		& 1.0000 & 1.0000 & 1.0010 & 1.0440 \\
		& SLap-GFT-I
		& 0.8874 & 0.8874 & 0.8911 & 0.9188
		& 0.8874 & 0.8874 & 0.8911 & 0.9188
		& 0.8874 & 0.8874 & 0.8911 & 0.9188 \\
		\bottomrule
		\bottomrule
		
		\multicolumn{2}{c}{}
		& \multicolumn{4}{c}{3-NN}
		& \multicolumn{4}{c}{5-NN}
		& \multicolumn{4}{c}{7-NN} \\
		\cmidrule(lr){3-6} \cmidrule(lr){7-10} \cmidrule(lr){11-14}
		\multicolumn{2}{c}{PM-25} & \multicolumn{12}{c}{} \\
		& & $\sigma=0.1$ & $\sigma=0.2$ & $\sigma=0.3$ & $\sigma=0.4$
		& $\sigma=0.1$ & $\sigma=0.2$ & $\sigma=0.3$ & $\sigma=0.4$
		& $\sigma=0.1$ & $\sigma=0.2$ & $\sigma=0.3$ & $\sigma=0.4$ \\
		\midrule
		ISNR &
		& 12.6677 & 6.6471 & 3.1252 & 0.6265
		& 12.6677 & 6.6471 & 3.1252 & 0.6265
		& 12.6677 & 6.6471 & 3.1252 & 0.6265 \\
		\multirow{5}{*}{SNR}
		& \makecell{UGRM-GFT-I\\$(\alpha,k)$}
		& \makecell{\textbf{5.2896}\\(0.5039,0.9986)}
		& \makecell{\textbf{5.1219}\\(0.4954,0.9764)}
		& \makecell{\textbf{4.8512}\\(0.5066,0.9993)}
		& \makecell{\textbf{4.5004}\\(0.4860,0.9647)}
		& \makecell{\textbf{5.4810}\\(0.5231,0.9753)}
		& \makecell{\textbf{5.2882}\\(0.4839,0.9096)}
		& \makecell{\textbf{4.9857}\\(0.5169,0.9710)}
		& \makecell{\textbf{4.9857}\\(0.5169,0.9710)}
		& \makecell{\textbf{5.5639}\\(0.5013,0.9300)}
		& \makecell{\textbf{5.3686}\\(0.5362,0.9952)}
		& \makecell{\textbf{5.0612}\\(0.3837,0.7684)}
		& \makecell{\textbf{4.6657}\\(0.3856,0.7736)} \\
		& Lap-GFT-I
		& 5.2766 & 5.1091 & 4.8434 & 4.4970
		& 5.3023 & 5.1289 & 4.8547 & 4.8547
		& 5.3132 & 5.1409 & 4.8675 & 4.5115 \\
		& Adj-GFT-I
		& 4.5227 & 4.3612 & 4.1061 & 3.7738
		& 5.2937 & 5.1204 & 4.8473 & 4.8473
		& 5.0360 & 4.8716 & 4.6121 & 4.2742 \\
		& Id-GFT-I
		& 0.6934 & 0.6136 & 0.4840 & 0.3091
		& 1.1094 & 1.0212 & 0.8782 & 0.8782
		& 0.7123 & 0.6322 & 0.5021 & 0.3265 \\
		& SLap-GFT-I
		& 2.7300 & 2.5989 & 2.3893 & 2.1124
		& 1.6216 & 1.5255 & 1.3700 & 1.3700
		& 1.4615 & 1.3685 & 1.2180 & 1.0159 \\
		\cmidrule(lr){1-14}
		\multirow{5}{*}{BAE}
		& \makecell{UGRM-GFT-I\\$(\alpha,k)$}
		& \makecell{\textbf{1.0666}\\(0.5128,0.9308)}
		& \makecell{\textbf{1.0684}\\(0.5367,0.9761)}
		& \makecell{\textbf{1.0745}\\(0.5198,0.9442)}
		& \makecell{\textbf{1.0872}\\(0.5285,0.9643)}
		& \makecell{\textbf{1.0927}\\(0.4710,0.8340)}
		& \makecell{\textbf{1.0944}\\(0.3961,0.7410)}
		& \makecell{\textbf{1.1008}\\(0.5281,0.9172)}
		& \makecell{\textbf{1.1008}\\(0.5281,0.9172)}
		& \makecell{\textbf{0.9864}\\(0.4424,0.8114)}
		& \makecell{\textbf{0.9936}\\(0.5223,0.9299)}
		& \makecell{\textbf{1.0039}\\(0.4838,0.8667)}
		& \makecell{\textbf{1.0172}\\(0.3661,0.7256)} \\
		& Lap-GFT-I
		& 1.1082 & 1.1082 & 1.1136 & 1.1228
		& 1.1847 & 1.1838 & 1.1858 & 1.1858
		& 1.0398 & 1.0454 & 1.0578 & 1.0737 \\
		& Adj-GFT-I
		& 1.9781 & 1.9781 & 1.9780 & 1.9780
		& 1.9259 & 1.9261 & 1.9263 & 1.9263
		& 1.9259 & 1.9261 & 1.9263 & 1.9265 \\
		& Id-GFT-I
		& 1.5716 & 1.5716 & 1.5716 & 1.5716
		& 1.5716 & 1.5716 & 1.5716 & 1.5716
		& 1.5716 & 1.5716 & 1.5716 & 1.5716 \\
		& SLap-GFT-I
		& 1.9969 & 1.9968 & 1.9968 & 1.9967
		& 2.0000 & 2.0000 & 2.0000 & 2.0000
		& 2.0000 & 2.0000 & 2.0000 & 2.0000 \\
		\bottomrule
		
		\bottomrule
		
		\multicolumn{2}{c}{}
		& \multicolumn{4}{c}{3-NN}
		& \multicolumn{4}{c}{5-NN}
		& \multicolumn{4}{c}{7-NN} \\
		\cmidrule(lr){3-6} \cmidrule(lr){7-10} \cmidrule(lr){11-14}
		\multicolumn{2}{c}{COVID} & \multicolumn{12}{c}{} \\
		& & $\sigma=0.1$ & $\sigma=0.2$ & $\sigma=0.3$ & $\sigma=0.4$
		& $\sigma=0.1$ & $\sigma=0.2$ & $\sigma=0.3$ & $\sigma=0.4$
		& $\sigma=0.1$ & $\sigma=0.2$ & $\sigma=0.3$ & $\sigma=0.4$ \\
		\midrule
		ISNR &
		& 13.1879 & 7.1673 & 3.6454 & 1.1467
		& 13.1879 & 7.1673 & 3.6454 & 1.1467
		& 13.1879 & 7.1673 & 3.6454 & 1.1467 \\
		\multirow{5}{*}{SNR}
		& \makecell{UGRM-GFT-I\\$(\alpha,k)$}
		& \makecell{\textbf{2.9621}\\(0.3524,0.8925)}
		& \makecell{\textbf{2.8881}\\(0.3418,0.8711)}
		& \makecell{\textbf{2.7591}\\(0.3030,0.8107)}
		& \makecell{\textbf{2.5835}\\(0.2724,0.7681)}
		& \makecell{\textbf{3.9388}\\(0.9078,0.7550)}
		& \makecell{\textbf{3.7881}\\(0.9264,0.8419)}
		& \makecell{\textbf{3.5484}\\(0.9252,0.8546)}
		& \makecell{\textbf{3.2342}\\(0.9330,0.9179)}
		& \makecell{\textbf{5.3238}\\(0.5406,0.9079)}
		& \makecell{\textbf{5.1372}\\(0.5263,0.8848)}
		& \makecell{\textbf{4.8305}\\(0.5688,0.9578)}
		& \makecell{\textbf{4.4307}\\(0.5328,0.8985)} \\
		& Lap-GFT-I
		& 2.6526 & 2.5654 & 2.4236 & 2.2323
		& 4.0294 & 3.9048 & 3.7042 & 3.4380
		& 4.8136 & 4.6617 & 4.4196 & 4.1020 \\
		& Adj-GFT-I
		& 0.7520 & 0.6969 & 0.6066 & 0.4831
		& 0.6761 & 0.6220 & 0.5333 & 0.4123
		& 0.4221 & 0.3745 & 0.2961 & 0.1887 \\
		& Id-GFT-I
		& $-$0.0206 & $-$0.0819 & $-$0.1820 & $-$0.3183
		& 1.1541 & 1.0745 & 0.9450 & 0.7704
		& 2.3944 & 2.2884 & 2.1175 & 1.8895 \\
		& SLap-GFT-I
		& 0.1621 & 0.0998 & $-$0.0020 & $-$0.1406
		& 0.1576 & 0.0957 & $-$0.0056 & $-$0.1433
		& 0.1001 & 0.0385 & $-$0.0624 & $-$0.1997 \\
		\cmidrule(lr){1-14}
		\multirow{5}{*}{BAE}
		& \makecell{UGRM-GFT-I\\$(\alpha,k)$}
		& \makecell{\textbf{2.2194}\\(0.1997,0.7674)}
		& \makecell{\textbf{2.2208}\\(0.2115,0.7918)}
		& \makecell{\textbf{2.2251}\\(0.1922,0.7074)}
		& \makecell{\textbf{2.2280}\\(0.3347,0.9340)}
		& \makecell{\textbf{1.7342}\\(0.4546,0.9615)}
		& \makecell{\textbf{1.7434}\\(0.4248,0.9066)}
		& \makecell{\textbf{1.7545}\\(0.4239,0.9056)}
		& \makecell{\textbf{1.7676}\\(0.4014,0.8618)}
		& \makecell{\textbf{1.7917}\\(0.4852,0.8904)}
		& \makecell{\textbf{1.7937}\\(0.5263,0.8848)}
		& \makecell{\textbf{1.7981}\\(0.4567,0.8411)}
		& \makecell{\textbf{1.8033}\\(0.5057,0.9108)} \\
		& Lap-GFT-I
		& 2.7946 & 2.7967 & 2.7988 & 2.8009
		& 1.7707 & 1.7712 & 1.7738 & 1.7798
		& 1.8130 & 1.8185 & 1.8252 & 1.8324 \\
		& Adj-GFT-I
		& 3.8521 & 3.8518 & 3.8515 & 3.8512
		& 3.8625 & 3.8685 & 3.8747 & 3.8811
		& 3.9079 & 3.9189 & 3.9301 & 3.9413 \\
		& Id-GFT-I
		& 4.0000 & 4.0000 & 4.0000 & 4.0000
		& 4.0000 & 4.0000 & 4.0000 & 4.0000
		& 4.0000 & 4.0000 & 4.0000 & 4.0000 \\
		& SLap-GFT-I
		& 3.9542 & 3.9549 & 3.9559 & 3.9571
		& 3.9672 & 3.9683 & 3.9695 & 3.9707
		& 3.9717 & 3.9726 & 3.9736 & 3.9746 \\
		\thickrule
	\end{tabular}
\end{table*}

\begin{table*}[htbp]
	\centering
	\tiny
	\caption{The Bandlimiting SNR and BAE for UGRM-GFT-II and Baseline Methods with Different Noise Levels $\sigma \in \{0.1, 0.2, 0.3, 0.4\}$ and Graph Connectivities (3-NN, 5-NN, 7-NN) on the SST, PM-25, and COVID Datasets}
	\label{tab4}
	\begin{tabular}{cccccccccccccc}
		\thickrule
		\multicolumn{2}{c}{}
		& \multicolumn{4}{c}{3-NN}
		& \multicolumn{4}{c}{5-NN}
		& \multicolumn{4}{c}{7-NN} \\
		\cmidrule(lr){3-6} \cmidrule(lr){7-10} \cmidrule(lr){11-14}
		\multicolumn{2}{c}{SST} & \multicolumn{12}{c}{} \\
		& & $\sigma=0.1$ & $\sigma=0.2$ & $\sigma=0.3$ & $\sigma=0.4$
		& $\sigma=0.1$ & $\sigma=0.2$ & $\sigma=0.3$ & $\sigma=0.4$
		& $\sigma=0.1$ & $\sigma=0.2$ & $\sigma=0.3$ & $\sigma=0.4$ \\
		\midrule
		ISNR &
		& 16.6280 & 10.6074 & 7.0856 & 4.5868
		& 16.6280 & 10.6074 & 7.0856 & 4.5868
		& 16.6280 & 10.6074 & 7.0856 & 4.5868 \\
		\multirow{5}{*}{SNR}
		& \makecell{UGRM-GFT-II\\$(\alpha,k)$}
		& \makecell{\textbf{9.3188}\\(0.4605,0.9498)}
		& \makecell{\textbf{9.1572}\\(0.4605,0.9498)}
		& \makecell{\textbf{8.9896}\\(0.1843,0.6556)}
		& \makecell{\textbf{8.6792}\\(0.2280,0.7005)}
		& \makecell{\textbf{9.5254}\\(0.3709,0.8545)}
		& \makecell{\textbf{9.3690}\\(0.2989,0.7556)}
		& \makecell{\textbf{9.1093}\\(0.4172,0.9289)}
		& \makecell{\textbf{8.5886}\\(0.1629,0.6289)}
		& \makecell{\textbf{9.3158}\\(0.4737,0.9742)}
		& \makecell{\textbf{9.4123}\\(0.2564,0.7070)}
		& \makecell{\textbf{8.8937}\\(0.2344,0.6620)}
		& \makecell{\textbf{8.8024}\\(0.2564,0.7070)} \\
		& Lap-GFT-II
		& 9.2691 & 9.1065 & 8.8442 & 8.4999
		& 9.2665 & 9.1040 & 8.8418 & 8.4976
		& 9.2660 & 9.1034 & 8.8413 & 8.4971 \\
		& Adj-GFT-II
		& 1.0546 & 1.0367 & 1.0071 & 0.9662
		& 1.0569 & 1.0390 & 1.0095 & 0.9685
		& 1.0579 & 1.0400 & 1.0104 & 0.9694 \\
		& Id-GFT-II
		& 0.5200 & 0.4882 & 0.4357 & 0.3633
		& 0.5200 & 0.4882 & 0.4357 & 0.3633
		& 0.5200 & 0.4882 & 0.4357 & 0.3633 \\
		& SLap-GFT-II
		& 3.5364 & 3.4890 & 3.4109 & 3.3036
		& 3.5359 & 3.4886 & 3.4104 & 3.3032
		& 3.5357 & 3.4883 & 3.4102 & 3.3029 \\
		\cmidrule(lr){1-14}
		\multirow{5}{*}{BAE}
		& \makecell{UGRM-GFT-II\\$(\alpha,k)$}
		& \makecell{\textbf{0.5190}\\(0.3096,0.7267)}
		& \makecell{\textbf{0.5319}\\(0.3255,0.7451)}
		& \makecell{\textbf{0.5517}\\(0.3255,0.7451)}
		& \makecell{\textbf{0.5888}\\(0.4277,0.8851)}
		& \makecell{\textbf{0.5190}\\(0.3097,0.7267)}
		& \makecell{\textbf{0.5319}\\(0.2047,0.6309)}
		& \makecell{\textbf{0.5517}\\(0.4871,0.9821)}
		& \makecell{\textbf{0.5888}\\(0.4277,0.8851)}
		& \makecell{\textbf{0.5191}\\(0.4871,0.9821)}
		& \makecell{\textbf{0.5319}\\(0.2581,0.6766)}
		& \makecell{\textbf{0.5517}\\(0.3255,0.7451)}
		& \makecell{\textbf{0.5888}\\(0.4277,0.8851)} \\
		& Lap-GFT-II
		& 0.5199 & 0.5331 & 0.5524 & 0.5921
		& 0.5199 & 0.5331 & 0.5524 & 0.5921
		& 0.5199 & 0.5331 & 0.5524 & 0.5921 \\
		& Adj-GFT-II
		& 0.9691 & 0.9691 & 0.9691 & 0.9745
		& 0.9691 & 0.9691 & 0.9691 & 0.9745
		& 0.9691 & 0.9691 & 0.9691 & 0.9745 \\
		& Id-GFT-II
		& 1.0000 & 1.0000 & 1.0020 & 1.0740
		& 1.0000 & 1.0000 & 1.0020 & 1.0740
		& 1.0000 & 1.0000 & 1.0020 & 1.0740 \\
		& SLap-GFT-II
		& 0.9313 & 0.9317 & 0.9333 & 0.9357
		& 0.9313 & 0.9317 & 0.9333 & 0.9357
		& 0.9313 & 0.9317 & 0.9333 & 0.9357 \\
		\doublerule
		
		\multicolumn{2}{c}{}
		& \multicolumn{4}{c}{3-NN}
		& \multicolumn{4}{c}{5-NN}
		& \multicolumn{4}{c}{7-NN} \\
		\cmidrule(lr){3-6} \cmidrule(lr){7-10} \cmidrule(lr){11-14}
		\multicolumn{2}{c}{PM-25} & \multicolumn{12}{c}{} \\
		& & $\sigma=0.1$ & $\sigma=0.2$ & $\sigma=0.3$ & $\sigma=0.4$
		& $\sigma=0.1$ & $\sigma=0.2$ & $\sigma=0.3$ & $\sigma=0.4$
		& $\sigma=0.1$ & $\sigma=0.2$ & $\sigma=0.3$ & $\sigma=0.4$ \\
		\midrule
		ISNR &
		& 12.6677 & 6.6471 & 3.1252 & 0.6265
		& 12.6677 & 6.6471 & 3.1252 & 0.6265
		& 12.6677 & 6.6471 & 3.1252 & 0.6265 \\
		\multirow{5}{*}{SNR}
		& \makecell{UGRM-GFT-II\\$(\alpha,k)$}
		& \makecell{\textbf{5.5244}\\(0.4473,0.9561)}
		& \makecell{\textbf{5.3757}\\(0.4691,0.9973)}
		& \makecell{\textbf{5.1350}\\(0.4691,0.9973)}
		& \makecell{\textbf{4.8199}\\(0.4691,0.9973)}
		& \makecell{\textbf{5.6750}\\(0.2381,0.6584)}
		& \makecell{\textbf{5.5102}\\(0.4871,0.9821)}
		& \makecell{\textbf{5.2610}\\(0.4590,0.9229)}
		& \makecell{\textbf{4.9239}\\(0.3885,0.8212)}
		& \makecell{\textbf{5.6418}\\(0.4902,0.9678)}
		& \makecell{\textbf{5.4709}\\(0.4539,0.9032)}
		& \makecell{\textbf{5.2001}\\(0.4815,0.9529)}
		& \makecell{\textbf{4.8492}\\(0.4483,0.8976)} \\
		& Lap-GFT-II
		& 5.1810 & 5.0350 & 4.8025 & 4.4974
		& 5.6697 & 5.5102 & 5.2567 & 4.9253
		& 5.6138 & 5.4483 & 5.1854 & 4.8420 \\
		& Adj-GFT-II
		& 4.1070 & 3.9976 & 3.8233 & 3.5919
		& 4.8246 & 4.7022 & 4.5076 & 4.2502
		& 4.7720 & 4.6501 & 4.4558 & 4.1987 \\
		& Id-GFT-II
		& 0.4704 & 0.3924 & 0.2655 & 0.0940
		& 0.4470 & 0.3713 & 0.2480 & 0.0813
		& 0.4470 & 0.3713 & 0.2480 & 0.0813 \\
		& SLap-GFT-II
		& 2.9952 & 2.8617 & 2.6489 & 2.3684
		& 1.6149 & 1.5215 & 1.3703 & 1.1672
		& 1.4988 & 1.4085 & 1.2622 & 1.0655 \\
		\cmidrule(lr){1-14}
		\multirow{5}{*}{BAE}
		&\makecell{UGRM-GFT-II\\$(\alpha,k)$}
		& \makecell{\textbf{1.1425}\\(0.3468,0.7088)}
		& \makecell{\textbf{1.1403}\\(0.5340,0.9731)}
		& \makecell{\textbf{1.1396}\\(0.1944,0.5927)}
		& \makecell{\textbf{1.1449}\\(0.3484,0.7113)}
		& \makecell{\textbf{1.1476}\\(0.3996,0.8355)}
		& \makecell{\textbf{1.1482}\\(0.3769,0.8009)}
		& \makecell{\textbf{1.1502}\\(0.3996,0.8355)}
		& \makecell{\textbf{1.1540}\\(0.3996,0.8355)}
		& \makecell{\textbf{1.0398}\\(0.4022,0.8312)}
		& \makecell{\textbf{1.0409}\\(0.4954,0.9764)}
		& \makecell{\textbf{1.0440}\\(0.3464,0.7582)}
		& \makecell{\textbf{1.0524}\\(0.4165,0.8461)} \\
		& Lap-GFT-II
		& 1.1834 & 1.1840 & 1.1845 & 1.1865
		& 1.1522 & 1.1533 & 1.1546 & 1.1583
		& 1.0424 & 1.0434 & 1.0465 & 1.0554 \\
		& Adj-GFT-II
		& 2.0000 & 2.0000 & 2.0000 & 2.0000
		& 2.0000 & 2.0000 & 2.0000 & 2.0000
		& 2.0000 & 2.0000 & 2.0000 & 2.0000 \\
		& Id-GFT-II
		& 2.0000 & 2.0000 & 2.0000 & 2.0000
		& 2.0000 & 2.0000 & 2.0000 & 2.0000
		& 2.0000 & 2.0000 & 2.0000 & 2.0000 \\
		& SLap-GFT-II
		& 1.9930 & 1.9929 & 1.9929 & 1.9928
		& 2.0000 & 2.0000 & 2.0000 & 2.0000
		& 2.0000 & 2.0000 & 2.0000 & 2.0000 \\
		\bottomrule
		\bottomrule
		
		\multicolumn{2}{c}{}
		& \multicolumn{4}{c}{3-NN}
		& \multicolumn{4}{c}{5-NN}
		& \multicolumn{4}{c}{7-NN} \\
		\cmidrule(lr){3-6} \cmidrule(lr){7-10} \cmidrule(lr){11-14}
		\multicolumn{2}{c}{COVID} & \multicolumn{12}{c}{} \\
		& & $\sigma=0.1$ & $\sigma=0.2$ & $\sigma=0.3$ & $\sigma=0.4$
		& $\sigma=0.1$ & $\sigma=0.2$ & $\sigma=0.3$ & $\sigma=0.4$
		& $\sigma=0.1$ & $\sigma=0.2$ & $\sigma=0.3$ & $\sigma=0.4$ \\
		\midrule
		ISNR &
		& 13.1879 & 7.1673 & 3.6454 & 1.1467
		& 13.1879 & 7.1673 & 3.6454 & 1.1467
		& 13.1879 & 7.1673 & 3.6454 & 1.1467 \\
		\multirow{5}{*}{SNR}
		& \makecell{UGRM-GFT-II\\$(\alpha,k)$}
		& \makecell{\textbf{3.1033}\\(0.2434,0.6749)}
		& \makecell{\textbf{3.0173}\\(0.2922,0.7300)}
		& \makecell{\textbf{2.8794}\\(0.2298,0.6636)}
		& \makecell{\textbf{2.6921}\\(0.3303,0.7712)}
		& \makecell{\textbf{4.2525}\\(0.2625,0.7730)}
		& \makecell{\textbf{4.1218}\\(0.2756,0.8021)}
		& \makecell{\textbf{3.9624}\\(0.2651,0.7816)}
		& \makecell{\textbf{3.7214}\\(0.3325,0.9844)}
		& \makecell{\textbf{5.8404}\\(0.4485,0.9505)}
		& \makecell{\textbf{5.6613}\\(0.3520,0.8025)}
		& \makecell{\textbf{5.3812}\\(0.3881,0.8550)}
		& \makecell{\textbf{5.0194}\\(0.3610,0.8185)} \\
		& Lap-GFT-II
		& 3.0755 & 2.9885 & 2.8458 & 2.6530
		& 4.0524 & 3.9348 & 3.7452 & 3.4930
		& 5.7838 & 5.6020 & 5.3174 & 4.9496 \\
		& Adj-GFT-II
		& 0.6073 & 0.5680 & 0.5026 & 0.4125
		& 0.5480 & 0.5045 & 0.4326 & 0.3337
		& 0.5056 & 0.4627 & 0.3917 & 0.2942 \\
		& Id-GFT-II
		& $-$0.0211 & $-$0.0837 & $-$0.1860 & $-$0.3251
		& $-$0.0206 & $-$0.0816 & $-$0.1814 & $-$0.3173
		& $-$0.0210 & $-$0.0832 & $-$0.1849 & $-$0.3233 \\
		& SLap-GFT-II
		& 0.2245 & 0.1654 & 0.0687 & $-$0.0631
		& 0.1879 & 0.1273 & 0.0282 & $-$0.1068
		& 0.1337 & 0.0727 & $-$0.0272 & $-$0.1633 \\
		\cmidrule(lr){1-14}
		\multirow{5}{*}{BAE}
		&\makecell{UGRM-GFT-II\\$(\alpha,k)$}
		& \makecell{\textbf{2.1579}\\(0.3486,0.8093)}
		& \makecell{\textbf{2.1590}\\(0.3486,0.8093)}
		& \makecell{\textbf{2.1604}\\(0.4544,0.9834)}
		& \makecell{\textbf{2.1624}\\(0.3708,0.8442)}
		& \makecell{\textbf{1.7163}\\(0.3897,0.8777)}
		& \makecell{\textbf{1.7276}\\(0.2647,0.7145)}
		& \makecell{\textbf{1.7394}\\(0.2880,0.7398)}
		& \makecell{\textbf{1.7514}\\(0.3373,0.7949)}
		& \makecell{\textbf{1.6761}\\(0.3897,0.8777)}
		& \makecell{\textbf{1.6824}\\(0.1380,0.5954)}
		& \makecell{\textbf{1.6894}\\(0.2776,0.7300)}
		& \makecell{\textbf{1.6968}\\(0.4247,0.9416)} \\
		& Lap-GFT-II
		& 2.2986 & 2.2971 & 2.2957 & 2.2942
		& 1.7727 & 1.7779 & 1.7854 & 1.7946
		& 1.7265 & 1.7291 & 1.7338 & 1.7389 \\
		& Adj-GFT-II
		& 3.9904 & 3.9903 & 3.9903 & 3.9902
		& 3.9624 & 3.9620 & 3.9623 & 3.9637
		& 3.9178 & 3.9197 & 3.9236 & 3.9282 \\
		& Id-GFT-II
		& 4.0000 & 4.0000 & 4.0000 & 4.0000
		& 4.0000 & 4.0000 & 4.0000 & 4.0000
		& 4.0000 & 4.0000 & 4.0000 & 4.0000 \\
		& SLap-GFT-II
		& 3.9620 & 3.9618 & 3.9616 & 3.9614
		& 3.9729 & 3.9728 & 3.9727 & 3.9726
		& 3.9735 & 3.9734 & 3.9733 & 3.9735 \\
		\thickrule
	\end{tabular}
\end{table*}

   \section{Conclusion and future work}
   This study proposed an SVD-based UGRM-GFT methodology for signal analysis on directed graphs and subsequently presented two variants for directed product graphs, UGRM-GFT-I and UGRM-GFT-II. This method employs a unified parameterized matrix representation, allowing for the flexible integration of various traditional graph representations and addressing the challenges of asymmetry in directed graphs via the numerically stable SVD. Theoretical analysis established approximation error bounds and characterized the spectral behavior of the UGRM-GFT with respect to its parameters, providing a solid basis for its application and interpretation. Experiments on real-world datasets demonstrated the proposed method's enhanced efficacy in denoising tasks. Compared to conventional techniques that utilize fixed Laplacian or adjacency matrices, our approach consistently achieved a higher SNR, and a lower BAE, largely due to its superior energy compaction capabilities.
   
   Future work could explore several directions. First, developing a principled, data-driven strategy for selecting the optimal UGRM parameters $(\alpha, k)$ beyond heuristic search would be valuable. Second, extending the UGRM framework to other graph operations, such as graph filtering and wavelet design, could further broaden its applicability. Finally, investigating the application of UGRM-GFT to other complex graph structures, such as time-varying or multi-layer graphs, represents a promising avenue for research.
\label{sec:chapter7} 

\appendices

\section{Proof of Theorem \ref{thm1}}
\label{app:A}

\textbf{Case $\boldsymbol{\beta < 0}$.} By \eqref{16}, we have
\begin{align}
	\|(\mathbf{P}^{\alpha,k})^T \mathbf{x}\|_2^2 
	&= \mathbf{x}^T \mathbf{U}^{\alpha,k} (\boldsymbol{\Sigma}^{\alpha,k})^2 (\mathbf{U}^{\alpha,k})^T \mathbf{x} \notag \\
	&= \sum_{t=0}^{N-1} (\sigma_t^{\alpha,k})^2 ((\mathbf{u}_t^{\alpha,k})^T \mathbf{x})^2 \notag \\
	&\geq (\sigma_{M-1}^{\alpha,k})^2 \sum_{t=M}^{N-1} ((\mathbf{u}_t^{\alpha,k})^T \mathbf{x})^2 
	\label{58}
\end{align}
and
\begin{align}
	\|\mathbf{P}^{\alpha,k}\mathbf{x}\|_2^2 
	&= \mathbf{x}^T \mathbf{V}^{\alpha,k} (\boldsymbol{\Sigma}^{\alpha,k})^2 (\mathbf{V}^{\alpha,k})^T \mathbf{x} \notag \\
	&\geq (\sigma_{M-1}^{\alpha,k})^2 \sum_{t=M}^{N-1} ((\mathbf{v}_t^{\alpha,k})^T \mathbf{x})^2. 
	\label{59}
\end{align}

From \eqref{19} and \eqref{28}, it follows that
\begin{equation}
	\begin{aligned}
		\|\mathbf{x} - \mathbf{x}_M^{\alpha,k}\|_2 
		&= \frac{1}{2} \left\| \sum_{t=M}^{N-1} \Big( \mathbf{u}_t^{\alpha,k} (\mathbf{u}_t^{\alpha,k})^T + \mathbf{v}_t^{\alpha,k} (\mathbf{v}_t^{\alpha,k})^T \Big) \mathbf{x} \right\|_2  \\
		&\leq \frac{1}{2} \!\left( \sum_{t=M}^{N-1} \left|(\mathbf{u}_t^{\alpha,k})^T \mathbf{x}\right|^2 \right)^{\!1/2} \!\! \\
		&+ \frac{1}{2} \!\left( \sum_{t=M}^{N-1} \left|(\mathbf{v}_t^{\alpha,k})^T \mathbf{x}\right|^2 \right)^{\!1/2}\!. 
	\end{aligned}
	\label{60}
\end{equation}

This together with (\ref{58}) and (\ref{59}) gives \eqref{29}.

\medskip
\textbf{Case $\boldsymbol{\beta \ge 0}$.}
The bound~\eqref{60} still holds by the same orthonormality 
argument. That is, since $\{\mathbf{u}_t^{\alpha,k}\}$ and 
$\{\mathbf{v}_t^{\alpha,k}\}$ are orthonormal bases of $\mathbb{R}^N$ 
regardless of the sign of $\beta$, we have
\begin{equation}
	\begin{aligned}
		\|\mathbf{x} - \mathbf{x}_M^{\alpha,k}\|_2
		&\leq
		\frac{1}{2}
		\left( \sum_{t=M}^{N-1} \left( (\mathbf{u}_t^{\alpha,k})^T \mathbf{x} \right)^2 \right)^{1/2} \\
		&\quad +
		\frac{1}{2}
		\left( \sum_{t=M}^{N-1} \left( (\mathbf{v}_t^{\alpha,k})^T \mathbf{x} \right)^2 \right)^{1/2}.
	\end{aligned}
	\label{61}
\end{equation}
Applying the operator consistency condition 
$(\mathbf{u}_t^{\alpha,k})^T\mathbf{x} \leq C_1\sigma_t^{\alpha,k}$ 
and $(\mathbf{v}_t^{\alpha,k})^T\mathbf{x} \leq C_2\sigma_t^{\alpha,k}$ 
for all $t \geq M$, we obtain
\begin{equation}
	\sum_{t=M}^{N-1} \left( (\mathbf{u}_t^{\alpha,k})^T \mathbf{x} \right)^2
	\leq C_1^2 \sum_{t=M}^{N-1} (\sigma_t^{\alpha,k})^2,
	\label{62}
\end{equation}
\begin{equation}
	\sum_{t=M}^{N-1} \left((\mathbf{v}_t^{\alpha,k})^T\mathbf{x} \right)^2
	\leq C_2^2\sum_{t=M}^{N-1}(\sigma_t^{\alpha,k})^2.
	\label{63}
\end{equation}
Substituting \eqref{62} and \eqref{63} into \eqref{60},
\begin{align}
	\|\mathbf{x} - \mathbf{x}_M^{\alpha,k}\|_2
	&\leq
	\frac{1}{2}\left(C_1^2\sum_{t=M}^{N-1}(\sigma_t^{\alpha,k})^2
	\right)^{1/2}
	+
	\frac{1}{2}\left(C_2^2\sum_{t=M}^{N-1}(\sigma_t^{\alpha,k})^2
	\right)^{1/2} \notag\\
	&= C\left(\sum_{t=M}^{N-1}(\sigma_t^{\alpha,k})^2\right)^{1/2},
\end{align}
which is exactly~\eqref{30}. \hfill$\square$

\section{Proof of Theorem \ref{thm2}}
\label{app:B}

\textbf{Case $\boldsymbol{\beta < 0}$.} By the SVD \eqref{31},
\begin{align}
	\|\mathbf{P}_{\boxtimes}^{\alpha,k}\mathbf{x}\|_2^2 
	&= \sum_{t=0}^{N-1} (\sigma_t^{\alpha,k})^2 \big((\mathbf{v}_t^{\alpha,k})^T \mathbf{x}\big)^2 \nonumber \\
	&\geq (\sigma_{M-1}^{\alpha,k})^2 \sum_{t=M}^{N-1} \big((\mathbf{v}_t^{\alpha,k})^T \mathbf{x}\big)^2 \label{65} 
\end{align}
and
\begin{align}
	\|(\mathbf{P}_{\boxtimes}^{\alpha,k})^T \mathbf{x}\|_2^2 
	&\geq (\sigma_{M-1}^{\alpha,k})^2 \sum_{t=M}^{N-1} \big((\mathbf{u}_t^{\alpha,k})^T \mathbf{x}\big)^2. \label{66}
\end{align}

From \eqref{37} and \eqref{38},
\begin{align}
	\|\mathbf{x} - \mathbf{x}_{M,\boxtimes}^{\alpha,k}\|_2 
	&\leq \frac{1}{2} \!\left( \sum_{t=M}^{N-1} \big((\mathbf{u}_t^{\alpha,k})^T \mathbf{x}\big)^2 \right)^{\!1/2} \nonumber \\
	&\quad + \frac{1}{2} \!\left( \sum_{t=M}^{N-1} \big((\mathbf{v}_t^{\alpha,k})^T \mathbf{x}\big)^2 \right)^{\!1/2}\!. \label{67}
\end{align}

Substituting \eqref{65}--\eqref{66} into \eqref{67} yields the first inequality. By the triangle inequality and $\mathbf{P}_{\boxtimes}^{\alpha,k} = \mathbf{P}_1^{\alpha,k} \otimes \mathbf{I}_{N_2} + \mathbf{I}_{N_1} \otimes \mathbf{P}_2^{\alpha,k}$,
\begin{equation}
		\|\mathbf{P}_{\boxtimes}^{\alpha,k}\mathbf{x}\|_2 
		\leq \|(\mathbf{P}_1^{\alpha,k} \otimes \mathbf{I}_{N_2})\mathbf{x}\|_2 + \|(\mathbf{I}_{N_1} \otimes \mathbf{P}_2^{\alpha,k})\mathbf{x}\|_2, 
	    \label{68}
\end{equation}
\begin{equation}
		\|(\mathbf{P}_{\boxtimes}^{\alpha,k})^T \mathbf{x}\|_2 
		\leq \|((\mathbf{P}_1^{\alpha,k})^T \!\otimes \mathbf{I}_{N_2})\mathbf{x}\|_2 + \|(\mathbf{I}_{N_1} \!\otimes (\mathbf{P}_2^{\alpha,k})^T)\mathbf{x}\|_2, 
\end{equation}
which gives the second inequality.

\medskip
\textbf{Case $\boldsymbol{\beta \ge 0}$.}
The error decomposition~\eqref{67} holds independently of the 
sign of $\beta$, since it follows solely from the perfect 
reconstruction identity~\eqref{37} and the orthonormality of 
$\{\mathbf{u}_t^{\alpha,k}\}$ and $\{\mathbf{v}_t^{\alpha,k}\}$:
\begin{equation}
	\begin{aligned}
	\|\mathbf{x} - \mathbf{x}_{M,\boxtimes}^{\alpha,k}\|_2
	&\leq
	\frac{1}{2}
	\left(\sum_{t=M}^{N-1}\left((\mathbf{u}_t^{\alpha,k})^T\mathbf{x}
	\right)^2\right)^{1/2} \\
	&+
	\frac{1}{2}
	\left(\sum_{t=M}^{N-1}\left((\mathbf{v}_t^{\alpha,k})^T\mathbf{x}
	\right)^2\right)^{1/2}.
	\end{aligned}
	\label{70}
\end{equation}
Applying the operator consistency condition 
$(\mathbf{u}_t^{\alpha,k})^T\mathbf{x} \leq C_1\sigma_t^{\alpha,k}$ 
and $(\mathbf{v}_t^{\alpha,k})^T\mathbf{x} \leq C_2\sigma_t^{\alpha,k}$ 
for all $t \geq M$,
\begin{equation}
	\begin{aligned}
	&\sum_{t=M}^{N-1}
	\left((\mathbf{u}_t^{\alpha,k})^T\mathbf{x}\right)^2
	\leq C_1^2\sum_{t=M}^{N-1}(\sigma_t^{\alpha,k})^2,
	\quad \\
	&\sum_{t=M}^{N-1}
	\left((\mathbf{v}_t^{\alpha,k})^T\mathbf{x}\right)^2
	\leq C_2^2\sum_{t=M}^{N-1}(\sigma_t^{\alpha,k})^2.
	\end{aligned}
	\label{71}
\end{equation}
Substituting these into \eqref{70},
\begin{align}
	\|\mathbf{x} - \mathbf{x}_{M,\boxtimes}^{\alpha,k}\|_2
	&\leq
	\frac{1}{2}\left(C_1^2\sum_{t=M}^{N-1}(\sigma_t^{\alpha,k})^2
	\right)^{1/2} \\
	&+
	\frac{1}{2}\left(C_2^2\sum_{t=M}^{N-1}(\sigma_t^{\alpha,k})^2
	\right)^{1/2} \notag\\
	&= C\left(\sum_{t=M}^{N-1}(\sigma_t^{\alpha,k})^2\right)^{1/2},
\end{align}
which is~\eqref{40}. \hfill$\square$

\section{Proof of Theorem \ref{thm3}}
\label{app:C}

\textbf{Case $\boldsymbol{\beta < 0}$.} By \eqref{41}, we have
\begin{align}
	\|(\mathbf{P}_1^{\alpha,k} \otimes \mathbf{I}_{N_2})\mathbf{x}\|_2^2 
	&= \sum_{i=0}^{N_1-1} \sum_{j=0}^{N_2-1} (\sigma_{1,i}^{\alpha,k})^2 \big((\mathbf{v}_{i,j}^{\alpha,k})^T \mathbf{x}\big)^2 \label{73}
\end{align}
and
\begin{align}
	\|(\mathbf{I}_{N_1} \otimes \mathbf{P}_2^{\alpha,k})\mathbf{x}\|_2^2 
	&= \sum_{i=0}^{N_1-1} \sum_{j=0}^{N_2-1} (\sigma_{2,j}^{\alpha,k})^2 \big((\mathbf{v}_{i,j}^{\alpha,k})^T \mathbf{x}\big)^2. \label{74}
\end{align}

This implies that
\begin{align}
	&\big(\|(\mathbf{P}_1^{\alpha,k} \otimes \mathbf{I}_{N_2})\mathbf{x}\|_2 + \|(\mathbf{I}_{N_1} \otimes \mathbf{P}_2^{\alpha,k})\mathbf{x}\|_2\big)^2 \notag \\
	&\quad \geq \sum_{i=0}^{N_1-1} \sum_{j=0}^{N_2-1} (\sigma_{1,i}^{\alpha,k} + \sigma_{2,j}^{\alpha,k})^2 \big((\mathbf{v}_{i,j}^{\alpha,k})^T \mathbf{x}\big)^2 \notag \\
	&\quad \geq (\mu_{M-1}^{\alpha,k})^2 \sum_{(i,j) \notin \mathcal{S}_M} \big((\mathbf{v}_{i,j}^{\alpha,k})^T \mathbf{x}\big)^2. \label{75}
\end{align}

Similarly, we obtain from \eqref{41} that
\begin{align}
	&\big(\|((\mathbf{P}_1^{\alpha,k})^T \otimes \mathbf{I}_{N_2})\mathbf{x}\|_2 + \|(\mathbf{I}_{N_1} \otimes (\mathbf{P}_2^{\alpha,k})^T)\mathbf{x}\|_2\big)^2 \notag \\
	&\quad \geq (\mu_{M-1}^{\alpha,k})^2 \sum_{(i,j) \notin \mathcal{S}_M} \big((\mathbf{u}_{i,j}^{\alpha,k})^T \mathbf{x}\big)^2. \label{76}
\end{align}

From \eqref{45} and \eqref{46} it follows that
\begin{align}
	\|\mathbf{x} - \mathbf{x}_{M,\otimes}^{\alpha,k}\|_2 
	&\leq \bigg(\sum_{(i,j) \notin \mathcal{S}_M} \big((\mathbf{u}_{i,j}^{\alpha,k})^T \mathbf{x}\big)^2 \bigg)^{1/2} \notag \\
	&\quad + \bigg(\sum_{(i,j) \notin \mathcal{S}_M} \big((\mathbf{v}_{i,j}^{\alpha,k})^T \mathbf{x}\big)^2 \bigg)^{1/2}. \label{77}
\end{align}

Substituting \eqref{75}--\eqref{76} into \eqref{77} gives \eqref{49}.

\medskip
\textbf{Case $\boldsymbol{\beta \ge 0}$.}
From \eqref{77}, which holds by the perfect reconstruction 
identity~\eqref{45}--\eqref{46} and the orthonormality of 
$\{\mathbf{u}_{i,j}^{\alpha,k}\}$ and $\{\mathbf{v}_{i,j}^{\alpha,k}\}$:
\begin{equation}
	\begin{aligned}
		\|\mathbf{x} - \mathbf{x}_{M,\otimes}^{\alpha,k}\|_2
		&\leq  \left( \sum_{(i,j)\notin S_M} \left((\mathbf{u}_{i,j}^{\alpha,k})^T\mathbf{x}\right)^2 \right)^{1/2} \\
		&\quad + \left( \sum_{(i,j)\notin S_M} \left((\mathbf{v}_{i,j}^{\alpha,k})^T\mathbf{x}\right)^2 \right)^{1/2}.
	\end{aligned}
	\label{78}
\end{equation}
Applying the operator consistency condition: for all $(i,j)$ 
corresponding to index $t \geq M$ with combined singular value 
$\mu_t^{\alpha,k} = \sigma_{1,i}^{\alpha,k} + \sigma_{2,j}^{\alpha,k}$,
\begin{equation}
	(\mathbf{u}_{i,j}^{\alpha,k})^T\mathbf{x}
	\leq C_1\mu_t^{\alpha,k},
	\qquad
	(\mathbf{v}_{i,j}^{\alpha,k})^T\mathbf{x} 
	\leq C_2\mu_t^{\alpha,k},
\end{equation}
we obtain
\begin{equation}
	\begin{aligned}
	&\sum_{(i,j)\notin S_M}((\mathbf{u}_{i,j}^{\alpha,k})^T\mathbf{x})^2
	\leq C_1^2\sum_{t=M}^{N-1}(\mu_t^{\alpha,k})^2, \\
	&\sum_{(i,j)\notin S_M}
	((\mathbf{v}_{i,j}^{\alpha,k})^T\mathbf{x})^2
	\leq C_2^2\sum_{t=M}^{N-1}(\mu_t^{\alpha,k})^2.
	\end{aligned}
\end{equation}
Thus,~\eqref{78} becomes
\begin{align}
	\|\mathbf{x} - \mathbf{x}_{M,\otimes}^{\alpha,k}\|_2
	&\leq
	\left(C_1^2\sum_{t=M}^{N-1}(\mu_t^{\alpha,k})^2\right)^{1/2}\\
	&+
	\left(C_2^2\sum_{t=M}^{N-1}(\mu_t^{\alpha,k})^2\right)^{1/2} \notag\\
	&= C\left(\sum_{t=M}^{N-1}(\mu_t^{\alpha,k})^2\right)^{1/2},
\end{align}
which completes the proof. \hfill$\square$

\end{document}